\documentclass[]{aa}
\usepackage{amsmath}
\usepackage{longtable}
\usepackage{natbib}
\usepackage{graphicx}
\usepackage{txfonts}
\bibpunct{(}{)}{;}{a}{}{,}

\begin{document}

\title{A brown dwarf desert for intermediate mass stars in Sco~OB2?\thanks{Based on observations collected at the European Southern Observatory, Chile. Program 073.D-0354(A)}}

 \author{M.B.N. Kouwenhoven\inst{1}
          \and
          A.G.A. Brown\inst{2}
          \and
          L. Kaper\inst{1}
          }

 \offprints{M.B.N. Kouwenhoven \email{t.kouwenhoven@sheffield.ac.uk},  
 current address: Department of Physics and Astronomy, Hicks Building,
 Hounsfield Road, Sheffield S3~7RH, United Kingdom}

 \institute{Astronomical Institute `Anton Pannekoek',
   University of Amsterdam,
   Kruislaan 403, 1098 SJ Amsterdam, The Netherlands
   \\\email{kouwenho@science.uva.nl, lexk@science.uva.nl}
   \and
   Leiden Observatory, University of Leiden,
   P.O. Box 9513, 2300 RA
   Leiden, The Netherlands \\\email{brown@strw.leidenuniv.nl} }

 \date{Received ---; accepted ---}

\authorrunning{Kouwenhoven et al.}


\abstract{

We present $JHK_S$ observations of 22 intermediate-mass stars in the
Scorpius-Centaurus OB~association, obtained with the NAOS/CONICA system at the
ESO Very Large Telescope. This survey was performed to determine the status of (sub)stellar candidate companions of Sco~OB2 member stars of spectral type A and
late-B. The distinction between companions and background stars 
is made on the basis of a comparison to isochrones and additional statistical arguments. We are sensitive to companions with 
an angular separation of $0.1''-11''$ ($13-1430$~AU) and the detection limit is $K_S=17$~mag.
 We detect 62 stellar components of which 18 turn out to be physical companions, 11~candidate companions, and 33~background stars. Three of the 18 confirmed companions were previously undocumented
as such. The companion masses are in the range $0.03~{\rm M}_\odot
\leq M \leq 1.19~{\rm M}_\odot$, corresponding to mass ratios $0.06 \leq q \leq
0.55$.  
We include in our sample a subset of 9~targets with multi-color ADONIS observations from \cite{kouwenhoven2005}. In the ADONIS survey secondaries with $K_S < 12$~mag were classified as companions; those with $K_S > 12$~mag as background stars. The multi-color analysis in this paper demonstrates that the simple $K_S=12$~mag criterion correctly classifies the secondaries in $\sim 80\%$ of the cases. We reanalyse the total sample (i.e. NAOS/CONICA and ADONIS) and conclude that of the 176~secondaries, 25~are physical companions, 55~are candidate companions, and 96~are background stars. Although we are sensitive (and complete) to brown dwarf companions as faint as $K_S=14$~mag in the semi-major axis range $130-520$~AU, we detect only one, corresponding to a brown dwarf companion fraction of $0.5 \pm 0.5\%$ ($M \ga 30~{\rm M_J}$). 
However, the number of brown dwarfs is consistent with an extrapolation of the (stellar) companion mass distribution into the brown dwarf regime. 
This indicates that the physical mechanism for the formation of brown dwarf companions around intermediate mass stars is similar to that of stellar companions, and that the embryo ejection mechanism does not need to be invoked in order to explain the small number of brown dwarf companions among intermediate mass stars in the Sco~OB2 association.


}

\maketitle


\section{Introduction}

The predominance of star formation in binary or multiple systems inside
stellar clusters makes the binarity and multiplicity of newly born
stars one of the most sensitive probes of the process of star and star
cluster formation \citep[see][ and references
therein]{blaauw1991}. Ideally one would like to have detailed
knowledge of the binary population at the time that the stars are being
formed. However, this is difficult to achieve in practice and therefore we have
embarked on a project to characterize the observationally better accessible
``primordial binary population'', which is defined as {\em the
population of binaries as established just after the gas has been removed from
the forming system, i.e., when the stars can no longer accrete gas from their
surroundings} \citep{kouwenhoven2005}. We chose to focus our efforts on the
accurate characterization of the binary population in nearby OB
associations. The youth and low stellar density of OB~associations ensure that their
binary population is very similar to the primordial binary population. We
refer to \cite{kouwenhoven2005} and \cite{kouwenhoventhesis} for a more extensive discussion and
motivation of this project.

\begin{table}[btp]
  \setlength{\tabcolsep}{0.7\tabcolsep}
  \begin{tabular}{l cc cccc cc c}
    \hline
          & $D$  & Age  & $S$ & $B$ & $T$ & $>3$ & $F_{\rm M}$ & $F_{\rm NS}$ & $F_{\rm C}$ \\
          & (pc) &(Myr) & \\
    \hline
    US    & 145  & 5--6     & 64    & 44    & 8     & 3     & 0.46  & 0.67  & 0.61\\
    UCL   & 140  & 15--22   & 132   & 65    & 19    & 4     & 0.40  & 0.61  & 0.52\\
    LCC   & 118  & 17--23   & 112   & 57    & 9     & 1     & 0.37  & 0.56  & 0.44\\
    \hline
    all   &      &          & 308   & 166   & 36    & 8     & 0.41  & 0.61  & 0.51\\
    \hline
  \end{tabular}
  \caption{Multiplicity among \textit{Hipparcos} members of the three subgroups of Sco~OB2. 
    The columns
    show the subgroup name (Upper Scorpius; Upper Centaurus Lupus; Lower
    Centaurus Crux), the distance \citep[see][]{dezeeuw1999}, the age
    (\cite{degeus1989,preibisch2002} for US; \cite{mamajek2002} for UCL and LCC), the number
    of known single stars, binary stars, triple systems and $N>3$ systems, and
    the binary statistics (see \S~\ref{section: binarystatistics}), after inclusion of the
    new results presented in this paper. \label{table: statistics} }
\end{table}

Our initial efforts are concentrated on the Sco~OB2 association. Sco~OB2 consists of the three
subgroups Upper Scorpius (US), Upper Centaurus Lupus (UCL), and Lower Centaurus Crux (LCC). The properties of the subgroups are listed in Table~\ref{table: statistics}. 
Its stellar
population is accurately known down to late A-stars thanks to the
\textit{Hipparcos} catalogue \citep{dezeeuw1999}, and extensive
literature data is available on its binary population \citep{brown2001}. In addition
Sco~OB2 has recently been the target of an adaptive optics survey of its
\textit{Hipparcos} B-star members \citep{shatsky2002}. 
We have conducted our own adaptive optics survey of 199 A-type
and late-B type stars in this association \citep{kouwenhoven2005} using 
the ADONIS instrument, which was mounted on the ESO 3.6~meter telescope at La~Silla, Chile. We performed these observations in the $K_S$-band (and for a subset of the targets additionally in the $J$ and $H$ band). We detected 151 stellar
components other than the target stars and used a simple brightness criterion
to separate background stars\footnote{When mentioning ``background star'',
we refer to any stellar object that does not belong to the system, including
foreground stars.} from physical companions. All
components fainter than $K_S=12$~mag were considered background stars; all brighter components were identified as candidate companion stars \citep[see also][]{shatsky2002}. Of the 74
candidate physical companions 33 were known already and 41 were new candidate companions.

In examining the binary properties of our sample of A and late B-stars we
noticed that at small angular separations ($\leq 4$~arcsec) no companions
fainter than $K_S \approx 12$~mag are present, assuming that the sources
fainter than $K_S \approx 14$~mag are background stars (as we had no information on their colors). The absence of companions with $K_S > 12$~mag and $\rho < 4''$ is clearly visible in Figure~3 of \cite{kouwenhoven2005}. 
This result implies that A and B stars do not
have close companions with masses less than about 0.08~M$_{\odot}$, unless the
assumed background stars {\em are} physical companions. In the latter case the
close faint sources would be brown dwarfs \cite[which are known to be present
in Sco~OB2; see][]{martin2004} and a gap would exist in the companion mass
distribution. In either case a peculiar feature would be present in the mass
distribution of companions which has to be explained by the binary
formation history.

We decided to carry out follow-up multi-color observations in order (1) to determine the reliability of our $K_S=12$~mag criterion to separate companions and background stars, (2) to investigate the potential gap or lower limit of the companion mass distribution, and (3) to search for additional close and/or faint companions.

These follow-up near-infrared observations were conducted with NAOS/CONICA (NACO) on the ESO Very Large Telescope at Paranal, Chile. We obtained $JHK_S$ photometric observations of 22~A and late-B members in Sco~OB2 and their secondaries\footnote{We use the term ``secondary'' for any stellar component in the field near the target star. A secondary can be a companion star or a background star.}. In Section~\ref{section: observationsanddatareduction} we describe our NACO sample, the observations, the data reduction procedures, and photometric accuracy of the observations. 
In Section~\ref{section: detectionlimits} we describe the detection limit and completeness limit of the ADONIS and NACO observations. 
In Section~\ref{section: status} we determine the status (companion or background star) of the secondaries with multi-color observations. We perform the analysis for the secondaries around the 22~NACO targets, and for those around the 9~targets in the ADONIS sample for which we have multi-color observations. In Section~\ref{section: status} we also analyze the background star statistics, and we evaluate the accuracy of the $K_S=12$~mag separation criterion. In Section~\ref{section: massfunction} we derive for each companion its mass and mass ratio.
In Section~\ref{section: gap} we discuss the lack of brown dwarf companions with separations between $1''$ and $4''$ (130$-$520~AU) in our sample, and discuss whether or not the brown dwarf desert exists for A~and late-B type members of Sco~OB2. 
Finally, we present updated binary statistics of the Sco~OB2 association in Section~\ref{section: binarystatistics} and summarize our results in Section~\ref{section: conclusion}.


\section{Observations and data reduction} \label{section: observationsanddatareduction}

\subsection{Definition of the NACO sample}\label{section: sample}

\begin{table}[tbp]
  \begin{tabular}{|cc|cc|cll|}
    \hline
    HIP \#         &    HD \#         & $\pi$ & $\sigma_\pi$ & $K_S$ & Type & Group \\
       &          & (mas) & (mas) & (mag) &  & \\
\hline
\multicolumn{7}{|l|}{NAOS/CONICA targets} \\ 
\hline  
59502       &   106036   &   10.26   &   0.49   &   6.87   &   A2V   &   LCC   \\
60851       &   108501   &   9.63   &   0.50   &   6.06   &   A0Vn   &   LCC   \\
61265       &   109197   &   7.50   &   0.48   &   7.46   &   A2V   &   LCC   \\
62026       &   110461   &   9.20   &   0.45   &   6.31   &   B9V   &   LCC   \\
63204       &   112381   &   9.07   &   0.49   &   6.78   &   A0p   &   LCC   \\
67260       &   119884   &   8.15   &   0.49   &   6.98   &   A0V   &   LCC   \\
67919       &   121040   &   9.64   &   0.49   &   6.59   &   A9V   &   LCC   \\
68532       &   122259   &   8.09   &   0.43   &   7.02   &   A3\,IV/V   &   UCL   \\
69113       &   123445   &   5.92   &   0.41   &   6.37   &   B9\,V   &   UCL   \\
73937       &   133652   &   8.17   &   0.46   &   6.23   &   Ap\,Si   &   UCL   \\
78968       &   144586   &   5.87   &   0.53   &   7.42   &   B9\,V   &   US    \\
79098       &   144844   &   7.28   &   0.45   &   5.69   &   B9\,V   &   US    \\
79410       &   145554   &   7.00   &   0.52   &   7.09   &   B9\,V   &   US    \\
79739       &   146285   &   6.79   &   0.52   &   7.08   &   B8\,V   &   US    \\
79771       &   146331   &   6.86   &   0.51   &   7.10   &   B9\,V   &   US    \\
80142       &   147001   &   5.82   &   0.44   &   6.66   &   B7\,V   &   UCL   \\
80474       &   147932   &   7.20   &   0.51   &   5.80   &   B5\,V   &   US    \\
80799       &   148562   &   7.91   &   0.52   &   7.45   &   A2\,V   &   US    \\
80896       &   148716   &   7.77   &   0.57   &   7.44   &   F3\,V   &   US    \\
81949       &   150645   &   6.21   &   0.52   &   7.33   &   A3\,V   &   UCL   \\
81972       &   150742   &   5.36   &   0.40   &   5.87   &   B3\,V   &   UCL   \\
83542       &   154117   &   5.00   &   0.53   &   5.38   &   G8/K0\,III   &   US    \\
\hline  \multicolumn{7}{|l|}{ADONIS multi-color subset} \\ 
\hline  
53701       &   95324   &   7.93   &   0.58   &   6.48   &   B8\,IV   &   LCC   \\
76071       &   138343   &   5.96   &   0.56   &   7.06   &   B9\,V   &   US    \\
77911       &   142315   &   6.87   &   0.49   &   6.68   &   B9\,V   &   US    \\
78530       &   143567   &   7.11   &   0.48   &   6.87   &   B9\,V   &   US    \\
78809       &   144175   &   7.20   &   0.51   &   7.51   &   B9\,V   &   US    \\
78956       &   144569   &   5.55   &   0.50   &   7.57   &   B9.5\,V   &   US    \\
79124       &   144925   &   6.41   &   0.53   &   7.13   &   A0\,V   &   US    \\
79156       &   144981   &   6.21   &   0.53   &   7.61   &   A0\,V   &   US    \\
80238       &   147432   &   7.64   &   0.68   &   7.34   &   A1\,III/IV   &   US    \\

    \hline
  \end{tabular}
  \caption{We have obtained follow-up multi-color observations with NACO for 22~targets in the Sco~OB2 association. We include in our analysis 9~targets with multi-color observations in the ADONIS sample. All targets listed above are known to have secondaries in the ADONIS survey.  The table lists for each star the parallax and error \cite[taken from][]{debruijne1999}, the $K_S$ magnitude, and the spectral type of the primary star. The $K_S$ magnitudes are those derived in this paper for the 22~stars observed with NACO, and are taken from \cite{kouwenhoven2005} for the other nine stars. The last column shows the subgroup membership of each star (US~= Upper Scorpius; UCL~= Upper Centaurus Lupus; LCC~= Lower Centaurus Crux), taken from
    \cite{dezeeuw1999}. \label{table: sample}}
\end{table}

The major goals of our NACO follow-up observations are to determine the validity of the $K_S=12$~mag criterion that we used to separate companions and background stars in our ADONIS sample \citep{kouwenhoven2005}, to study the companion mass distribution near the stellar-substellar boundary, and to search for additional faint and/or close companions.

Our NACO sample consists of 22~member stars (listed in Table~\ref{table: sample}) in the Sco~OB2 association: 10~of spectral type B, 10~of spectral type~A, and one each of spectral type~F and~G. The targets are more or less equally distributed over the three subgroups of Sco~OB2: 9~in US, 6~in UCL, and 7~in LCC. All 22~targets are known to have secondaries in the ADONIS survey.

We included in our sample all seven target stars with faint ($K_S > 14$~mag) and close ($\leq 4$~arcsec) candidate background stars: HIP61265, HIP67260, HIP73937, HIP78968, HIP79098, HIP79410, and HIP81949. The other 15~targets all have candidate companion stars, for which we will use the multi-color data to further study their nature. Priority was given to target stars with multiple secondaries (candidate companions and candidate background stars) and targets close to the Galactic plane (HIP59502, HIP60851, HIP80142, and HIP81972) because of the larger probability of finding background stars.

There are 9~targets in the ADONIS dataset with (photometric) multi-color observations. These targets are also listed in Table~\ref{table: sample} and all have secondaries. \cite{kouwenhoven2005} use only the $K_S$ magnitude to determine the status of a secondary, including the secondaries around the 9~targets with $JHK_S$ observations. Later, in Section~\ref{section: status}, we will combine the data of the 22~NACO targets and the 9~ADONIS targets with multi-color observations, and determine the status of the secondaries of these 31~targets using their $JHK_S$ magnitudes. In the remaining part of Section~\ref{section: observationsanddatareduction} we describe the NACO observations, data reduction procedures, and photometric accuracy.

\begin{figure*}[btp]
  \begin{tabular}{ccc}
    \includegraphics[width=0.32\textwidth,height=!]{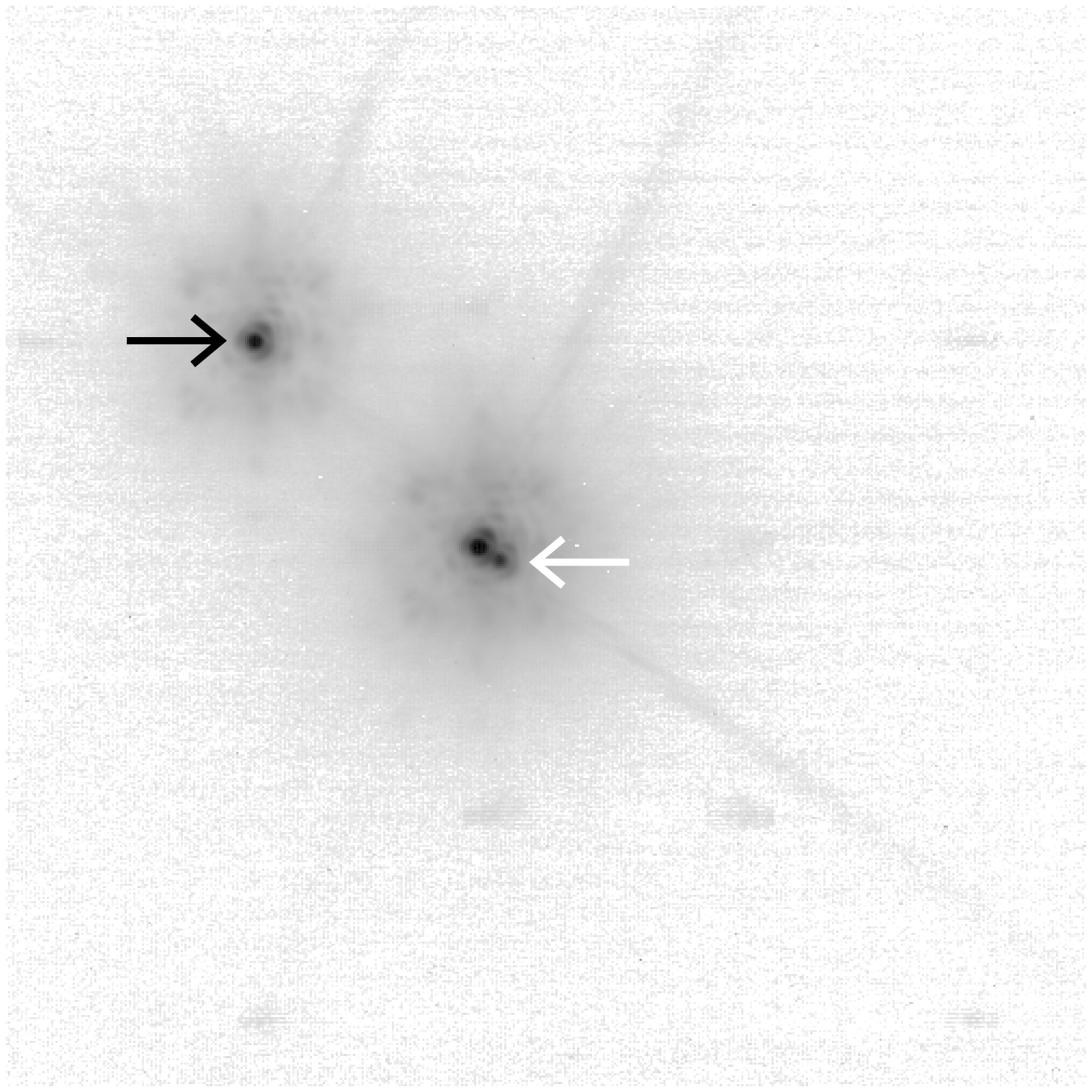} &
    \includegraphics[width=0.32\textwidth,height=!]{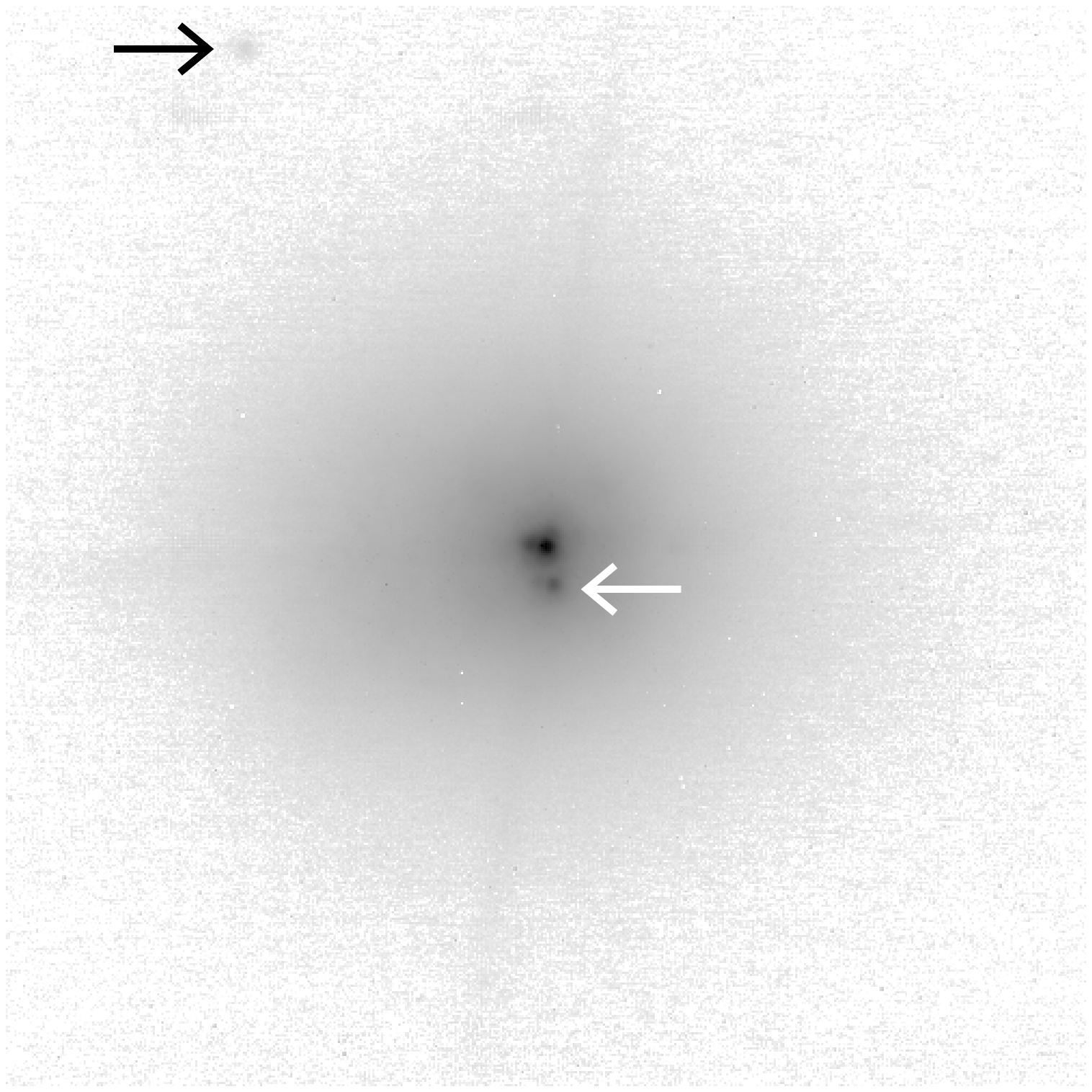} &
    \includegraphics[width=0.32\textwidth,height=!]{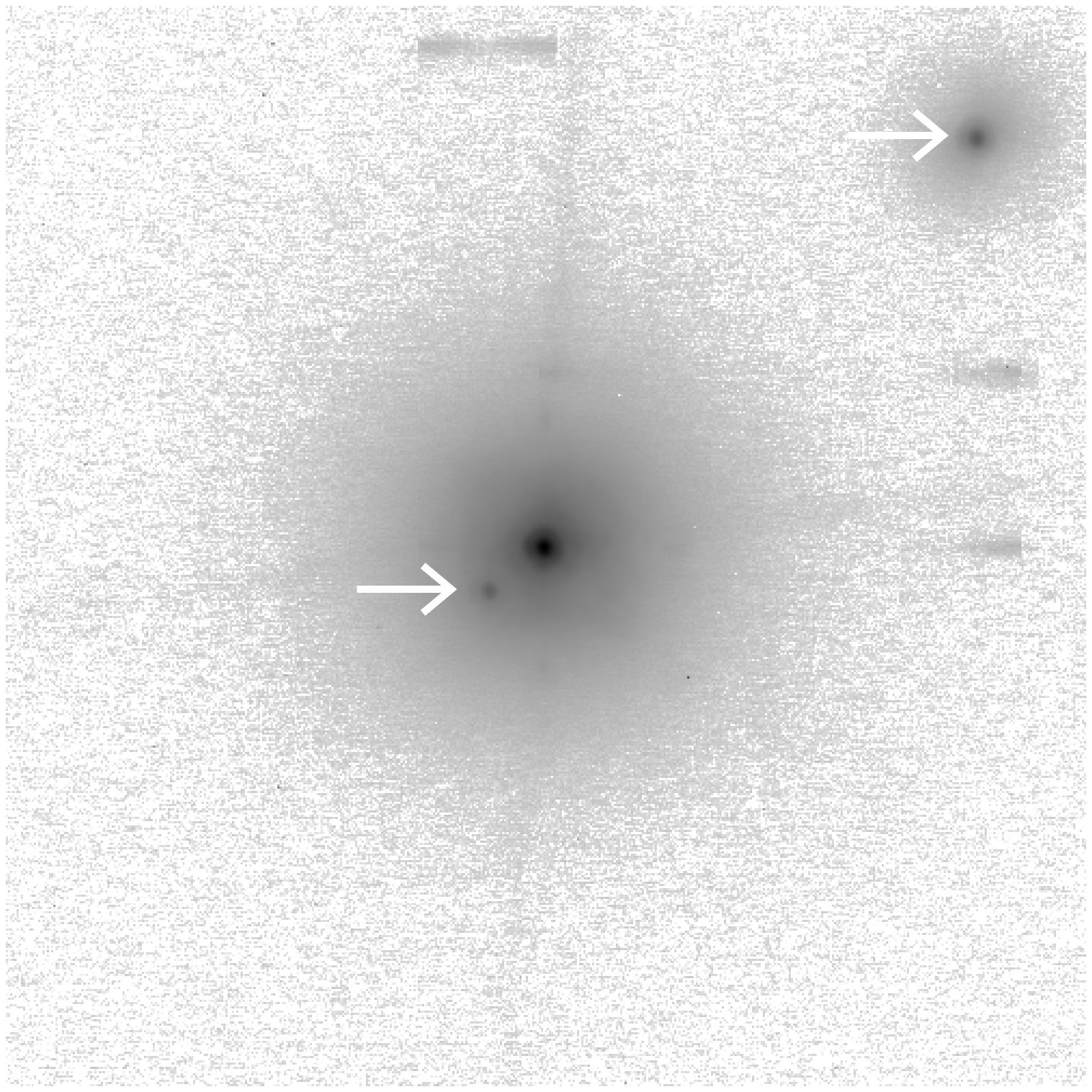} \\
  \end{tabular}
  \caption{With our NACO survey we find three close companions (shown in the
    figure),
    which were not detected in our ADONIS survey. The panels
    ($6.6'' \times 6.6'')$ are centered on the primary
    stars. {\em Left:} The binary HIP63204 in $K_S$, with a companion
    at angular separation $0.15''$ and a background star 
    at angular separation $1.87''$. {\em
    Middle:} The binary HIP73937 in $K_S$, with a close companion at $\rho=0.24''$ and a 
    background star at
    $\rho=3.56''$. {\em Right:} HIP79771 in $K_S$, with two companion stars at
    $\rho=0.44''$ and $\rho=3.67''$. Companions are indicated with white
    arrows and background stars with black arrows.
    Several artifacts are visible in the fields of HIP63204 and HIP79771, which can easily be recognized as such. The panels show a subset of the total field of
    view for each observation, which is $14'' \times 14''$. For these three targets we observe no
    stellar components other than those shown in the panels.  
    \label{figure: triples}}     
\end{figure*}

\subsection{NACO observations} \label{section: observations}

The observations were performed using the NAOS/CONICA system, consisting of
the near-infrared camera CONICA \citep{lenzen1998} and the adaptive optics
system NAOS \citep{rousset2000}. NAOS/CONICA is installed at the Nasmyth~B
focus of UT4 at the ESO Very Large Telescope on Paranal, Chile. The
observations were carried out in Service~Mode on the nights of April~6,
April~28-30, May~4-5, June~8, June~19, June~25, June~27-28, July~3, July~24,
and September~10, 2004. Some representative images are shown in Figure~\ref{figure: triples}.

The targets were imaged using the S13 camera, which has a pixel scale of
13.27~mas/pixel, and a field of view of $14''\times 14''$. The CONICA detector
was an Alladin~2 array in the period April~6 to May~5, 2004. The detector was
replaced by an Alladin~3 array in May~2004, which was used for the remaining
observations. We used the readout~mode Double\_RdRstRd and the detector mode
HighDynamic. For both detectors, the rms readout noise was $46.2\ e^-$ and the
gain was $\approx 11 e^-$/ADU. The full-well capacity of the Alladin~2 array
is 4300~ADU, with a linearity limit at about 50\% of this value. For the
Alladin~3 array the full-well capacity is 15\,000~ADU, with the linearity limit
at about two-thirds of this value.

Each observation block corresponding to a science target includes six
observations. The object is observed with the three broad band filters $J$
($1.253~\mu$m), $H$ ($1.643~\mu$m), and $K_S$ ($2.154~\mu$m). Since our
targets are bright, several of them will saturate the detector, even with the
shortest detector integration time. For this reason we also obtained
measurements in $J$, $H$, and $K_S$ with the short-wavelength neutral density
filter (hereafter NDF). The NDF transmissivity is about 1.4\% in the
near-infrared. The observations {\em with} NDF allow us to study the primary
star and to obtain an accurate point spread function (PSF), while the observations {\em without}
NDF allow us to analyze the faint companions in detail. In order to
characterize the attenuation of the NDF we observed the standard stars GSPC~S273-E and
GSPC~S708-D. These are LCO/Palomar NICMOS Photometric Standards
\citep{persson1998}. The near-infrared magnitudes of these stars are
consistent with spectral types G8V and G1V, respectively. By comparing the
detected $JHK_S$ fluxes, with and without the NDF, we determined the
attenuation of the NDF in the three filters.

Each observation consisted of six sequential exposures of the form $OSSOOS$,
where $O$ is the object, and $S$ is a sky observation. Each exposure was
jittered using a jitter box of 4~arcsec. We used a sky offset of 15~arcsec,
and selected a position angle such that no object was in the sky field.

Each exposure consists of 5 to 35 short observations of integration times in
the range $0.35-5.3$~seconds, depending on the brightness of the source. For
the target observations without NDF we chose the minimum integration time of
0.35~seconds. For all standard star observations and target observations with
NDF we chose the integration time such that the image does not saturate for
reasonable Strehl ratios. We optimized the integration time to obtain the
desired signal-to-noise ratio. The short integrations are combined by taking the
median value.

Visual wavefront sensing was performed directly on the target stars, which
minimized the effects of anisoplanatism. For a subset of the target stars, the
observation block was carried out multiple times. The target stars were
usually positioned in the center of the field. Occasionally we observed the
target off-center to be able to image a companion at large angular separation.

The observations on the nights of 29~April, 28~June, and 3~July, 2004, were
obtained under bad weather or instrumental conditions and were removed from
the dataset. These observations were repeated under better conditions later on
in the observing run. 
The observations on the nights of 6~April and 27~June, 2004,
were partially obtained under non-photometric conditions, and were calibrated using the targets themselves (see \S~\ref{section: photometry}).
All other observations were performed under photometric
conditions. Most observations (85\%) were obtained with a seeing between
0.5~and 1.5~arcsec. For a large fraction of the remaining observations the
seeing was between 1.5~and 2.0~arcsec. The majority (65\%) of the observations were
obtained at an airmass of less than 1.2, and for 98\% of the observations the
airmass was less than 1.6.

\subsection{Data reduction procedures}

The primary data reduction was performed with the ECLIPSE package
\citep{devillard1997}. Calibration observations, including dark images, flat
field images, and standard star images, were provided by ESO Paranal. Twilight
flat fields were used to create a pixel sensitivity map. For several
observations, no twilight flats were available. In these cases we used the
lamp flats. The dark-subtracted observations were flatfielded and
sky-subtracted. Finally, the three jittered object observations were combined.

\subsection{Component detection} \label{section: componentdetection}

The component detection is performed with the STARFINDER package
\citep{diolaiti2000}. The PSF of the target star is
extracted from the background subtracted image. The flux of the primary star
is the total flux of the extracted PSF. A scaled-down version of the PSF is
compared to the other signals in the field with a peak flux larger than
$2.5-3$ times the background noise\footnote{Note that a peak flux of
$2.5-3$~times the background noise corresponds to a total flux with a much
larger significance, since the flux of a faint companion is spread out over
many pixels. All detected components in our survey have a signal-to-noise ratio larger than $12.8$.}. 
The profile of these signals is then cross-correlated with the
PSF. Only those signals with a profile very similar to that of the PSF star
(i.e., a correlation coefficient larger than $\approx 0.7$) are considered
as real detections. Finally, the angular separation, the position angle, and the
flux of the detected component are derived.

\begin{figure}[btp]
  \centering
  \includegraphics[width=0.5\textwidth,height=!]{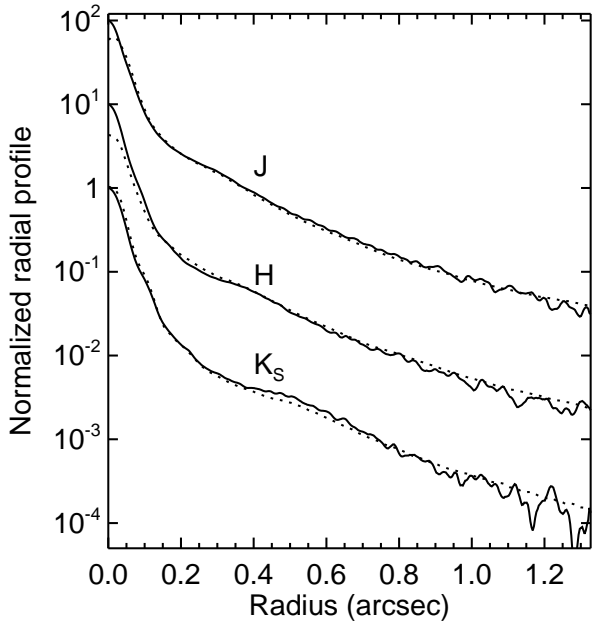}
  \caption{The radial profile of the PSF for the target star HIP78968. The
    observations are obtained using the NACO system in the night of May~5,
    2004 in $J$, $H$, and $K_S$. The corresponding Strehl ratios for these
    observations are 6.5\% in $J$, 15.6\% in $H$, and 23.6\% in $K_S$.
    Observations are obtained with neutral density filter (NDF; solid curves) 
    and without NDF 
    (dotted curves). The profiles are normalized such that the
    peak flux for the images obtained with NDF is 100 in $J$, 10 in $H$, and 1
    in $K_S$. The observations with NDF allow us to extract the PSF and
    measure the flux of the primary star. The images obtained without NDF are
    much deeper, but the primary is usually saturated. Assuming the PSF is
    similar, we use the non-saturated PSF to analyze the secondaries in the
    saturated image. \label{figure: psfprofile}}
\end{figure}

As discussed in \S~\ref{section: observations}, we observe each target in
$JHK_S$ {\em with} NDF to obtain an accurate PSF template, and {\em without}
NDF to do accurate photometry on the faint companions. The observations
without NDF are often saturated, which makes PSF extraction impossible. None
of the observations with NDF are saturated. Since these observations are
carried out close in time and close in airmass, we assume that the PSFs of the
observations with and without NDF are not significantly different. This is
illustrated in Figure~\ref{figure: psfprofile}, where we plot the radial
profiles of the extracted PSF of HIP78968. For the saturated images we use the
PSF that was extracted from the corresponding non-saturated image for analysis
of the secondaries.

Artifacts are present in the image when observing a bright
object. The location of the artifacts is well-defined (CONICA manual), and
since the artifacts look significantly non-circular, they can be easily
recognized as non-stellar components (see Figure~\ref{figure: triples}). Additionally, observations obtained with
the NDF show a faint artifact at $\sim 2$~arcsec to the North-East of the
target star. Care was taken that the extracted PSF and the analysis of
companions in that area were not affected by the presence of this artifact.

\subsection{Photometry} \label{section: photometry}

Observations of standard stars from the \cite{persson1998} catalog were
provided by ESO. The standard stars are used to determine the magnitude
zeropoints for each night and each filter individually. No standard stars are
available in the nights of 28~April and 10~September, 2004. For 28~April we
determine the zero point magnitude using 2MASS as a reference system, and the
target stars HIP67260, HIP67919, HIP68532 as substitute standard stars. For
10~September, we use the 2MASS data for HIP83542 to calibrate $J$ and $H$. For
the $K_S$ filter we use the HIP83542 measurement of \cite{kouwenhoven2005}
since its 2MASS $K_S$ magnitude is inaccurate due to confusion with 
the diffraction spike of a nearby star. 
The $H$ magnitude of HIP59502 was
obtained under non-photometric conditions on 6~April, 2004, so instead we used
the corresponding 2MASS measurement of this star. All observations in the
night of 27~June, 2004 were obtained under non-photometric conditions; the
fluxes of HIP80474 and HIP81972 are therefore calibrated using 2MASS.
We derive the calibrated magnitudes using the mean
extinction coefficients for Paranal: $k_J=0.11$, $k_H=0.06$, and
$k_{K_S}=0.07$.

The attenuation $m_{\rm NDF}$ (in magnitudes) of the NDF is determined for
$J$, $H$, and $K_S$ using the standard stars GSPC~S273-E and GSPC~S708-D and
the non-saturated target stars. All observations are performed in pairs (with
and without NDF), which allows us to determine the NDF attenuation. For each
filter $m_{\rm NDF}$ is calculated as the median
difference in magnitude:
$m_{\rm NDF} = \left< m_{\star,{\rm without\ NDF}} - m_{\star,{\rm with\ NDF}} \right>
$.
We find $m_{\rm NDF,J} = 4.66\pm 0.03$ mag, $m_{\rm
NDF,H} = 4.75\pm 0.02$ mag, and $m_{\rm NDF,K_S} = 4.85\pm 0.02$ mag, respectively. We
do not measure a significant difference between the values found for the
early-type program stars and the late-type standard stars. All observations
done with the NDF are corrected with the values mentioned above.


\subsection{Photometric precision and accuracy of the NACO observations} \label{section: photometricaccuracy}

We estimate the photometric uncertainty of the NACO observations using simulations 
(see \S~\ref{section: starfinderprecision}) and the comparison with other datasets (\S~\ref{section:
2masscomparison}--\ref{section: tokovinin}). For primaries, the external $1\sigma$ error
in $J$, $H$, and $K_S$ is $\sim 0.04$~mag, corresponding to an error of $\sim
0.06$~mag in the colors. Typical $1\sigma$ external errors in magnitude and color for
the bright companions ($8 \la K_S/{\rm mag} \la 13$) are 0.08~mag and
0.11~mag, respectively. For the faintest sources ($K_S \ga 13$~mag) the
errors are 0.12~mag in magnitude and 0.17~mag in color. In the following subsections we will 
discuss the analysis of our photometric errors.

\subsubsection{Algorithm precision} \label{section: starfinderprecision}

The instrumental magnitudes of all objects are obtained using STARFINDER. We
investigate the precision of the STARFINDER algorithm using simulated
observations. We create simulations of single and binary systems with varying
primary flux ($10^4 - 10^6$~counts), flux ratio ($\Delta K_S = 0-10$~mag),
angular separation ($0''-13''$), Strehl ratio ($1\%-50\%$), and position
angle. We estimate the flux error by comparing the
input flux with the flux measured by STARFINDER for several realizations.

Using the simulations we find that the precision of the STARFINDER fluxes is
$\sim 1\%$ ($\sim 0.01$~mag) for most primary stars in our sample, for all
relevant Strehl ratios and as long as the PSF of the primary star is not
significantly influenced by the presence of a companion. The 1\% error is due to the
tendency of STARFINDER to over-estimate the background underneath bright
objects \citep{diolaiti2000}. For the fainter primaries in our sample (flux
between $5 \times 10^4$ and $5 \times 10^5$ counts), the flux error is $1-3\%$
($\sim 0.01-0.03$~mag).

For secondaries outside the PSF-halo of the primary, the error is typically
$\sim 0.01$~mag if the flux difference with the primary is less than
5~mag. Fainter secondaries have a larger flux error, ranging from $0.01$~mag
to $0.1$~mag, depending on the brightness of primary and companion.

If the secondary is in the halo of the primary, its flux error is somewhat
larger. For example, for a companion at $\rho=2''$ which is less than
4~magnitudes fainter than the primary, the flux error is 4\% ($\sim
0.04$~mag) or smaller. Deblending the PSF of a primary and close secondary does not
introduce a much larger error, as long as the magnitude difference is less than $\sim 5$~magnitudes. No close companions with a magnitude difference larger than 5~mag are detected in our NACO observations.

For several fields the observations without NDF are saturated. In order to
analyze the faint companions in the field we use the PSF of the corresponding
non-saturated observation obtained with the NDF (see \S~\ref{section:
observations}). These observations are performed close in airmass and time, so that
their PSFs are similar. We estimate the flux error by
comparing PSFs corresponding to non-saturated images obtained with NDF and
without NDF. These comparisons show that the resulting error ranges from 0.02 to
0.5~magnitudes, depending on the brightness of the secondary. We therefore
minimize flux calculations using this method, and only use measurements
obtained with the PSF of the non-saturated image when no other measurements
are available. In the latter case, we place a remark in Table~\ref{table:
longtable}.

\subsubsection{Comparison with 2MASS} \label{section: 2masscomparison}

We compare the near-infrared measurements of the 22~targets in our NACO survey with the measurements in 2MASS
\citep{2mass} to get an estimate of the external errors. We only select those
measurements in 2MASS that are not flagged. Since the resolution in our
observations is higher than the $\approx 4''$ resolution of 2MASS, we combine
the observed fluxes of the primaries and close companions before the comparison with
2MASS. For the observations {\em not} calibrated with the 2MASS measurements,
the rms difference between our measurements and those of 2MASS are $0.055$~mag in $J$, $0.040$~mag in $H$, and $0.049$~mag in $K_S$.

\subsubsection{Comparison with the ADONIS survey of \cite{kouwenhoven2005}} \label{section: adoniscomparison}

We detect all but two of the stellar components found by
\cite{kouwenhoven2005} around the 22 target stars in our NACO survey. We
do not observe the faint companions of HIP80142 at $\rho=8.54''$ and HIP81949
at $\rho=9.70''$ because they are not within our NACO field of view. 
We find three bright companions at small angular separation of HIP63204 ($\rho=0.15''$), HIP73937 ($\rho=0.34''$), and
HIP79771 ($\rho=0.44''$). Since these objects are not found in the ADONIS observations of 
\cite{kouwenhoven2005}, their fluxes and those of the corresponding primaries
are summed for comparison with \cite{kouwenhoven2005}.

The rms difference between the 22 primaries observed with NACO and those described 
in \cite{kouwenhoven2005} is 0.055~mag in $K_S$. HIP69113 and HIP78968
additionally have multi-color observations in \cite{kouwenhoven2005}, which
are in good agreement with the measurements presented in this paper. The $J$
and $H$ measurements of HIP80474 and HIP80799 are flagged ``non-photometric''
in \cite{kouwenhoven2005}, and are not discussed here.

Our dataset and that of \cite{kouwenhoven2005} have 35 stellar components
other than the target stars in common. The rms difference between the $K_S$
magnitude of these objects in the two papers is 0.26~magnitudes. The
differences are similar for the objects that have common $J$ and $H$
measurements in both papers.

\subsubsection{Comparison with  \cite{shatsky2002}} \label{section: tokovinin}

Three targets in our NACO survey are also included in the binarity survey amongst B-stars in
Sco~OB2 by \cite{shatsky2002}: HIP79098, HIP80142, and HIP81972. Seven
secondaries are detected both their survey and in our NACO survey (1 for HIP79098; 2 for HIP80142; 4 for
HIP81972). \cite{shatsky2002} classify these seven secondaries all as `definitely optical'
or `likely optical'. They performed their observations
in both coronographic and non-coronographic mode, and were therefore able to find five~faint secondaries which do not appear in our NACO sample.

The $J$ and $K_S$ magnitudes of HIP79098 and HIP80142 and their companions are
in good agreement with our measurements. Our measurements of HIP81972 are in
good agreement with those in 2MASS as well as the measurements in
\cite{kouwenhoven2005}, but there is a discrepancy between our measurements of
HIP81972 and those in \cite{shatsky2002}. The magnitude difference between
HIP81972 and its companions in our observations and in \cite{shatsky2002} are
similar. The observations of HIP81972 are flagged `likely photometric' in
\cite{shatsky2002}, but since they disagree with those in this paper and those
in 2MASS, we assume they are non-photometric, and ignore them for the
magnitude comparison.

\subsection{General properties of the NACO observations}\label{section: generalproperties}

In the fields around the 22~targets we observed with NACO, we find 62 components other than the target stars. The properties of these targets and their secondaries are listed in Table~\ref{table: longtable}.  The 22~primaries have $5.5~{\rm mag} < J < 7.8~{\rm mag}$, $5.0~{\rm mag} < H < 7.7~{\rm mag}$, and $4.9~{\rm mag} < K_S < 7.7~{\rm mag}$. The brightest companions observed are $\sim 7.5$~mag in the three filters, while the faintest secondaries found have $J=16.6$~mag, $H=17.3$~mag, and $K_S=17.3$~mag. 

With NACO we detect components in the angular separation range $0.15'' < \rho < 11.8''$. The lower limit on $\rho$ depends on the Strehl ratio and the magnitude difference between primary and companion (see also \S~\ref{section: detectionlimits}). The upper limit is determined by the size of the field-of-view. The median formal error in angular separation is 4~mas for bright components ($8~{\rm mag} \la K_S \la 13~{\rm mag}$) and 10~mas for faint components ($K_S \ga 13~{\rm mag}$). Position angles are measured from North to East. The median formal error in the position angle is $0^\circ.007$. 

We find 27~stellar components that are not detected by
\cite{kouwenhoven2005}. Three close secondaries are found at small angular
separation from HIP63204-2 ($\rho=0.15''$; $K_S=8.40$~mag), HIP73937-1 ($\rho=0.34''$; $K_S=8.37$~mag), and HIP79771-2 ($\rho=0.44''$; $K_S=11.42$~mag). The former two have been reported as candidate companions \citep{wds1997}; the latter was previously undocumented. The other 25 new secondaries are all faint ($K_S \ga 12$~mag). Two of these 25~secondaries were also reported by \cite{shatsky2002}.


\section{The completeness and detection limit of the ADONIS and NACO surveys} \label{section: detectionlimits}

\begin{figure}[tbp]
  \centering
  \includegraphics[width=0.5\textwidth,height=!]{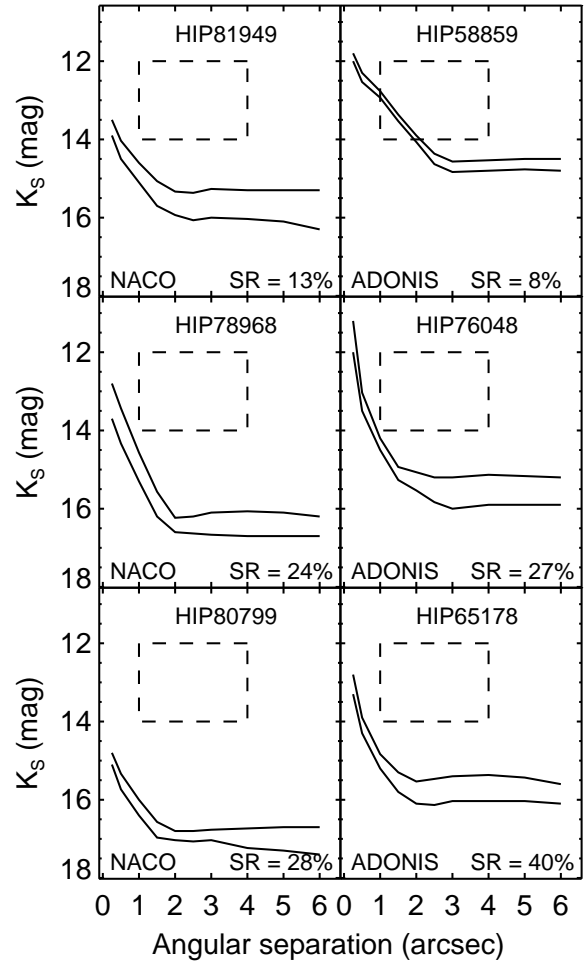}
  \caption{The detection limit and completeness limit for several targets in our ADONIS survey \citep{kouwenhoven2005} and NACO survey (this paper). The target star, Strehl ratio, and instrument are indicated in each panel. The lower and upper curve show the detection limit and the completeness limit, respectively. The detection and completeness limits shown above are representative for the ADONIS and NACO observations.
    The target stars have $K_S$ magnitudes of 7.33~mag (HIP81949), 7.42~mag (HIP78968), 7.45~mag (HIP80799), 6.52~mag (HIP58859), 6.26~mag (HIP76048), and 6.71~mag (HIP65178). The completeness limit is generally $\sim 0.3$~mag brighter than the detection limit. 
    At close angular separation a higher Strehl ratio results in a fainter detection limit. The dashed rectangle encloses the region with $12~\mbox{mag} \leq K_S \leq 14~\mbox{mag}$ and $1'' \leq \rho \leq 4''$, which is relevant for our analysis of the substellar population in Sco~OB2 (see \S~\ref{section: gap}). In this region the NACO observations are complete and ADONIS observations are more than 95\% complete.
    \label{figure: detectionlimits}
  }
\end{figure}

We cannot detect sources fainter than a certain magnitude because of the background noise in the images. The faintest detectable magnitude additionally depends on the angular distance to the primary star, the primary star magnitude, and the Strehl ratio. For a correct interpretation of the results of the survey it is therefore important to characterize the limiting magnitude of the observations (the detection limit) and the magnitude at which a star is likely detected (the completeness limit).

We study the completeness limit and detection limit as a function of angular separation from the primary for six stars. These are HIP58859, HIP65178, and HIP76048 from our ADONIS survey and HIP80799, HIP78968, and HIP81949 from our NACO survey. These stars are selected to cover the range in Strehl ratio of the observations, so that the completeness and detection limits are representative for the other targets in the ADONIS and NACO surveys.

For each observation STARFINDER extracts the PSF from the image (see \S~\ref{section: componentdetection}). We simulate observations by artificially adding a scaled and shifted copy of the PSF to the observed image. We reduce the simulated image as if it were a real observation. We repeat this procedure twenty times for simulated secondaries with different angular separation and magnitude. We define the detection limit and completeness limit as the magnitude (as a function of angular separation) at which respectively 50\% and 90\% of the simulated secondaries are detected.
The curves in Figure~\ref{figure: detectionlimits} show the completeness and detection limit for the six stars mentioned above. Due to our sampling the magnitude error of the completeness and detection limit is $~\sim 0.15$~mag. 
The figure clearly shows that a high Strehl ratio facilitates the detection of closer and fainter objects as compared to observations with lower Strehl ratio. For all stars the completeness limit is $\sim 0.3$~mag above the detection limit.

In Section~\ref{section: gap} we analyze the substellar companion population in Sco~OB2 in the angular separation range $1'' \leq \rho \leq 4''$ and magnitude range $12~\mbox{mag} \leq K_S \leq 14~\mbox{mag}$.  The NACO observations of 22~targets are complete in this region. Only a few targets in the ADONIS sample are incomplete in this region. Assuming a flat semi-major axis distribution, we estimate that  about 5\% of the faint companions at small angular separation are undetected for the 177~targets that are {\em only} observed with ADONIS (see  Figure~\ref{figure: detectionlimits}). For the combined NACO and ADONIS sample this means that we are more than 95\% complete in the region $12~\mbox{mag} \leq K_S \leq 14~\mbox{mag}$ and $1'' \leq \rho \leq 4''$.

With NACO we are sensitive down to brown dwarfs and massive planets. To estimate the mass corresponding to the faintest detectable magnitude as a function of angular separation, we use the models of \cite{chabrier2000}. We assume a distance of 130~pc, the mean distance of Sco~OB2, and an age of 5~Myr for the US subgroup and 20~Myr for the UCL and LCC subgroups (cf. \S~\ref{section: cmd}). At a distance of 130~pc, the brightest brown dwarfs have an apparent magnitude of $K_S \approx 12$~mag.
With NACO we are able to detect brown dwarfs at an angular separation larger than $\approx 0.3''$. The magnitude of the faintest detectable brown dwarf increases with increasing angular separation between $\rho = 0.3''-2''$. 
The minimum detectable mass is a function of age due to the cooling of the brown dwarfs. For angular separations larger than $\approx 2''$ the halo of the primary PSF plays a minor role. For these angular separations we are sensitive (but not complete) down to $K_S \approx 16.5$~mag with NACO, corresponding to planetary masses possibly as low as $\sim 5~{\rm M_J}$ for US and $\sim 10~{\rm M_J}$ for UCL and LCC. However, because of the large number of background stars and the uncertain location of the isochrone for young brown dwarfs and planets, we will not attempt to identify planetary companions in this paper (see \S~\ref{section: separation} for a further discussion).

Several secondaries are not detected in one or two filters. For the filter(s) in which the secondary is not observed we determine a lower limit on the magnitude of the missing star using simulations. We perform the simulations as described in \cite{kouwenhoven2005}. The position of the secondary is known from observations in other filters. This position is assigned to the simulated companion, so that the detection limit as a function of angular separation is taken into account. Images are created with simulated secondaries of different magnitude. We reduce the images as if they were real observations. The lower limit on the magnitude is then determined by the faintest detectable simulated secondary. Two secondaries (HIP73937-1 and HIP76071-1) have a lower limit in $J$ because they are unresolved in the wings of the primary star PSF. The other secondaries with a lower magnitude limit for a filter have a flux below the background noise.


\section{Status of the stellar components} \label{section: status}

In this section we determine the status (companion star or background star) of the secondaries. We analyze the secondaries detected around the 22~targets observed with NACO, as well as the secondaries around the 9~targets with multi-color observations in the ADONIS sample. For the 31~targets analyzed in this section we detect 72~secondaries in total. Sco~OB2 members and their companions should be located near the isochrone in the color-magnitude diagram, while the background stars should show a much larger spread. We use this property in \S~\ref{section: separation} to separate companions and background stars. As a consistency check we study in \S~\ref{section: backgroundstarpopulation} how our results compare to the expected number of background stars in our observations.

\subsection{Color-magnitude diagram and isochrones} \label{section: cmd}

\begin{figure*}[btp]
  \begin{tabular}{cc} 
    \includegraphics[width=0.48\textwidth,height=!]{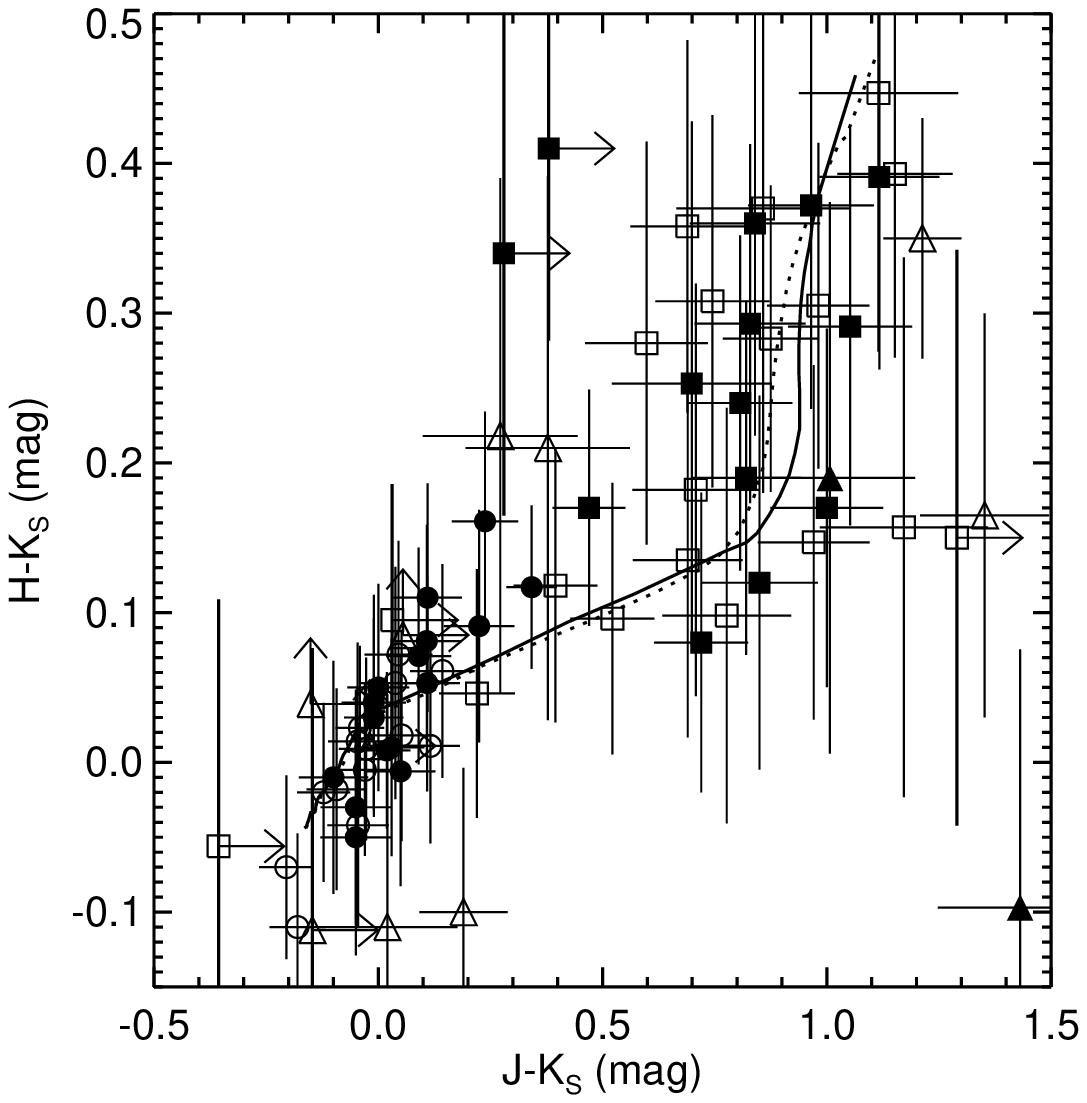} &
    \includegraphics[width=0.48\textwidth,height=!]{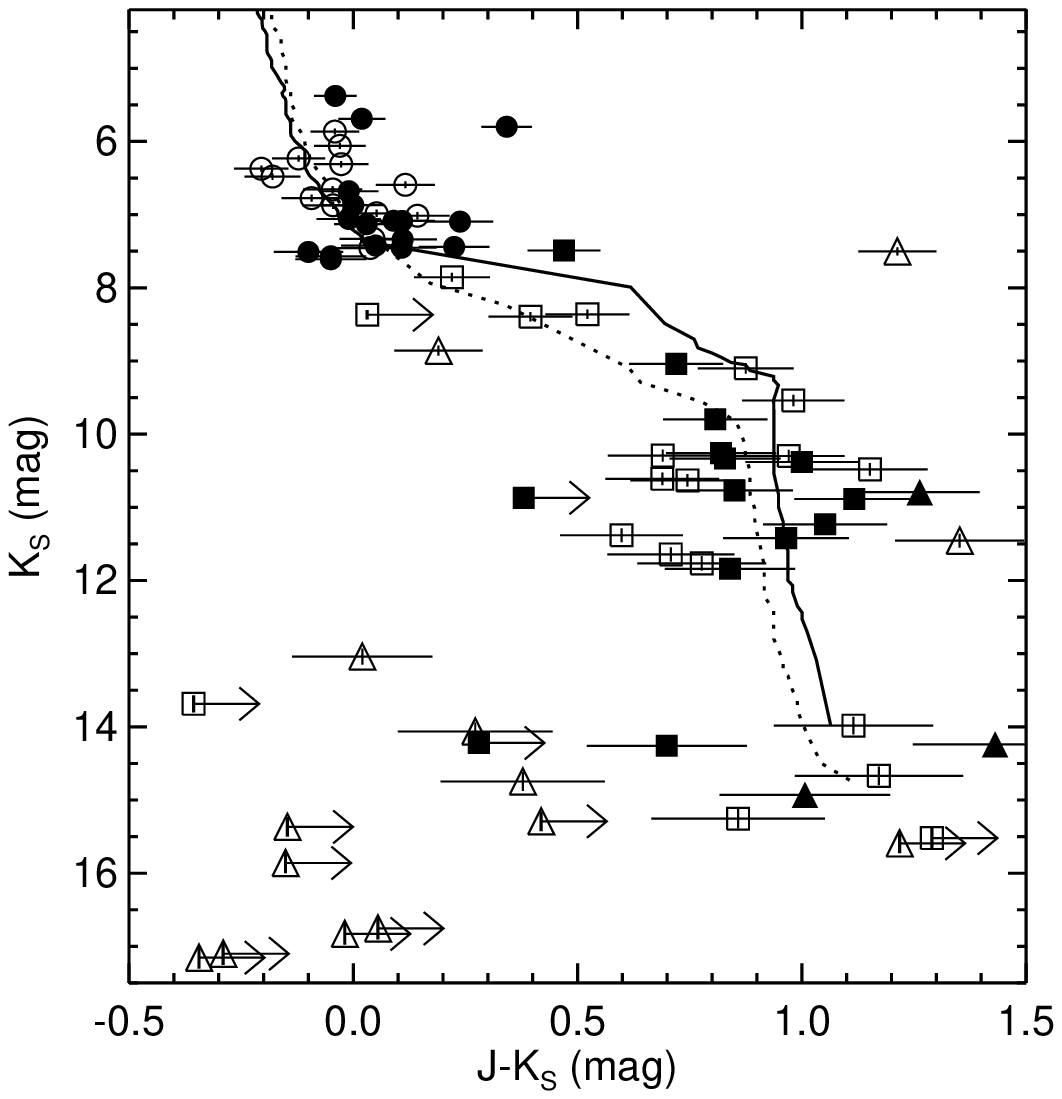} \\
  \end{tabular}
  \caption{
    The color-color diagram ({\em left}) and color-magnitude diagram ({\em right}) of the objects in our sample. Measurements are shown for the 22~targets observed with NACO and for the 9~targets with multi-color observations in the ADONIS sample. Both panels show target stars (circles), confirmed and candidate companions (squares), and background stars (triangles). The target stars and secondaries in the US subgroup are indicated with filled symbols; those from UCL and LCC are indicated with open symbols. 
The $1\sigma$ error bars are indicated for all data points. Lower limits are given for objects that are not detected in all three filters. Several detected objects are outside the ranges of the figures; these are all background stars. The status (companion or background star) of the secondaries is discussed in \S~\ref{section: separation}. The 5~Myr isochrone for US and the 20~Myr isochrone for UCL and LCC are indicated with the solid and dotted curves, respectively. 
    \label{figure: ccdiagrams}}     
\end{figure*}

For the 22~targets in the NACO sample and the 9~targets in the ADONIS multi-color subset we have magnitudes in three filters, as well as for most of their secondaries. Several of the faintest secondaries are undetected in one or two filters, as they are below the detection limit. Color-color and color-magnitude diagrams with the 31~targets and 72~secondaries are shown in Figure~\ref{figure: ccdiagrams}. Lower limits are indicated for objects that are not detected in all three filters. Several secondaries in our sample are either very red or very blue. These secondaries fall outside the plots in Figure~\ref{figure: ccdiagrams}, and are all background stars.

For our analysis we adopt the isochrones described in \cite{kouwenhoven2005}, which consist of models from \cite{chabrier2000} for $0.02~\mbox{M}_\odot \leq M < 1~\mbox{M}_\odot$, \cite{palla1999} for $1~\mbox{M}_\odot \leq M < 2~\mbox{M}_\odot$, and  \cite{girardi2002} for $M > 2~\mbox{M}_\odot$.  For members of the US subgroup we use the 5~Myr isochrone, and for UCL and LCC members we use the 20~Myr isochrone. 

Absolute magnitudes $M_J$, $M_H$, and $M_{K_S}$ are derived from the apparent magnitudes using for each star individually the parallax and interstellar extinction $A_V$ from \cite{debruijne1999}; see \cite{kouwenhoven2005} for details. The error on the parallax is $5-10\%$ for all targets, and can therefore be used to derive reliable distances to individual stars \citep[e.g.,][]{brown1997}.
The median fractional error on the distance is 6.5\%, which introduces an additional error of 0.15~mag on the absolute magnitudes. Combining this error with the error in the apparent magnitude (\S~\ref{section: photometricaccuracy}) we obtain $1\sigma$ error estimates for the absolute magnitudes of $0.16$~mag for the primaries, $0.17$~mag for the bright companions, and $0.19$~mag for the faint companions. The colors are directly calculated from the apparent magnitudes, and are not affected by parallax errors.

\begin{figure*}[btp]
  \centering
  \includegraphics[width=\textwidth,height=!]{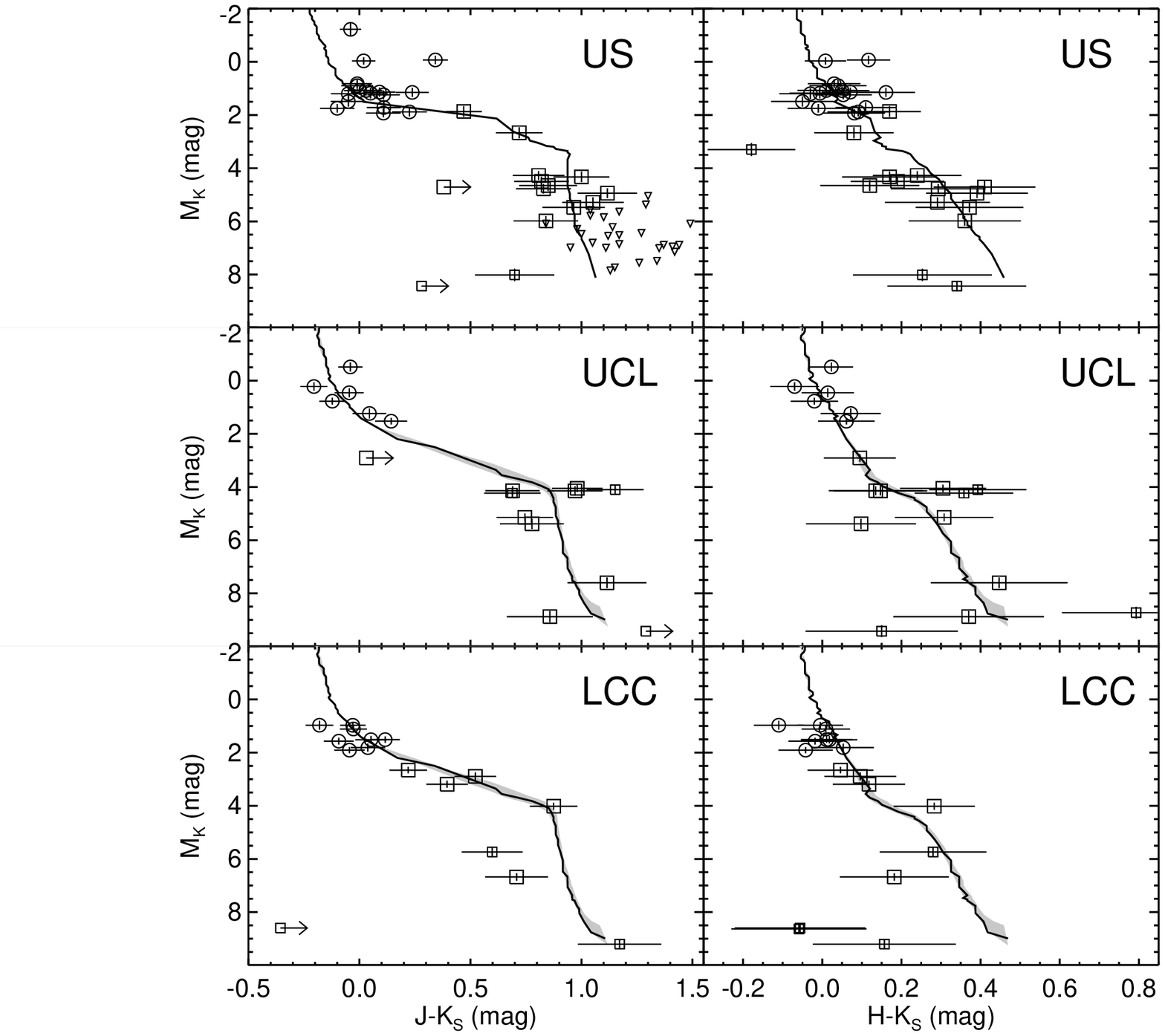}
  \caption{The (absolute) color-magnitude diagrams for the 22~targets in our NACO sample and the 9~targets with multi-color observations in the ADONIS sample. 
The results are split into the three subgroups US ({\em top}), UCL ({\em middle}), and LCC ({\em bottom}). The primary stars are indicated with circles; the confirmed companions with large squares, and the candidate companions with small squares. The $M_{K_S}$ magnitude is derived from the $K_S$ magnitude by correcting for distance and extinction for each target individually. The solid curves represent isochrones of 5~Myr (for US) and 20~Myr (for UCL and LCC).  The 15~Myr and 23~Myr isochrones enclose the gray-shaded area   and represent the uncertainty in the age of the    UCL and LCC subgroups. For each data point we indicate the $1\sigma$ errors. 
The photometry of the observed objects cannot be used to distinguish between the subgroups of Sco~OB2, due to the errors and the small sample.   
The free-floating brown dwarfs in US    identified by \cite{martin2004} are indicated with triangles, adopting a distance of 145~pc. 
    \label{figure: hrdiagram2}}
\end{figure*}

The color-magnitude diagrams (with {\em absolute} magnitudes) for the subgroups are shown in Figure~\ref{figure: hrdiagram2}. The measurements for US are in the top panels, and those of UCL and LCC are in the middle and bottom panels, respectively. The curves represent the 5~Myr (for US) and 20~Myr (for UCL and LCC) isochrones. The gray-shaded area indicates the inaccuracy in isochrone placement due to the age uncertainty in the subgroups ($\sim 1$~Myr for US members, $\sim 4$~Myr for UCL/LCC members; see Table~\ref{table: statistics}). Due to our small sample and the errors in the photometry we cannot see a difference between the magnitude and color distributions of the three subgroups.
For comparison we have included the free-floating brown dwarf candidates in US reported by \cite{martin2004}.

\subsection{Distinction between companions and background stars} \label{section: separation}

Individual distances to Sco~OB2 member stars are available \citep{debruijne1999}, as well as the ages of the three subgroups of Sco~OB2. A companion of a Sco~OB2 member star has (by definition) practically the same distance as its primary. It is very likely that a primary and companion in Sco~OB2 are coeval. The probability that a companion is captured dynamically is very small, since this involves either a multiple-star interaction or significant tidal dissipation.
It is even less likely that the companion is a captured field star, i.e., that the primary and companion have a different age. Background stars generally have a different age, distance, or luminosity class than Sco~OB2 members. In principle it is possible that an ADONIS or NACO field contains two members of the Sco~OB2 association at different distances due to projection effects, but \cite{kouwenhoven2005} showed that this effect can be neglected. 
Physical companion stars thus have the same age and distance as their primary, and therefore should fall on the isochrone for the subgroup to which the primary belongs, contrary to background stars. We use this property to separate physical companions and background stars.

For each stellar component we determine in the color-magnitude diagram (Figure~\ref{figure: hrdiagram2}) the point on the isochrone that corresponds best to the measured position. The differences in color and magnitude of the star and the nearest point on the isochrone are then compared to the observational errors. We use the $\chi^2$ test to determine how compatible the observed color and magnitude of a secondary are with the isochrone.
For example, if the best-fitting value in the $(J-K_S,M_{K_S})$-diagram deviates by $\Delta(J-K_S)$ in color and by $\Delta M_{K_S}$ in magnitude from the isochrone, the $\chi^2$ value is given by
\begin{equation} 
\chi^2 = \frac{ [\, \Delta(J-K_S) \, ]^2 }{  \sigma^2_{J-K_S} + \sigma^2_{J-K_S,{\rm iso}}}  
         + \frac{ \Delta M^2_{K_S}}{ \sigma^2_{M_{K_S}} + \sigma^2_{M_{K_S},{\rm iso}} } 
\end{equation}
where $\sigma_{J-K_S}$ and $\sigma_{M_{K_S}}$ are the observational errors in color and absolute magnitude, respectively. The errors on the location of the isochrone (due to age and metallicity uncertainty) are denoted with $\sigma_{J-K_S,{\rm iso}}$ and $\sigma_{M_{K_S},{\rm iso}}$. 
The age uncertainty is $\sim 1$~Myr for members of US and $\sim 4$~Myr for members of UCL and LCC (see Table~\ref{table: statistics}).  The error in the placement of the corresponding isochrones due to age uncertainty is shown in Figure~\ref{figure: hrdiagram2}, and is small compared to the photometric errors. 
We assume a solar metallicity for all observed stellar components. The metallicity $[M/H]$ of Sco~OB2 has not been studied in detail. In their metallicity study of Ori~OB2, \cite{cunha1994} found a metallicity slightly ($\sim 0.2$~dex) lower than solar for this association. Using the models of \cite{siess2000} we estimate that the metallicity uncertainty $\Delta [M/H]=0.2$ results in an additional isochronal error of $0.05$~mag in $JHK_S$ and $0.06$~mag in the colors of low-mass ($M \la 0.5~{\rm M}_\odot$) companions. 

The $\chi^2$ values for the secondaries in each of the three color-magnitude diagrams are listed in Table~\ref{table: criteria} (as long as they are available). For the classification into companions and background stars we consider the {\em largest} of the three $\chi^2$ values available for each secondary. We choose this strategy (instead of, e.g., taking the average $\chi^2$ value), because background stars may be consistent with the isochrone for e.g. $J-K_S$, but not for $H-K_S$. The physical companions, however, should be consistent with the isochrone in all color-magnitude diagrams (within the error bars).

\begin{table}
  \begin{tabular}{lcp{0.6cm}p{0.1cm}p{0.9cm}}
    \hline
    Secondary status    & Symbol   & \multicolumn{3}{l}{Criterion}\\
    \hline
    Confirmed companion &c &         &$ \chi^2$&$\leq 2.30$ \\
    Candidate companion &? & $2.30<$ &$\chi^2$&$\leq 11.8$ \\
    Background star     &b & $11.8<$ &$\chi^2$&  \\
    \hline
  \end{tabular}
  \caption{Criteria adopted to separate the secondaries with multi-color observations into confirmed companions, candidate companions, and background stars. The $\chi^2$ values of 2.30 and 11.8 correspond to the $1\sigma$ and $3\sigma$ levels. This means that (statistically) 68.3\% of the physical companions have $\chi^2 < 2.30$ and 99.73\% of the physical companions have $\chi^2 < 11.8$. A secondary with $\chi^2 < 2.30$ is very likely a companion star. A secondary with $\chi^2 > 11.8$ is almost certainly a background star. The secondaries with $2.30 < \chi^2 < 11.8$ may be companions or background stars.  \label{table: classification}}
\end{table}


Table~\ref{table: classification} lists the criteria we adopt to classify the secondaries into three groups: confirmed companions, candidate companions, and background stars. The $\chi^2$ value of 2.30 corresponds to the $1\sigma$ confidence level, which means that statistically 68.3\% of the companion stars have $\chi^2 < 2.30$. Similarly, 99.73\% of the companions have $\chi^2 < 11.8$ ($3\sigma$ confidence level). The above confidence levels are for a dataset with two degrees of freedom, under the assumption that the errors are Gaussian \citep{numericalrecipies}.

For several objects we have lower limits on the magnitudes in one or two of the filters. Using the lower limits on $J$, $H$, or $K_S$ we calculate upper or lower limits on $\chi^2$ and are able to classify several additional objects as background stars.

Even though we are sensitive (although not complete) to massive planets around our NACO targets, we do not classify the very faint secondaries in this paper, for two reasons. First, many faint background stars are expected in the ADONIS and NACO field of view (see \S~\ref{section: backgroundstarpopulation}). Due to the larger error bars for faint secondaries, several background stars may be consistent with the isochrone; the vast majority of the ``candidate planets'' are likely background stars. Second, the presently available evolutionary models for massive planets are not very reliable for young ages \citep[see e.g.,][for a review]{chabrier2005}. Throughout our analysis we do not consider objects with a mass below $0.02~{\rm M}_\odot$. Consequently, all secondaries with an inferred mass smaller than $0.02~{\rm M}_\odot$ are classified as background stars.

\begin{table*}
  \begin{tabular}{| l | rrr rr | rrr r | cl |}
  \hline
  Star & \multicolumn{1}{c}{$J$} & \multicolumn{1}{c}{$H$} & \multicolumn{1}{c}{$K_S$} & \multicolumn{1}{c}{$\rho$} & \multicolumn{1}{c}{PA}  & \multicolumn{1}{c}{$M_J$}  & \multicolumn{1}{c}{$M_H$}  & \multicolumn{1}{c}{$M_{K_S}$}  & \multicolumn{1}{c}{Mass}      & \multicolumn{1}{c}{Status} &  \multicolumn{1}{c|}{Remarks} \\
  \hline
       & \multicolumn{1}{c}{mag} & \multicolumn{1}{c}{mag} & \multicolumn{1}{c}{mag}   & \multicolumn{1}{c}{arcsec} & \multicolumn{1}{c}{deg}   & \multicolumn{1}{c}{mag}    & \multicolumn{1}{c}{mag}    &  \multicolumn{1}{c}{mag}       & \multicolumn{1}{c}{M$_\odot$} &        &          \\
  \hline
  \multicolumn{12}{|l|}{NACO targets} \\ 
  \hline
  HIP59502    -1   &   12.35   &   11.83   &   11.64   &   2.94   &   26.39   &   7.39   &   6.86   &   6.68   &   0.14   &   c   &          \\
HIP62026    -1   &   8.08   &   7.90   &   7.86   &   0.23   &   6.34   &   2.88   &   2.71   &   2.66   &   1.19   &   c   &          \\
HIP63204    -2   &   8.79   &   8.51   &   8.40   &   0.15   &   236.56   &   3.59   &   3.31   &   3.19   &   1.06   &   c   &          \\
HIP67260    -1   &   8.88   &   8.46   &   8.36   &   0.42   &   229.46   &   3.42   &   2.99   &   2.90   &   1.10   &   c   &          \\
HIP67919    -1   &   9.98   &   9.38   &   9.10   &   0.69   &   296.56   &   4.89   &   4.30   &   4.02   &   0.75   &   c   &          \\
HIP68532    -1   &   10.52   &   9.85   &   9.54   &   3.05   &   288.50   &   5.03   &   4.36   &   4.05   &   0.73   &   c   &          \\
HIP68532    -2   &   11.38   &   10.94   &   10.63   &   3.18   &   291.92   &   5.89   &   5.45   &   5.14   &   0.39   &   c   &          \\
HIP69113    -1   &   10.98   &   10.43   &   10.29   &   5.34   &   65.15   &   4.83   &   4.28   &   4.14   &   0.77   &   c   &          \\
HIP69113    -2   &   11.27   &   10.45   &   10.30   &   5.52   &   67.17   &   5.12   &   4.29   &   4.15   &   0.72   &   c   &          \\
HIP73937    -1   &   $>8.40$   &   8.46   &   8.37   &   0.24   &   190.58   &   $>2.94$   &   3.00   &   2.91   &   1.11   &   c   &          \\
HIP79739    -1   &   12.28   &   11.52   &   11.23   &   0.96   &   118.33   &   6.34   &   5.58   &   5.29   &   0.16   &   c   &          \\
HIP79771    -1   &   12.00   &   11.28   &   10.89   &   3.67   &   313.38   &   6.06   &   5.33   &   4.94   &   0.19   &   c   &          \\
HIP79771    -2   &   12.39   &   11.79   &   11.42   &   0.44   &   128.59   &   6.44   &   5.85   &   5.47   &   0.13   &   nc   &          \\
HIP80799    -1   &   10.60   &   10.04   &   9.80   &   2.94   &   205.02   &   5.08   &   4.51   &   4.27   &   0.34   &   c   &          \\
HIP80896    -1   &   11.16   &   10.63   &   10.33   &   2.28   &   177.23   &   5.60   &   5.07   &   4.77   &   0.24   &   c   &          \\
HIP81972    -3   &   12.54   &   11.86   &   11.77   &   5.04   &   213.45   &   6.16   &   5.48   &   5.39   &   0.35   &   c   &   J       \\
HIP81972    -4   &   15.10   &   14.43   &   13.98   &   2.79   &   106.94   &   8.72   &   8.05   &   7.60   &   0.06   &   nc   &   JHK       \\
HIP81972    -5   &   16.11   &   15.63   &   15.26   &   7.92   &   229.27   &   9.73   &   9.25   &   8.88   &   $\approx$0.03   &   nc   &   JHK       \\
\hline  \multicolumn{12}{|l|}{ADONIS multi-color subset} \\ \hline  HIP76071    -1   &   $>11.25$   &   11.28   &   10.87   &   0.69   &   40.85   &   $>5.09$   &   5.12   &   4.71   &   0.23   &   c   &          \\
HIP77911    -1   &   12.68   &   12.20   &   11.84   &   7.96   &   279.25   &   6.82   &   6.34   &   5.98   &   0.09   &   c   &          \\
HIP78809    -1   &   11.08   &   10.45   &   10.26   &   1.18   &   25.67   &   5.32   &   4.69   &   4.50   &   0.30   &   c   &          \\
HIP78956    -1   &   9.76   &   9.12   &   9.04   &   1.02   &   48.67   &   3.39   &   2.75   &   2.67   &   1.16   &   c   &          \\
HIP79124    -1   &   11.38   &   10.55   &   10.38   &   1.02   &   96.18   &   5.33   &   4.50   &   4.33   &   0.33   &   c   &          \\
HIP79156    -1   &   11.62   &   10.89   &   10.77   &   0.89   &   58.88   &   5.50   &   4.77   &   4.65   &   0.27   &   c   &          \\
HIP80238    -1   &   7.96   &   7.66   &   7.49   &   1.03   &   318.46   &   2.34   &   2.04   &   1.87   &   1.67   &   c   &          \\

  \hline
  \end{tabular}
  \caption{Properties of the 25~confirmed companion stars found around the 22~members in our NACO survey and the 9~members with multi-color observations in the ADONIS survey.
The columns show the secondary designation, the $J$, $H$, and $K_S$ magnitudes, the angular separation, and the position angle (measured from North to East). Magnitude lower limits are given if a secondary is not detected in a filter. We list the absolute magnitude and corresponding mass in columns $7-10$. 
The 11th column lists the status of the object (c = confirmed companion star, nc = new confirmed companion star). The last column shows additional remarks. A ``J'', ``H'', or ``K'' means that the secondary flux in this filter was obtained from the image obtained {\em without} the NDF, using the PSF from the corresponding image that was obtained {\em with} NDF (see \S~\ref{section: componentdetection}). Properties of the observed primaries, candidate companions, and background stars are not shown here; these are listed in Table~\ref{table: longtable}.  \label{table: companionstable}} 
\end{table*}

\begin{table}
  \begin{tabular}{lrrrr}
    \hline
    & $\rho < 1''$ & $1''\leq\rho\leq 4''$ & $\rho > 4''$ & Total \\
    \hline
    $K_S < 12$~mag & 9 \ ($-$)   & 10 \ (2)  & 4 \ (2)  & 23 \ (4) \\ 
    $12 \leq K_S \leq 14$~mag &  $-$ \ ($-$) &  1 \ ($-$)  & $-$ \ (1)  & 1 \ (1) \\ 
    $K_S > 14$~mag &  $-$ \ ($-$) &  $-$ \ (3) &  1 \ (3) & 1 \ (6) \\
    no $K_S$ available & $-$ \ ($-$) & $-$ \ ($-$) & $-$ \ (1) & $-$ \ (1) \\
    \hline
    Total & 9 \ ($-$) & 11 \ (5) & 5 \ (7) & 25 \ (12) \\
    \hline
  \end{tabular}
  \caption{The distribution of companion stars over angular separation $\rho$ and $K_S$ magnitude for our sample of 31~targets with multi-color observations. Each entry lists the number of confirmed companion stars, i.e., the secondaries with $\chi^2 < 2.30$. Between brackets we list the number of candidate companion stars, which have $2.30 < \chi^2 < 11.8$. We have also included the candidate companion HIP80142-2 ($\rho=5.88''$), for which no $K_S$ measurement is available. Several candidate companions are likely to be background stars, especially faint candidates at large separation. In the region $1''\leq \rho \leq 4''$ and $12~{\rm mag} \leq K_S \leq 14$~mag (which we will study in \S~\ref{section: gap}) we find one confirmed companion and no candidate companions.  \label{table: 13sigma}}
\end{table}

The status given to each secondary is listed in Table~\ref{table: criteria}. The secondaries with $\chi^2 < 2.30$ are very likely companions because of their proximity to the isochrone. As statistically only 1~out of 370 companions have $\chi^2 > 11.8$, we claim with high confidence that the secondaries with $\chi^2>11.8$ are background stars. We cannot confirm the status of the secondaries with $2.30 < \chi^2 < 11.8$. These secondaries are consistent with the isochrone within the $3\sigma$ error bars. However, due to the large number of faint background stars, it is likely that several background stars also satisfy this criterion. As we cannot confirm either their companion status or their background star status, we will refer to the secondaries with  $2.30 < \chi^2 < 11.8$ as candidate companions.

The properties of the 25~confirmed companion stars found around the 22~members in the NACO survey and the 9~members with multi-color observations in the ADONIS survey are listed in Table~\ref{table: companionstable}. 
Table~\ref{table: 13sigma} lists the distribution of confirmed and candidate companions over angular separation $\rho$ and $K_S$ magnitude. The largest fraction of candidate companions (relative to the number of confirmed companions) is seen for faint secondaries at large angular separation. In Section~\ref{section: gap} we will study the virtual absence of companions with   $1''\leq\rho\leq 4''$ and $12 \leq K_S \leq 14$~mag. Table~\ref{table: 13sigma} shows that only one confirmed companion is detected in this region; no candidate companions are found.

\subsection{The background star population} \label{section: backgroundstarpopulation}

Our method to separate companions and background stars is based on a comparison between the location of the secondaries in the color-magnitude diagrams and the isochrone. The number of background stars identified with this method should be comparable to the {\em expected} number of background stars in the fields around the targets. In this section we make a comparison between these numbers, where the expected number of background stars is based on (1) the Besan\c{c}on model of the Galaxy, and (2) the background star study in Sco~OB2 performed by \cite{shatsky2002}.

We use the Besan\c{c}on model of the Galaxy \citep{besancon} to characterize the statistical properties of the background star population. We obtain star counts in the direction of the centers of the three subgroups, as well as for $(l,b)=(300^\circ,0^\circ)$, where LCC intersects the Galactic plane. We include objects of any spectral type, luminosity class, and population, up to a distance of 50~kpc, and convert the model $K$ magnitude to $K_S$ magnitude.
As expected, the Besan\c{c}on model shows a strong variation in the number of background stars with Galactic latitude. Most background stars are found in the direction of the Galactic plane.
For a given numerical value of the magnitude limit, more background stars are expected to be found in the $K_S$ band than in the $J$ and $H$ bands. 
Although the Besan\c{c}on model is in good agreement with the observed properties of the Galaxy, it cannot be used to make accurate predictions for individual lines of sight. For example, the high variability of the interstellar extinction with the line-of-sight, which is known to be important for the background star statistics \citep{shatsky2002}, is not taken into account in the Besan\c{c}on model. A high interstellar extinction reduces the observed number of background stars significantly, which is especially important for the US subgroup, which is located near the $\rho$~Oph star forming region.

\begin{figure}[btp]
  \centering
  \includegraphics[width=0.5\textwidth,height=!]{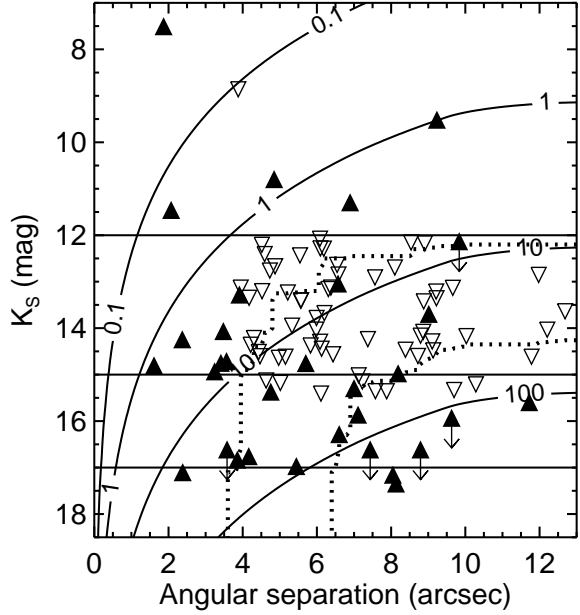}
  \caption{The {\em expected} number of background stars brighter than $K_S$ at angular separation smaller than $\rho$ for the combined sample of 199~targets (solid contours). Overplotted are the background stars detected around the 22~NACO targets (filled triangles), and those found around the 177~targets observed with ADONIS {\em only} (open triangles). The {\em observed} cumulative number of background stars is indicated with the dotted contours, for values of 10 and 50~background stars, respectively. In \cite{kouwenhoven2005} we classify secondaries brighter than $K_S=12$~mag (upper horizontal line) as candidate companions, and those fainter than $K_S=12$~mag as probable background stars. The faintest objects we detect in our ADONIS and NACO surveys have $K_S$ magnitude of approximately 15~and 17~magnitudes, respectively (lower horizontal lines). 
    \label{figure: besancon}}
\end{figure}

Let $F(K_S)$ be the number of background stars brighter than $K_S$, per unit of surface area. Above we mentioned that the number of background stars in the Besan\c{c}on model, i.e., the normalization of $F(K_S)$, varies strongly with Galactic coordinates. The profile of $F(K_S)$, however, is very similar for different lines of sight, and can be approximated with a function of the form $F(K_S) = C_i  \times 10^{\gamma \cdot K_S}$, with $\gamma = 0.32 \pm 0.01$ for $5 \leq K_S \leq 20$~mag. The constant $C_i$ defines the normalization of $F(K_S)$, which depends on the Galactic coordinates.

The number of background stars within a certain angular separation $\rho$ is proportional to the enclosed area $A(\rho)$ in the field of view within that angular separation. For our NACO observations we have a square detector of size $L_{\rm NACO}=14$~arcsec; for the observing strategy we used for the ADONIS observations we effectively have $L_{\rm ADO}=\frac{3}{2} \cdot 12.76$~arcsec \citep[see][]{kouwenhoven2005}. For a given $\rho$ (in arcsec) the enclosed area (in arcsec$^2$) is then given by
\begin{equation} \label{equation: area}
  A_{i}(\rho) =  \left\{ 
  \begin{array}{lll}
    \rho^2 & \mbox{for} & \rho \leq L_i/2 \\
    \multicolumn{3}{l}{ 
      \pi \rho^2 - 4\rho^2 \arccos \left( L_i/2\rho \right) 
      + L_i \sqrt{4\rho^2 - L_i^2} 
      } \\
    \quad\quad\quad\quad\quad\quad\quad\quad 
    & \mbox{for} & L_i/2 < \rho \leq L_i/\sqrt{2} \\
    L_i^2 & \mbox{for} & \rho > L_i/\sqrt{2} \\
  \end{array}
  \right. \,,
\end{equation}
where the subscript $i$ refers to either the ADONIS or the NACO observations.
Here we make the assumption that the target star is always in the center of the field of view. In our NACO survey we occasionally observe the target star off-axis in order to study a secondary at angular separation larger than $L_{\rm NACO}/\sqrt{2} = 9.9''$, but we ignore this effect here.

We now have expressions for the quantity $N(K_S,\rho)$, the expected number of background stars with magnitude brighter than $K_S$ and angular separation smaller than $\rho$, as a function of $K_S$ and $\rho$:
\begin{equation} \label{equation: backgroundstars}
  N_{i}(K_S,\rho) = F(K_S) \cdot A_{i}(\rho)  = C_{i} \cdot 10^{\gamma \cdot K_S} \cdot  A_{i}(\rho) \,,
\end{equation}
where $\gamma= 0.32 \pm 0.01$, $A(\rho)$ is defined in Equation~\ref{equation: area}, and $C_{i}$ is a normalization constant.

\begin{table}[btp]
  \begin{tabular}{lcccc}
    \hline
     & \multicolumn{2}{l}{\ \quad ADONIS} & \multicolumn{2}{l}{\ \quad NACO} \\
    \hline
    Region          & stars per field          & $N_{\rm fields}$ & stars per field &  $N_{\rm fields}$ \\
    \hline
    US             & $0.06\quad^{+0.06}_{-0.02}$ & 51 &$0.10\quad^{+0.11}_{-0.04}$ & 9 \\
    UCL            & $0.27\quad^{+0.06}_{-0.02}$ & 64 &$0.47\quad^{+0.11}_{-0.10}$ & 3 \\
    LCC            & $0.43 \pm 0.08$        & 40 &$0.73 \pm 0.14$        & 5 \\
    GP             & $1.51 \pm 1.03$        & 22 &$2.59 \pm 1.76$        & 5 \\
    \hline
    Total          &                        & 177&                       & 22 \\
    \hline
  \end{tabular}
  \caption{The number of background stars {\em expected} in our ADONIS and NACO fields of view, based on the background star study of \cite{shatsky2002}. Column~1 lists the four regions for which we study the background statistics. The targets with Galactic latitude $|b|\leq 5^\circ$ are included in the region GP. The other targets are grouped according to their membership of US, UCL, and LCC. For the 177~targets {\em only} observed with ADONIS and the 22~targets observed with NACO we list the expected number of background stars per field of view, and the number of targets observed in the four regions. In total we expect to find $70.75 \pm 4.90$ background stars in the ADONIS sample, and $18.96 \pm 3.96$ in the NACO sample.  \label{table: backgroundstars}}
\end{table}

For the normalization of Equation~\ref{equation: backgroundstars} we compare our observations with the background star study of \cite{shatsky2002}. Apart from their target observations, the authors additionally obtained sky images centered at 21~arcsec from each target in order to characterize the background star population. 
From the background star study of \cite{shatsky2002} we derive the expected number of background stars in the ADONIS and NACO field of view. We assume a detection limit of $K_S = 15$~mag for ADONIS and $K_S=17$~mag for NACO \citep[which roughly corresponds to the completeness limit of the background star study of][]{shatsky2002}. Using the data from the background star study of \cite{shatsky2002} we calculate the expected number of background stars for each ADONIS and NACO field, for the three subgroups and for targets close to the Galactic plane ($|b|<5^\circ$). Table~\ref{table: backgroundstars} lists the expected number of background stars with corresponding Poisson errors. Additionally, we list the number of targets observed in each of the four regions. For the 177~targets observed with ADONIS {\em only} we expect $\approx 71$ background stars, and for the 22~targets observed with NACO we expect $\approx 19$ background stars. 
In our combined ADONIS and NACO dataset, the expected number of background stars is $90 \pm 6$. This gives normalization factors  $C_{\rm ADONIS}=1.74 \times 10^{-8}$ arcsec$^{-2}$ and  $C_{\rm NACO}=1.60 \times 10^{-8}$ arcsec$^{-2}$ for Equation~\ref{equation: backgroundstars}.  

Figure~\ref{figure: besancon} shows the expected number of background stars brighter than $K_S$ and closer than $\rho$ as a function of $K_S$ and $\rho$, for the combined ADONIS and NACO dataset. Since most of our targets are observed with ADONIS only, the shape of the expected background star density distribution is dominated by that of ADONIS. 
As we have a square field of size $L_{\rm ADO}=19.1$~arcsec, the cumulative number of background stars rises rapidly between $\rho=0''$ and $\rho=L_{\rm ADO}/\sqrt{2} =13.5''$ and becomes flat for larger separations. The background stars from the NACO survey are represented with the filled triangles, and those detected {\em only} in the ADONIS survey are represented with open triangles. Figure~\ref{figure: besancon} shows that the $(\rho,K_S)$ distribution of the observed 97~background stars is in good agreement with that of the expected 90~background stars. In the extreme case that all 12~candidate companions are actually background stars, the observed number of background stars is~110. In this extreme case there are 22\% more background stars than expected, which suggests that a significant part of the candidate companions may indeed be physical companions.

\begin{figure}[btp]
  \centering
  \includegraphics[width=0.5\textwidth,height=!]{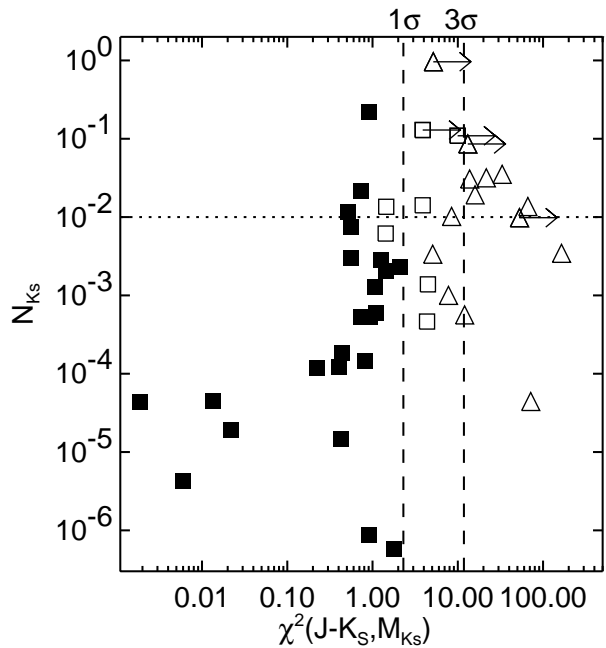}
  \caption{The $\chi^2$ distance to the isochrone in the $(J-K_S,M_{K_S})$ diagram versus $N(K_S,\rho)$, where $N(K_S,\rho)$ is the expected number of background stars brighter than $K_S$ and angular separation smaller than $\rho$.
The vertical dashed lines correspond to $\chi^2=2.30$ ($1\sigma$; {\em left}) and $\chi^2=11.8$ ($3\sigma$; {\em right}). The symbols represent the confirmed companions (filled squares), the candidate companions (open squares), and background stars (triangles). Note that the classification of a secondary is based on the $\chi^2$ values for the different color-magnitude diagrams; not only for the $\chi^2$ of the $(J-K_S,M_{K_S})$-diagram, which is shown above. The expected number of background stars in a field (vertical axis) is used as a consistency check. The horizontal dotted line represents the 1\%~filter used by \cite{poveda1982} to separate companion stars (below the line) and background stars (above the line). The 1\% filter is a reasonable method when multi-color observations are not available, but is not used in our study. The triangle in the lower-right quadrant represents HIP63204-1 (see~\S~\ref{section: individualnotes}). \label{figure: bgchanceandisochrone}}
\end{figure}

Background stars are by definition not associated with the target star, and therefore generally have (1) a random position in the field of view of the observation, and (2) a position in the color-magnitude diagram that is likely to be inconsistent with the isochrone.
Figure~\ref{figure: bgchanceandisochrone} shows the relation between $N(K_S,\rho)$ and the $\chi^2$ value derived from the location of the secondary in the ($J-K_S,M_{K_S}$) diagram with the isochrone . In other words, Figure~\ref{figure: bgchanceandisochrone} shows the probability of detecting a background star at separation $\rho$ (or smaller) and magnitude $K_S$ (or brighter) versus how far away the secondary is from the isochrone.
The vertical dashed lines in Figure~\ref{figure: bgchanceandisochrone} are at $\chi^2=2.30$ and $\chi^2=11.8$, the values used to determine the status of the companions (see Table~\ref{table: classification}).
This correlation provides additional support to the method we use to separate companions and background stars.

\cite{poveda1982} performed a statistical study of binary stars in the Index Catalogue of Visual Double Stars. They showed that it is statistically plausible to assume that components with $N(\rho,m_2) > 0.01$ are background stars, where $m_2$ is the magnitude of the secondary. This technique is referred to as the ``1\%~filter''. The horizontal line in Figure~\ref{figure: bgchanceandisochrone} represents the 1\%~filter used by \cite{poveda1982}. Secondaries below this line would be classified as companions using the 1\%~filter, and those above would be classified as background stars. Figure~\ref{figure: bgchanceandisochrone} shows that the 1\%~filter is a reasonable technique, but not as accurate as the multi-color technique used in this paper.

\subsection{Notes on some individual secondaries} \label{section: individualnotes}

We detect two hierarchical triple systems: HIP68532 and HIP69113. The two companions of HIP68532 \citep[previously reported in][]{kouwenhoven2005} have a mass of 0.73~M$_\odot$ and 0.39~M$_\odot$, respectively. HIP68532 has a companions-to-primary mass ratio of $(0.73+0.39)/1.95 = 0.57$. The companions are separated $0.23''$ ($\sim28$~AU) from each other and $3.11''$ ($\sim385$~AU) from the primary, giving an estimate of 0.073 for the ratio between the semi-major axes of the inner and outer orbits. The two companions of HIP69113 \citep[previously reported in][]{huelamo2001} have a mass of 0.77~M$_\odot$ and 0.72~M$_\odot$, respectively, corresponding to a companions-to-primary mass ratio of 0.39. The companions are separated $0.26''$ ($\sim44$~AU) from each other and $5.43''$ ($\sim917$~AU) from the primary, giving an estimate of 0.048 for the ratio between the semi-major axes of the inner and outer orbits.

For HIP62026-1 we find a significant difference in position angle between the ADONIS and NACO observations. With the ADONIS observations, obtained on 8~June 2001, we find $(\rho,\varphi)=(0.22'',12.5^\circ)$. With NACO we measure $(\rho,\varphi)=(0.23'',6.34^\circ)$, 1033~days later. As the angular separation between HIP62026-1 and its primary is small, the observed position angle difference may well be the result of orbital motion. Assuming a circular, face-on orbit, we estimate an orbital period of 165~year for the system HIP62026. The total mass of the system (taken from Table~\ref{table: longtable}) is $3.64\pm 0.25~{\rm M}_\odot$, which gives via Kepler's law a semi-major axis of 46~AU. This value is of the same order of magnitude as the (projected) semi-major axis of $\sim 24$~AU derived from the angular separation of $0.22''$ between the components (adopting a distance of 109~pc to HIP62026).

HIP63204-1 is a bright and red object separated only $1.87$~arcsec from the LCC member HIP63204. The isolated location of HIP63204-1 in the bottom-right quadrant of Figure~\ref{figure: bgchanceandisochrone} shows that the probability of finding a background star of this magnitude (or brighter) at this angular separation (or closer) is small. According to its location in the color-magnitude diagrams, HIP63204-1 is a background star and hence we classify it as such. HIP63204 and its companion HIP63204-2 at $\rho=0.15''$ have masses of $2.05~{\rm M}_\odot$ and $1.06~{\rm M}_\odot$, respectively. If HIP63204-1 (at $\rho=1.87''$) would be a companion, its mass would be approximately $1~{\rm M}_\odot$, in which case HIP63204 would be an unstable triple system.  The colors of HIP63204-1 are consistent with a $0.075~{\rm M}_\odot$ brown dwarf with an age of 10~Gyr at a distance of 60~pc, and are also consistent with those of an M5~III giant at a distance of $\sim 5.6$~kpc \citep[using the models of][]{allen2000,chabrier2000}.

HIP81972-3, HIP81972-4, and HIP81972-5 fall on the 20~Myr isochrone in all three color-magnitude diagrams. These objects are likely low-mass companions of HIP81972 (see \S~\ref{section: massfunction}).  HIP81972-5 is the topmost companion (black square) in Figure~\ref{figure: bgchanceandisochrone}. HIP81972-5 is the faintest companion in our sample, and the expected number of background stars with a similar or brighter magnitude and a similar or smaller separation is large ($\sim 16$ for the ADONIS sample). The secondary HIP81972-2 (at $\rho=7.02''$) was reported before as a possible companion of HIP81972 in \cite{wds1997}, but the secondary is classified as a background star by \cite{shatsky2002}. With our NACO multi-color observations we cannot determine the nature of this secondary with high confidence. The LCC member HIP81972 is located close to the Galactic equator ($b=+3^\circ 11'$), so care should be taken; background stars with a magnitude similar to that of the secondaries of HIP81972 are expected in the field around this star.

\cite{kouwenhoven2005} identified seven ``close background stars'' ($K_S > 14~\mbox{mag}; \rho < 4''$), for which the background star status was derived using the $K_S$ magnitude only. These are objects next to the targets HIP61265, HIP67260, HIP73937, HIP78958, HIP79098, HIP79410, and HIP81949. The ADONIS observations are incomplete in this region (see Figure~\ref{figure: detectionlimits}). More low-mass companions with $K_S > 14$~mag and $1'' < \rho < 4''$ may be present for the 177~A and late-B stars {\em only} observed with ADONIS.

With our NACO multi-color observations we confirm that five of the ``close background stars'' (HIP61265-2, HIP73937-2, HIP79098-1, 79410-1, and HIP81949-2) are background stars. 
For the other two secondaries, HIP67260-3 and HIP78969-1, we cannot determine whether they are background stars or brown dwarf companions. As many background stars with similar magnitudes are expected in the field, these are likely background stars. However, follow-up spectroscopic observations are necessary to determine the true nature of these close secondaries.

\subsection{Accuracy of the $K_S=12$ separation criterion} \label{section: k12}

\begin{table}[btp]
  \begin{tabular}{c cc cc c}
    \hline
    Status & \multicolumn{2}{l}{$K_S < 12$~mag} & \multicolumn{2}{l}{$K_S > 12$~mag} & Total \\
    \hline
    c & 23 &(70\%) & 2& (6\%)& 25 \\
    ? & 4 &(12\%)& 7 &(19\%)& 11 \\
    b & 6 &(18\%)& 27& (75\%) & 33 \\
    \hline
    Total  & 33 &(100\%)& 36& (100\%)& 69 \\  
    \hline
  \end{tabular}
  \caption{Accuracy of the $K_S=12$~mag criterion to separate companions and background stars. This table contains 69~out of the 72~secondaries with multi-color observations in the ADONIS or NACO dataset. Three secondaries (1~candidate companion; 2~background stars) for which no $K_S$ magnitudes are available, are not included. The first column shows the status of the secondary (c = confirmed companion, ? = candidate companion, b = background star). Columns~2 to~5 list the distribution over status for secondaries with $K_S<12$~mag and $K_S>12$~mag. Depending on the true nature of the candidate companion stars, the $K_S=12$~mag criterion correctly classifies the secondaries in $\sim 80\%$ of the cases. \label{table: k12comparison}}
\end{table}

One of the goals of our study is to evaluate the accuracy of the $K_S=12$~mag criterion that we used to separate companions and background stars in the ADONIS survey. This is possible, now that we have performed a multi-color analysis of 72~secondaries around 31~members of Sco~OB2.
Table~\ref{table: k12comparison} shows the distribution of secondary status for the secondaries with $K_S <12$~mag and those with $K_S>12$~mag (three secondaries without $K_S$ measurements are not included). According to the $K_S=12$~criterion, all secondaries brighter than $K_S=12$~mag are companions, while all fainter secondaries are background stars.

If we consider only the confirmed companions and background stars, we see that the $K_S=12$~mag criterion correctly classifies the secondaries in $f = (23+27)/(25+33) = 86\%$ of the cases. If all candidate companions are indeed companions, we have $f=78\%$, while if all candidate companions are background stars, we have $f=82\%$. This indicates that  $\sim 80\%$ of the candidate companions identified by \cite{kouwenhoven2005} are indeed companion stars.

The $K_S=12$~mag criterion is accurate for the measured set of secondaries {\em as a whole}. It is obvious that for the lowest-mass companions the criterion is not applicable, as virtually all brown dwarf and planetary companions have $K_S > 12$~mag at the distance of Sco~OB2. Out of the 25~confirmed companions we find with our multi-color analysis, 23 indeed have $K_S < 12$~mag, but two have $K_S > 12$~mag. These are the brown dwarf companions of HIP81972 (see\S~\ref{section: separation} and~\ref{section: massfunction} for a discussion).  

Now that we have confirmed the validity of the $K_S=12$~mag criterion, many of the candidate companions found in the ADONIS survey very likely are physical companion stars. Table~\ref{table: adonisnaco} gives an overview of all candidate and confirmed companion stars identified in the ADONIS and NACO surveys.


\section{Masses and mass ratios} \label{section: massfunction}

For each primary and companion star we derive the mass using its color and magnitude. We find the best-fitting mass by minimizing the $\chi^2$ difference between the isochrone and the measurements, while taking into account the errors in the measurements:
\begin{equation}
\chi^2 = \left( \frac{\Delta (J-K_S)}{\sigma_{J-K_S}} \right)^2 +  \left( \frac{\Delta (H-K_S)}{\sigma_{H-K_S}} \right)^2 +  \left( \frac{\Delta M_{K_S}}{\sigma_{M_{K_S}}} \right)^2 .
\end{equation}
The masses of all primaries and confirmed companions are listed in Table~\ref{table: longtable}. We additionally list the masses of the candidate companions, assuming that they are indeed companions, but we do not include these masses in our analysis. We find primary masses between $1.1~{\rm M}_\odot$ and $4.9~{\rm M}_\odot$. The confirmed companion star masses range between $0.03~{\rm M}_\odot$  and $1.19~{\rm M}_\odot$, with mass ratios $0.006 < q < 0.55$.  The average error in the mass as a result of the error in the color and magnitude is $8.5\%$ for the primaries and $12\%$ for the companion stars. The average error in the mass ratio is $12.5\%$.
Although accurate $B$ and $B-V$ measurements are available for the primaries, we do not use these. The $B$ and $V$ measurements often include the flux of unresolved close companions, and therefore lead to overestimating the primary masses. 

\cite{kouwenhoven2005} derived masses from $K_S$ magnitudes only. For the primary stars these are close to those obtained from multi-color observations in our current analysis. The rms difference between the masses derived using the two methods is 6.6\%. No systematic difference is present for the primaries. \cite{kouwenhoven2005} overestimated the companion star masses with $\sim 2.2\%$ ($\sim 0.01~{\rm M}_\odot$) on average.

The companions with the lowest mass are those of HIP81972, which have masses of $0.35~{\rm M}_\odot$ ($370~{\rm M_J}$), $0.06~{\rm M}_\odot$ ($63~{\rm M_J}$), and $0.03~{\rm M}_\odot$ ($32~{\rm M_J}$). The latter two are likely brown dwarfs. With an angular separation of $\rho=2.79''$, the $63~{\rm M_J}$ component is the only observed brown dwarf in the $1''-4''$ angular separation interval (see~\S~\ref{section: gap}). The other two companions of HIP81972 have a larger angular separation.


\section{The lower end of the companion mass distribution} \label{section: gap}

\begin{figure}[btp]
  \centering
  \includegraphics[width=0.5\textwidth,height=!]{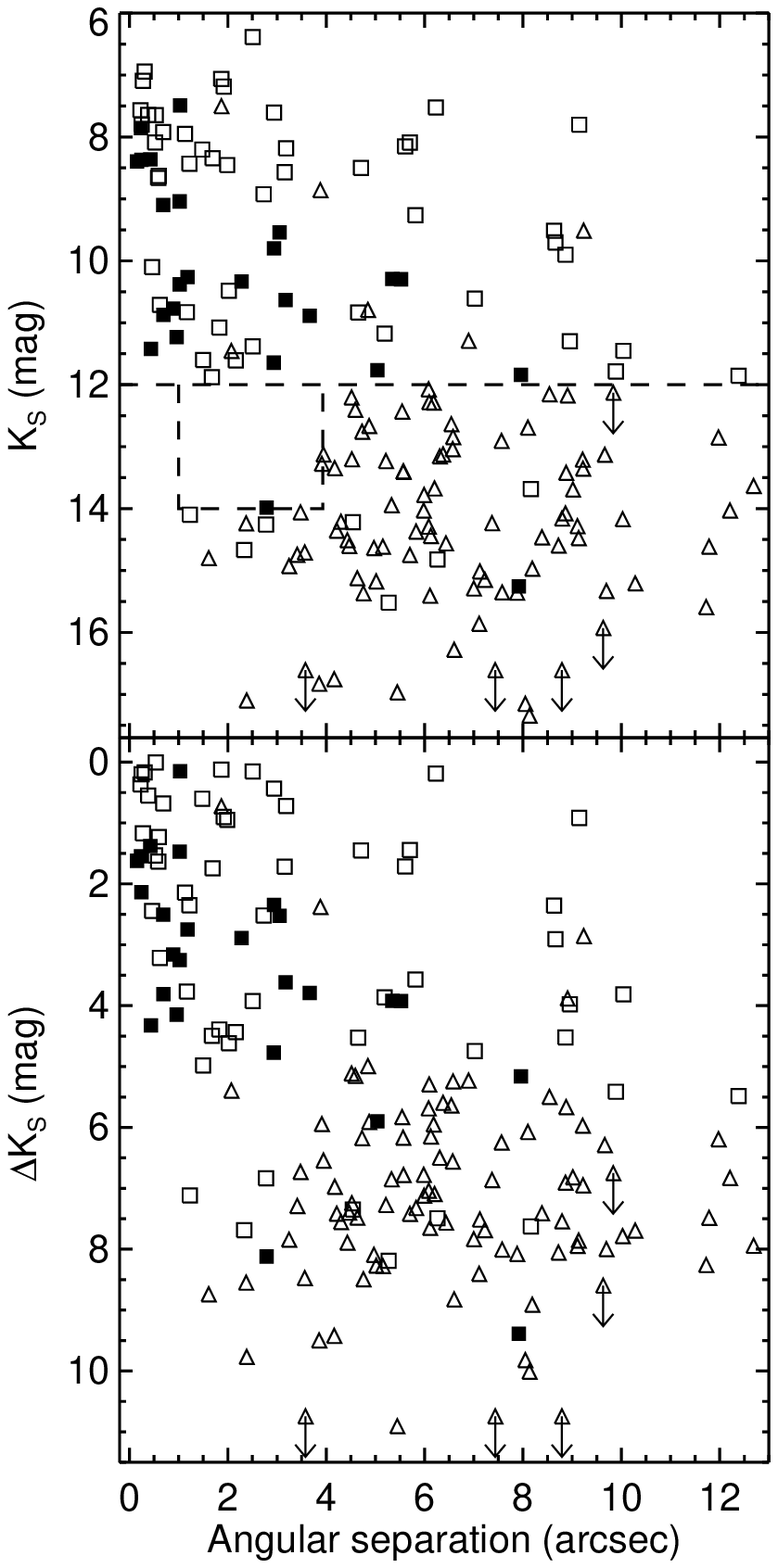}
  \caption{Companion star magnitude $K_S$ ({\em top}) and magnitude difference $\Delta K_S$ ({\em bottom}) versus angular separation for the combined ADONIS and NACO datasets.
The symbols represent confirmed companions (filled squares), candidate companions (open squares), and background stars (triangles); see Section~\ref{section: separation} for further information on this classification.
The horizontal line corresponds to $K_S=12$~mag, the criterion used by \cite{kouwenhoven2005} to separate companion stars and background stars. Typical detection and completeness limits corresponding to the observations are shown in Figure~\ref{figure: detectionlimits}. For a given $K_S$ magnitude, the number of background stars closer than angular separation $\rho$ is given by Equation~\ref{equation: backgroundstars}. This figure clearly shows the dearth of brown dwarf companions with $1'' < \rho < 4''$ around A and late-B stars in Sco~OB2 (region indicated with the dashed rectangle).  \label{figure: detlimk} } 
\end{figure}

\cite{kouwenhoven2005} discussed the potential lack of substellar companions to A and late-B members of Sco~OB2. With our NACO follow-up observations we confirm the very low number of brown dwarf companions with respect to the number of stellar companions found around these stars (\S~\ref{section: gapexistence}). In Section~\ref{section: realdesert} we will discuss whether a brown dwarf desert exists among A and late-B members of Sco~OB2. In Section~\ref{section: origin} we will briefly discuss the potential origin of such a brown dwarf desert. We will show that the small brown dwarf companion fraction among A- and B stars in Sco~OB2 can be explained by an extrapolation of the {\em stellar} companion mass distribution, i.e., there is no need to eject brown dwarf companions from binary systems at an early stage of the formation process \cite[the embryo ejection scenario;][]{reipurth2001}.

\subsection{The brown dwarf ``gap'' for $1'' \leq \rho \leq 4''$} \label{section: gapexistence}

\cite{kouwenhoven2005} observed a gap in the $(\rho,K_S)$ distribution of the stellar companions in the Sco~OB2 binary population: no secondaries with a magnitude $12~\mbox{mag} \leq K_S \leq 14~\mbox{mag}$ and an angular separations $\rho < 4''$ were detected. These secondaries should have been detected, had they existed, since the ADONIS survey is almost complete in this region (see Figure~\ref{figure: detectionlimits}). Figure~\ref{figure: detlimk} shows the distribution of $K_S$ and $\Delta K_S$ as a function of $\rho$ for the ADONIS and NACO observations combined. The ``gap'' in the $(\rho,K_S)$ distribution described above is clearly visible. With our NACO survey we detect one companion at the bottom of this region: the brown dwarf companion HIP81972-4 ($K_S = 13.98 \pm 0.12$~mag; see also \S~\ref{section: separation} and~\ref{section: massfunction}). No other secondaries are present in this region.

The stellar companion fraction is the fraction of stars with stellar companions. Among A and late-B stars in Sco~OB2 in the semi-major axis range $1''-4''$ ($130-520$~AU) we find a stellar companion fraction of $14\pm 3\%$. Similarly, the brown dwarf companion fraction is the fraction of stars with brown dwarf companions. The brown dwarf companion fraction for the stars in this separation range is $0.5 \pm 0.5\%$\footnote{Note that we only find one brown dwarf companion with $12~{\rm mag} \leq K_S \leq 14~{\rm mag}$. We find no background stars in this region, so accidental misclassification of companions as background stars is not an issue here.} (for brown dwarfs with $K_S < 14$~mag). The substellar-to-stellar companion ratio $R$ is defined as
\begin{equation} \label{equation: substel-to-stel}
R = \frac{ \mbox{number of brown dwarf companions} }{ \mbox{number of stellar companions} } \,.
\end{equation}
In our study we cannot calculate $R$ because we do not know how many faint ($K_S > 14$~mag) brown dwarf companions are missing. We therefore calculate the {\em restricted} substellar-to-stellar companion ratio $R_\star$, using only the brown dwarf companions brighter than $K_S=14$~mag\footnote{In this section we denote the {\em observed} quantities with a star as a subscript. For example, $R$ denotes the substellar-to-stellar companion ratio (including all brown dwarfs), while $R_\star$ indicates the {\em observed} substellar-to-stellar companion ratio, including only the brown dwarfs brighter than $K_S=14$~mag.}.  In the angular separation range $1''-4''$ we observe $R_{\rm \star,IM} = 0.036 \pm 0.036$ for intermediate mass stars in Sco~OB2.
We cannot make similar statements for companions with properties other than those described above. For angular separations smaller than $1''$ our survey is significantly incomplete for $12~\mbox{mag} \leq K_S \leq 14~\mbox{mag}$. For $\rho > 4''$ many objects with $12~\mbox{mag} \leq K_S~\leq 14~\mbox{mag}$ are likely background stars, of which the status still needs to be confirmed. Finally, the ADONIS survey is incomplete for $K_S > 14$~mag.

\subsection{A real brown dwarf desert?} \label{section: realdesert}

The brown dwarf desert is defined as a deficit (not necessarily a total absence) of brown dwarf companions, either relative to the frequency of companion stars or relative to the frequency of planetary companions \citep{mccarthy2004}. 
In this paper the brown dwarf desert for A~and late-B members of Sco~OB2 is characterized by a small number of observed companions $N_{\rm \star,BD}$ with $12~\mbox{mag} \leq K_S \leq 14$~mag and $1'' \leq \rho \leq 4''$ and a small substellar-to-stellar companion ratio $R_\star$ (where the star in the subscript refers to the {\em observed} quantities). In general, the quantities $N_{\rm BD}$ and $R$ depend on  (1) the mass distribution, (2) the pairing properties of the binary systems, and (3) the spectral type of the stars in the sample. Unlike $R$, the value of $N_{\rm BD}$ also depends on (4) the multiplicity fraction $F_{\rm M}$ (Equation~\ref{equation: binarystatistics}), and (5) the semi-major axis (or period) distribution. We use simulated observations and compare the observed values with those predicted for various models, in order to roughly estimate the mass distribution and pairing properties. For comparison between observations and simulations we only consider the companions brighter than $K_S=14$~mag in the angular separation range $1'' \leq \rho \leq 4''$.

\begin{table*}
  \begin{tabular}{llcccc}
    \hline
    \# & Model & $N_{\rm \star,BD,IM}$ & $R_{\rm \star,IM}$  & $N_{\rm \star,BD,LM}$ & $R_{\rm \star,LM}$ \\
    \hline
    0 & ADONIS/NACO observations                                                                    & $1   \pm 1  $    & $0.036 \pm 0.036$ & unknown             & unknown \\
    \hline
    1 &  extended Preibisch  MF, $\alpha=-0.9$, random pairing             & $      5.50\pm       0.48 $&$     0.34\pm     0.03 $&$       7.19\pm      0.17 $&$      0.50 \pm    0.01  $ \\ 
    2 &  extended Preibisch  MF, $\alpha=-0.3$, random pairing             & $      4.50\pm       0.41 $&$     0.24\pm     0.03 $&$       5.08\pm      0.13 $&$      0.30 \pm    0.01  $ \\ 
    3 &  extended Preibisch  MF, $\alpha=+2.5$, random pairing            & $      1.07\pm       0.18 $&$     0.05\pm     0.01 $&$       1.42\pm      0.07 $&$      0.07 \pm    0.01  $ \\ 
    4 &  Salpeter MF, random pairing                                      & $     15.31\pm       2.79 $&$     6.00\pm     2.90 $&$      17.18\pm      0.88 $&$      3.95 \pm    0.45  $ \\ 
    5 &  extended Preibisch MF, $\alpha=-0.9$, $f_q(q) \propto q^{-0.33}$    & $      0.72\pm       0.24 $&$     0.04\pm     0.01 $&$       3.42\pm      0.14 $&$      0.18 \pm    0.01  $ \\
    6 &  extended Preibisch MF, $\alpha=-0.3$, $f_q(q) \propto q^{-0.33}$    & $      0.71\pm       0.22 $&$     0.04\pm     0.01 $&$       3.35\pm      0.14 $&$      0.18 \pm    0.01  $ \\ 
    7 &  extended Preibisch MF, $\alpha=+2.5$, $f_q(q) \propto q^{-0.33}$   & $      1.19\pm       0.27 $&$     0.06\pm     0.01 $&$       3.30\pm      0.13 $&$      0.18 \pm    0.01  $ \\ 
    8 &  Salpeter MF, $f_q(q) \propto q^{-0.33}$                            & $      1.00\pm       0.57 $&$     0.05\pm     0.02 $&$       3.70\pm      0.57 $&$      0.20 \pm    0.03  $ \\ 
    \hline
   \end{tabular}
  \caption{The observed and expected number of brown dwarfs with $1'' \leq \rho \leq 4''$ and $12 \leq K_S \leq 14$~mag for the sample of 199~target stars. The left columns shows the various models for which we simulated observations. Each model has a semi-major axis distribution $f_a(a) \propto a^{-1}$ with $15~$R$_\odot \leq a \leq 5\times 10^6$~R$_\odot$, and a multiplicity fraction of $F_{\rm M} =100\%$. 
Columns 3 and 4 show for a survey of intermediate mass stars (late-B and A stars; $1.4~{\rm M}_\odot < M < 7.7~{\rm M}_\odot$) the expected number of brown dwarfs $N_{\rm \star,BD,IM}$ and the substellar-to-stellar companion ratio $R_{\rm \star,IM}$, both with $1\sigma$ errors. 
By comparing the predicted values of $N_{\rm \star,BD,IM}$ and $R_{\rm \star,IM}$ with the observations, we can exclude models 1, 2, and 4. In \cite{kouwenhoven2005} we exclude random pairing from the Preibisch mass distribution (models 1$-$3) since these models are inconsistent with the observed mass ratio distribution of stellar companions. We additionally list the values $N_{\rm \star,BD,LM}$ and $R_{\rm \star,LM}$ that are expected for a survey amongst 199~low-mass stars ($0.3~{\rm M}_\odot < M < 1.4~{\rm M}_\odot$) in columns 5 and 6. For models with $F_{\rm M} < 100\%$ the expected number of brown dwarfs reduces to $F_{\rm M} \times N_{\rm \star,BD}$, while $R$ remains unchanged. Models with a smaller semi-major axis range and models with the log-normal period distribution found by \cite{duquennoy1991} have a larger expected value of $N_{\rm \star,BD,IM}$, $N_{\rm \star,BD,LM}$. Under the assumption that companion mass and semi-major axis are uncorrelated, the values of  $R_{\rm \star,IM}$ and  $R_{\rm \star,LM}$ are equal to those listed above, if the log-normal period distribution is chosen.
    \label{table: modelbackgroundstars}}
\end{table*}

We simulate models using the STARLAB package \citep[e.g.,][]{ecology4}. The primary mass is drawn in the mass range $0.02~{\rm M}_\odot \leq M \leq 20~{\rm M}_\odot$, either the Salpeter mass distribution 
\begin{equation} \label{equation: imfsalpeter}
f_M(M) = \frac{dM}{dN} \propto M^{-2.35} \,,
\end{equation}
or from the the extended Preibisch mass distribution
\begin{equation} \label{equation: imfpreibisch}
  f_M(M) = \frac{dM}{dN} \propto \left\{
  \begin{array}{llll}
    M^{\alpha}   & {\rm for \quad } 0.02 & \leq M/{\rm M}_\odot & < 0.08 \\
    M^{-0.9}     & {\rm for \quad } 0.08 & \leq M/{\rm M}_\odot & < 0.6 \\
    M^{-2.8}     & {\rm for \quad } 0.6  & \leq M/{\rm M}_\odot & < 2   \\
    M^{-2.6}     & {\rm for \quad } 2    & \leq M/{\rm M}_\odot & < 20  \\
  \end{array}
  \right.\,.
\end{equation}
The extended Preibisch mass distribution  \citep[see][]{kouwenhoven2005} is based on the study by \cite{preibisch2002}, who derived $f_M(M)$ with $M > 0.1$~M$_\odot$ for the US subgroup of Sco~OB2.  Since our current knowledge about the brown dwarf
population in OB~associations (particularly Sco~OB2) is incomplete \citep[e.g., Table 2 in][]{preibisch2003}
we simulate associations with three different slopes for the mass distribution
in the brown dwarf regime. 
We extend the Preibisch mass distribution down to $0.02~{\rm M}_\odot$ with $\alpha=-0.9$, $\alpha=-0.3$, or $\alpha=+2.5$. The mass distribution with $\alpha=-0.9$ has the same slope in the brown dwarf regime as for the low-mass stars. The simulations with $\alpha=-0.3$ and $\alpha=+2.5$ bracket the values for $\alpha$ that are observed in various clusters and the field star population \citep[see][for a summary]{preibisch2003}.
The companion mass is obtained via randomly pairing the binary components from the mass distribution or via a mass ratio distribution of the form $f_q(q) \propto q^{-0.33}$ with $0 < q < 1$ and the requirement that any companion has a mass larger than $0.02~{\rm M}_\odot$. The latter mass ratio distribution was derived  from the observed mass ratio distribution in our ADONIS survey \citep{kouwenhoven2005}. For the models with random pairing, the primary star and companion mass are drawn independently from the mass distribution, and switched, if necessary, so that the primary is the most massive star.

Each simulated association consists of 100\,000 binaries, has a distance of 130~pc and a homogeneous density distribution with a radius of 20~pc, properties similar to those of the subgroups in Sco~OB2. We assume a thermal eccentricity distribution and a  semi-major axis distribution of the form $f_a(a) \propto a^{-1}$, which is equivalent to $f_{\log a}(\log a) = \mbox{constant}$ (\"{O}piks law). The lower limit of $a$ is set to $15~$R$_\odot$. The upper limit is set to $5\times 10^6~$R$_\odot \approx 0.1~\mbox{pc}$, the separation of the widest observed binaries in the Galactic disk \citep[e.g.,][]{close1990,chaname2004}.

Table~\ref{table: modelbackgroundstars} lists for eight models the predicted value of $N_{\rm \star,BD}$, the expected number of brown dwarfs with $1'' \leq \rho \leq 4''$ and $12 \leq K_S \leq 14$~mag, normalized to a sample of 199~stars. The table also lists $R_{\star}$, the ratio between the number of brown dwarf companions with $K_S \leq 14$~mag and the number of stellar companions in the separation range $1''-4''$.
Table~\ref{table: modelbackgroundstars2} lists $N_{\rm BD}$ (the intrinsic number of brown dwarfs with $1''\leq \rho \leq 4''$), and corresponding ratio $R$. In this table {\em all} companions are taken into account, including those with $K_S > 14$~mag. The values in Table~\ref{table: modelbackgroundstars} can be compared directly with the observations, while those in Table~\ref{table: modelbackgroundstars2} represent the intrinsic properties of each association model.
A brown dwarf with $K_S=14$~mag in the US subgroup has a mass of less than $0.02~$M$_\odot$  (21~${\rm M_J}$), and a brown dwarf with a similar brightness in the UCL and LCC subgroup has a mass of $\sim 0.038~$M$_\odot$ (40~${\rm M_J}$).

The definition of the brown dwarf desert given in the beginning of this section is generally used for binarity studies of late-type stars. In our study the primaries are intermediate mass stars, allowing companion stars over a larger mass range than for low mass primaries. This naturally leads to lower values of $N_{\rm BD}$ and $R$. 
We therefore also list the results for a simulated survey of 199~low mass stars. Tables~\ref{table: modelbackgroundstars} and~\ref{table: modelbackgroundstars2} show that indeed the expected values $N_{\rm BD}$ and $R$ for low-mass stars are higher than those for intermediate-mass stars by $\sim 30\%$ for the random pairing models, and by $\sim 250\%$ for the models with $f_q(q) \propto q^{-0.33}$.

\begin{table}
  \begin{tabular}{lcccc}
    \hline
    \# & $N_{\rm BD,IM}$ & $R_{\rm IM}$  & $N_{\rm BD,LM}$ & $R_{\rm LM}$ \\
    \hline
    0  &  unknown      & unknown & unknown            & unknown \\
    \hline
    1  & $      6.68 \pm      0.53  $&$      0.44 \pm     0.04  $&$       8.50 \pm      0.18  $&$      0.65 \pm     0.02  $\\ 
    2  & $      5.46 \pm      0.45  $&$      0.31 \pm     0.03  $&$       6.42 \pm      0.15  $&$      0.42 \pm     0.01  $\\ 
    3  & $      2.01 \pm      0.25  $&$      0.10 \pm     0.01  $&$       2.65 \pm      0.09  $&$      0.14 \pm     0.01  $\\
    4  & $     15.82 \pm      2.84  $&$      7.75 \pm     4.12  $&$      18.26 \pm      0.91  $&$      5.60 \pm     0.72   $\\ 
    5  & $      1.20 \pm      0.31  $&$      0.06 \pm     0.02  $&$       4.26 \pm      0.16  $&$      0.24 \pm     0.01   $\\ 
    6  & $      1.13 \pm      0.28  $&$      0.06 \pm     0.02  $&$       4.30 \pm      0.16  $&$      0.25 \pm     0.01   $\\ 
    7  & $      1.42 \pm      0.29  $&$      0.07 \pm     0.01  $&$       4.18 \pm      0.14  $&$      0.24 \pm     0.01   $\\ 
    8  & $      1.19 \pm      0.63  $&$      0.06 \pm     0.03  $&$       5.02 \pm      0.66  $&$      0.30 \pm     0.04   $\\
    \hline
   \end{tabular}
  \caption{The observed and expected number of brown dwarfs with $1'' \leq \rho \leq 4''$ for the sample of 199~target stars. In this table we show the results for the full brown dwarf mass range ($0.02$~M$_\odot \leq M \leq 0.08$~M$_\odot$), unlike in Table~\ref{table: modelbackgroundstars}, where we show the results for the brown dwarfs restricted to $12 \leq K_S \leq 14$~mag.
    \label{table: modelbackgroundstars2}}
\end{table}

A multiplicity fraction of $F_{\rm M}=100\%$  is assumed in each model. For A and late~B members of Sco~OB2 the {\em observed} multiplicity fraction $F_{\rm M}$ is $\approx 50\%$ \citep{kouwenhoven2005}. This is a lower limit of the {\em true} multiplicity fraction due to the presence of unresolved companions, and hence we have $50\% \la F_{\rm M} \leq 100\%$. For a multiplicity fraction smaller than $100\%$, the expected number of brown dwarfs is given by $F_{\rm M} \times N_{\rm \star,BD}$, while the values of $R$ remain unchanged. In each model we adopted \"{O}piks law, with $15~$R$_\odot \leq a \leq 5\times 10^6~$R$_\odot$. In reality, the upper limit for $a$ may be smaller, as Sco~OB2 is an expanding association \citep{blaauw1964A,brown1999}. If this is true, the values for $N_{\rm \star,BD}$ are underpredicted, as less companions are expected to have very large separations. 
Furthermore, instead of \"{O}piks law, it may also be possible that the log-normal period distribution found by \cite{duquennoy1991} holds. For a model with \"{O}piks law at a distance of 130~pc, 11\% of the companions have separations between $1''-4''$, while for a model with the log-normal period distribution, 13\% of the companions have separations between $1''-4''$. If the log-normal period distribution holds, the values for $N_{\rm \star,BD}$ in Tables~\ref{table: modelbackgroundstars} and~\ref{table: modelbackgroundstars2} are underpredicted. The value of $R$ (and $R_\star$) does not change for the possible adaptations described here, under the assumption that the stellar and substellar companions have the same semi-major axis (or period) distribution.

\begin{figure*}[btp]
  \centering
  \includegraphics[width=1\textwidth,height=!]{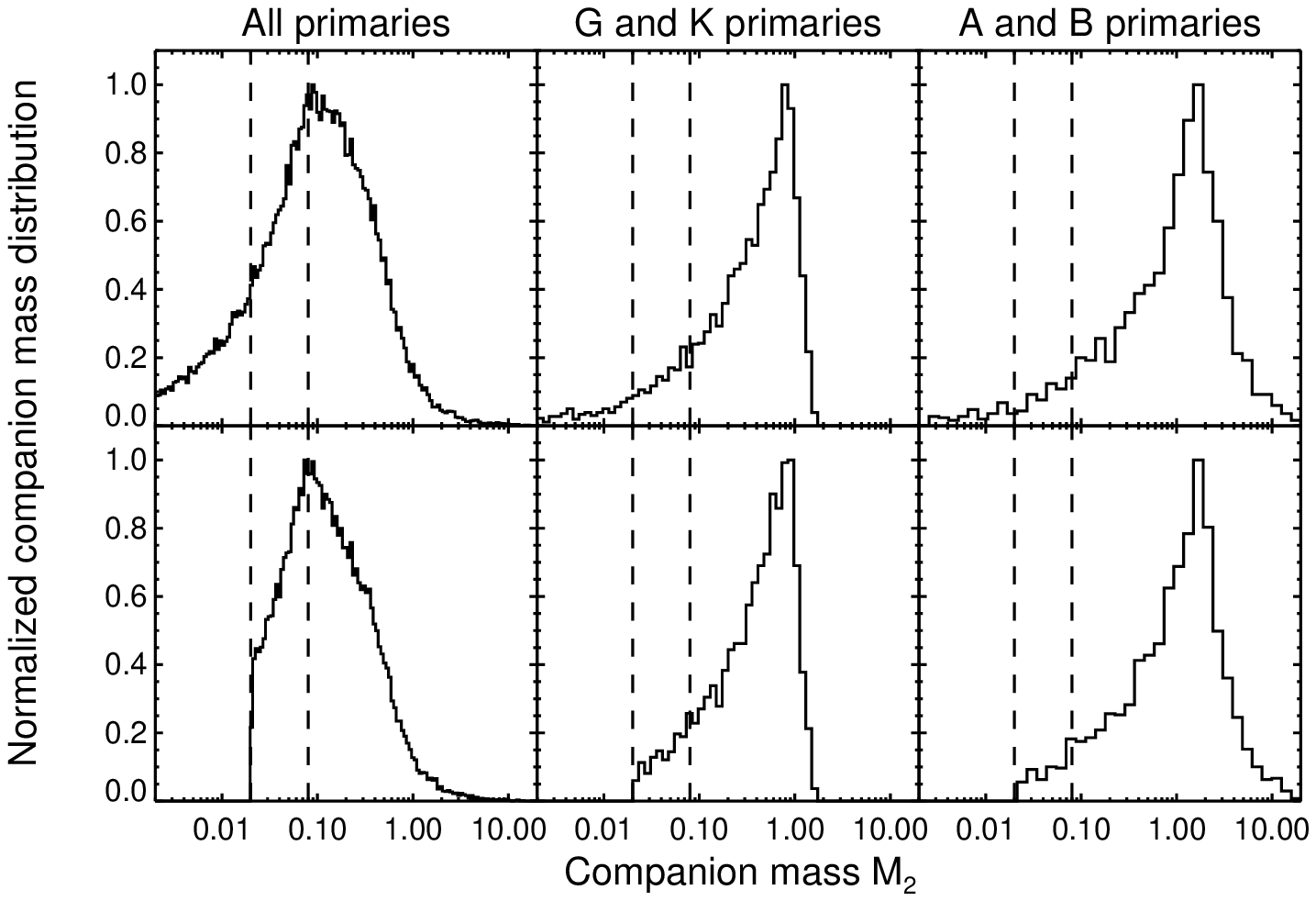}
  \caption{The companion mass distribution $f_{M_2}(M_2)$ for a simulated association. In each panel we show the distribution $f_{M_2}(M_2)$ of an association consisting of 50\,000 binaries for which the primary mass distribution is given by Equation~\ref{equation: imfpreibisch}. From left to right, the panels show the distribution of companion mass in all binaries with stellar primaries, for those of G and K primaries, and for those of A and B primaries, respectively. In the top panels we adopted a mass ratio distribution $f_q(q) \propto q^{-0.33}$ with $0 < q \leq 1$. In the bottom panels we adopt the same distribution, but with the additional constraint that $M_2 \geq 0.02$~M$_\odot$. The brown dwarf regime is indicated with the dashed lines. This figure shows that, even though the mass ratio distribution is strongly peaked to low values of $q$, the substellar-to-stellar companion ratio among intermediate- and high-mass stars is very low. 
  \label{figure: fm2_distribution} } 
\end{figure*}

Table~\ref{table: modelbackgroundstars} shows that the models with $f_q(q) \propto q^{-0.33}$ are in good agreement with our observations for any value of $\alpha$. The reason for this is that $N_{\rm \star,BD,IM}$ and $R_{\rm \star,IM}$ are independent of $\alpha$ for these models, as only the primary is chosen from the mass distribution.
For the models with random pairing, the two components of each binary system are {\em independently} chosen from the mass distribution. Only those models with a turnover in the mass distribution in the brown dwarf regime are consistent with the observations (for a multiplicity fraction of $0.5 \la F_{\rm M} \leq 1$).
However, in \cite{kouwenhoven2005} we excluded random pairing by studying the observed mass ratio distribution for stellar companions. The remaining models that are consistent with our observations have an extended Preibisch mass distribution and a mass ratio distribution of the form $f_q(q) \propto q^{-0.33}$. Although this distribution is peaked to low values of $q$, the number of brown dwarf companions is significantly smaller than the number of stellar companions. For example, for a sample of binaries with a primary mass of $3$~M$_\odot$, the substellar-to-stellar companion mass ratio $R$ (see Equation~\ref{equation: substel-to-stel}) resulting from $f_q(q) \propto q^{-0.33}$ is given by
\begin{equation}
R = \frac{\displaystyle\int_{0.02/3}^{0.08/3} f_q(q)\,dq}{\displaystyle\int_{0.08/3}^{1} f_q(q)\,dq} 
  = \frac{\displaystyle\left[\ q^{0.67}\ \right]_{\ 0.02/3}^{\ 0.08/3}}{\displaystyle\left[\ q^{0.67}\ \right]_{\ 0.08/3}^{\ 1}} 
  = 0.059 \,.
\end{equation}
Figure~\ref{figure: fm2_distribution} further illustrates that a small value for $R$ is expected among binaries with an intermediate-mass or solar-type primary, even if $f_q(q)$ is strongly peaked to low values of $q$. 
Among primaries with a mass of 1~M$_\odot$ and 3~M$_\odot$, this fraction is $\sim 14\%$ and $\sim 6\%$, respectively. For this mass ratio distribution, the number of brown dwarf companions is significantly smaller than the number of stellar companions, even if observational biases are not taken into account. If binary formation truly results in a mass ratio distribution similar to $f_q(q)\propto q^{-0.33}$, the brown dwarf desert (in terms of the ``deficit'' of brown dwarf companions relative to stellar companions) is a natural outcome of the star forming process for intermediate mass stars.

The observed number of brown dwarfs (with 12~mag $\leq K_S \leq 14$~mag) is $N_{\rm \star,BD,IM} = 1\pm 1$. After correction for unseen low-mass brown dwarfs (with $K_S>14$~mag) this translates to $N_{\rm BD,IM} = 1.6 \pm 1.6$ brown dwarfs (cf. Tables~\ref{table: modelbackgroundstars} and~\ref{table: modelbackgroundstars2}). If we assume a semi-major axis distribution of the form $f_a(a) \propto a^{-1}$ with $15~$R$_\odot < a < 5\times 10^6~$R$_\odot$ and a distance of 130~pc, we expect $\sim 11\%$ of the companions to be in the angular separation range $1''-4''$. Assuming that companion mass and semi-major axis are uncorrelated, this also means that $11\%$ of the brown dwarfs are in this range. Extrapolation gives an estimate of $(1.6 \pm 1.6)/0.11 = 14.5 \pm 14.5$~brown dwarf companions around the 199~target stars, or a brown dwarf companion fraction of $7.3 \pm 7.3\%$ for intermediate mass stars in Sco~OB2. On the other hand, if we assume the log-normal period distribution found by \cite{duquennoy1991}, we find a corresponding brown dwarf companion fraction of $6.2 \pm 6.2\%$ (for a model binary fraction of 100\%).
Note that if a mass ratio distribution $f_q(q)$ is adopted, these values are independent of the slope $\alpha$ of the mass distribution in the brown dwarf regime.

\cite{mccarthy2004} find a companion frequency of $0.7 \pm 0.7 \%$ for brown dwarf companions ($M > 30~{\rm M_J}$) to F, G, K, and M stars in the separation range $120-1200$~AU. Assuming a semi-major axis distribution of the form $f_a(a) \propto a^{-1}$, our brown dwarf companion frequency of $0.5 \pm 0.5\%$ ($M \ga 30~{\rm M_J}$) for the range $130-520$~AU translates to a value of $0.83 \pm 0.83\%$ for the range $120-1200$~AU, which is in good agreement with the frequency found by \cite{mccarthy2004}. This ``extrapolated'' brown dwarf companion frequency may underestimate the true value, if the brown dwarf desert does not exist at larger separations \citep[which may be the case for low-mass stars in the solar neighbourhood; e.g.,][]{gizis2001}.

\subsection{The origin of the brown dwarf desert} \label{section: origin}

Most stars are formed and reside in binary or multiple stellar systems. Knowledge about binary and multiple systems in young stellar groupings is of fundamental importance for our understanding of the star formation process. The formation of brown dwarfs and the dearth of brown dwarf companions has attained much interest over the last decade.
Theories have been developed, explaining the existence of the brown dwarf desert using migration \citep[][most effective at $a \la 5$~AU]{armitage2002} or ejection \citep{reipurth2001} of brown dwarfs.
The most popular theory that explains the brown dwarf desert is the {\em embryo ejection scenario}. 
This scenario predicts ejection of brown dwarfs soon after their formation by dynamical interactions in unstable multiple systems \citep{reipurth2001}. Hydrodynamical calculations \citep[e.g.,][]{bate2003} suggest that star formation is a highly dynamic and chaotic process. Brown dwarfs are ejected during or soon after their formation. In this theory brown dwarfs can be seen as failed stellar companions.

\begin{table}
  \begin{tabular}{ccccc}
    \hline
    \# & $F_{1''-4''}~(\%)$ & $F_{\rm IM}~(\%)$ & $N_{\rm BD,IM,i}$ & $N_{\rm BD,IM,i,total}$ \\
    \hline
    1  & 10.9 & 4.93  & $ 6.26  \pm 1.19  $  & $ 7.93 \pm 2.05 $\\
    2  & 10.9 & 5.34  & $ 6.78  \pm 1.29  $  & $ 8.45 \pm 2.11 $\\
    3  & 10.9 & 6.33  & $ 8.04  \pm 1.52  $  & $ 9.71 \pm 2.26 $\\
    4  & 10.9 & 0.40  & $ 0.51  \pm 0.10  $  & $ 2.18 \pm 1.67 $\\
    5  & 10.9 & 2.45  & $ 3.11  \pm 0.59  $  & $ 4.78 \pm 1.77 $\\
    6  & 10.9 & 2.70  & $ 3.43  \pm 0.65  $  & $ 5.10 \pm 1.79 $\\
    7  & 10.9 & 3.19  & $ 4.05  \pm 0.77  $  & $ 5.72 \pm 1.84 $\\
    8  & 10.9 & 0.20  & $ 0.25  \pm 0.05  $  & $ 1.92 \pm 1.67 $\\
    \hline
  \end{tabular}
  \caption{An estimate of the number of {\em primordial} binaries in Sco~OB2 with A and late-B primaries and brown dwarf companions  (with $0.02$~M$_\odot \leq M \leq 0.08$~M$_\odot$) in the angular separation range $1'' \leq \rho \leq 4''$. Columns~1 and~2 list the model number (cf. Table~\ref{table: modelbackgroundstars}), and the fraction of binaries with angular separation $1'' \leq \rho \leq 4''$ (assuming \"{O}pik's law). Column~3 lists the fraction $F_{\rm IM}$ of primaries in the simulated association that is of type A or late-B. Column 4 shows the contribution of brown dwarf companions in Sco~OB2 with A and late-B primaries in the angular separation range $1'' \leq \rho \leq 4''$, inferred from the 28~free-floating brown dwarfs in Sco~OB2 found by \cite{martin2004}, assuming that {\em all} brown dwarfs were formed as companions. Column 5 shows the total number of primordial brown dwarf companions with $1'' \leq \rho \leq 4''$ and A or late-B primaries, with the observed brown dwarf companions (corrected for unseen brown dwarfs with $K>14$~mag) included. The values of $N_{\rm BD,IM,i,total}$ are upper limits, as it is likely that not {\em all} free-floating brown dwarfs were formed as companions. \label{table: martin}}
\end{table}

In the section above we have shown that the small number of brown dwarfs among A and late-B members of Sco~OB2 can be explained with an extrapolation of the mass ratio distribution for stellar companions into the brown dwarf regime. There is thus no need for a mechanism to remove brown dwarfs.
On the other hand, the embryo ejection scenario predicts that (at least a fraction of) the free-floating brown dwarfs in Sco~OB2 have been formed as companions to association members. Below, we study the consequences in the case that embryo ejection has affected the binary population, making use of the detection of 28~free-floating brown dwarfs in Upper Scorpius by \cite{martin2004}.
Under the assumption that this is what happened, we roughly estimate the number of primordial binaries with an A or late-B primary and a brown dwarf companion. For comparison between model predictions and observations we consider only those companions with an angular separation $1'' \leq \rho \leq 4''$ and $K_S < 14$~mag, for which our ADONIS and NACO observations are complete.

\cite{martin2004} present a sample of 104~candidate very low mass members, based on DENIS $IJK$ photometry, in a region of 60~square degrees in US. The authors report spectroscopic observations of 40~of these candidates and show that 28~are indeed strong candidate members of the US subgroup. 
Under the assumption that \cite{martin2004} randomly selected their 40~observed targets out of the sample of 104~candidates, we estimate the total number of brown dwarfs in US to be $73 \pm 14$ in the 60~square degrees region in US.
The projected area of the three subgroups of Sco~OB2 is approximately 960~square degrees, which gives us an estimate of $1165 \pm 221$ free-floating brown dwarfs in Sco~OB2. 

For our estimate of the number of primordial binaries we assume that all free-floating brown dwarfs were once companions. The number of systems with an A or late-B primary constitutes a small fraction $F_{\rm IM}$ of the total number of binaries, depending on the mass distribution (see Table~\ref{table: martin}). Assuming a primordial semi-major axis distribution of the form $f_a(a) \propto a^{-1}$ with $15$~R$_\odot < a < 5\times 10^6$~R$_\odot$, about $11\%$ of the brown dwarf companions are in the angular separation range $1'' \leq \rho \leq 4''$. 

For each model in Table~\ref{table: martin} we calculate how many of the $1165 \pm 221$ free-floating have the properties $1'' \leq \rho \leq 4''$ and $K_S < 14$~mag, and obtain the contribution $N_{\rm BD,IM,i} = F_{1''-4''} \times F_{\rm IM} \times (1165 \pm 221)$ of the free-floating brown dwarfs found by \cite{martin2004} to the number of primordial binaries in Sco~OB2 with A and late-B primaries and brown dwarf companions in the angular separation range $1'' \leq \rho \leq 4''$. We estimate the total number of primordial brown dwarf companions $N_{\rm BD,IM,i,total}$ with $1'' \leq \rho \leq 4''$ of A and late-B primaries by adding the observed number of brown dwarf companions, corrected for unseen companions with $K_S> 14$~mag (i.e., $1.67 \pm 1.67$).

We have assumed that all free-floating brown dwarfs were once companion stars, and therefore obtained upper limits for $N_{\rm BD,IM,i,total}$. By comparing $N_{\rm BD,IM}$ in Table~\ref{table: modelbackgroundstars2} with $N_{\rm BD,IM,i,total}$ in Table~\ref{table: martin} we can derive which primordial mass and mass ratio distributions are consistent with the predictions. As not necessarily all free-floating brown dwarfs have their origin in a binary system, all models with $N_{\rm BD,IM} \la N_{\rm BD,IM,i,total}$ are consistent with the predictions (i.e., the {\em current} number of brown dwarf companions should be less or equal to the {\em primordial} number of brown dwarf companions). A comparison shows that all models are consistent, except model 4 (random pairing from the Salpeter mass distribution). Under the hypothesis that embryo ejection has affected Sco~OB2, the current mass ratio distribution is slightly shallower than the primordial mass ratio distribution, due to the ejection of brown dwarf companions.
The above derivation gave an estimate of the number of primordial binaries with brown dwarf companions, under the assumption that the origin of the free-floating brown dwarfs in Sco~OB2 can be explained with the embryo ejection scenario. With our observations we cannot exclude firmly that this happened for several of the free-floating brown dwarfs. 

However, if binary formation would result in a mass ratio distribution similar to $f_q(q) \propto q^{-0.33}$, the ``brown dwarf desert'', if defined as a deficit of brown dwarf companions relative to stellar companions, would be a {\em natural outcome of star formation}. The embryo ejection scenario is not necessary to explain the small observed brown dwarf companion fraction in this case.


\section{Binarity and multiplicity in Sco~OB2} \label{section: binarystatistics}

In \cite{kouwenhoven2005} we provided a census on binarity in Sco~OB2, consisting of all available data on visual, spectroscopic, eclipsing, and astrometric binaries and multiples. In Table~\ref{table: statistics} we present an update on the binary statistics in Sco~OB2. The statistics have been updated with the new results presented in this paper, as well as with the binaries recently discovered by \cite{nitschelm2004}, \cite{jilinski2006}, and \cite{chen2006}.

The multiple system fraction $F_{\rm M}$, the non-single star fraction  $F_{\rm NS}$, and companion star fraction $F_{\rm C}$ are defined as:
\begin{eqnarray} \label{equation: binarystatistics}
  F_{\rm M}       &=& (B+T+\dots) \ /\ (S+B+T+\dots);\\
  F_{\rm NS}      &=& (2B+3T+\dots) \ /\ (S+2B+3T+\dots);\\
  F_{\rm C}       &=& (B+2T+\dots) \ /\ (S+B+T+\dots),
\end{eqnarray}
where $S$, $B$, and $T$ denote the number of single systems, binary systems
and triple systems in the association. 
In the Sco~OB2 association at least $40\%$ of the systems are multiple. Of the individual stars at least $60\%$ is part of a multiple system. Each system contains on average $F_{\rm C} \approx 0.5$ known companion stars. The updated values of $F_{\rm M}$, $F_{\rm NS}$, and  $F_{\rm C}$ are slightly larger than the values mentioned in \cite{kouwenhoven2005}, respectively. Note that these frequencies are lower limits due to the presence of undiscovered companion stars.


\section{Conclusions} \label{section: conclusion}

We have carried out near-infrared $JHK_S$ observations of 22 A and late-B stars in the Sco~OB2 association. The observations were performed with the NAOS/CONICA system at the ESO Very Large Telescope at Paranal, Chile. The observations resulted from a follow-up program of our previous work \citep{kouwenhoven2005}, in which we surveyed 199~A and late-B Sco~OB2 members for binarity with ADONIS. The data were obtained with the goal of (1) determining the validity of the $K_S=12$~mag criterion we used in our ADONIS survey to separate companions and background stars, (2) studying the behaviour of the companion mass distribution in the low-mass regime, and (3) searching for additional companion stars.  We have included in our analysis the multi-color observations of 9~targets observed with ADONIS. In our ADONIS survey, these targets were analyzed using their $K_S$ magnitude only. The main results of our study are:

\begin{itemize}
\item[--] We detect 72~secondaries around the 31~target stars in our analysis. By comparing the near-infrared colors with the isochrones in the color-magnitude diagram, we find 25~confirmed companion stars, 12~candidate companion stars, and 35~background stars.   
\item[--] For most objects in our ADONIS survey \citep{kouwenhoven2005} only the $K_S$ magnitude was available. We used a magnitude criterion to separate companion stars ($K_S < 12$~mag) and background stars ($K_S > 12$~mag). With our analysis of the 22~NACO targets and 9~ADONIS targets with multi-color observations, we estimate the accuracy of the $K_S=12$~mag criterion. We find that the $K_S=12$~mag criterion is a very useful tool, correctly classifying the secondaries in $\sim 80\%$ of the cases. 
\item[--] We report two candidate brown dwarf companions of HIP81972. From their near-infrared magnitudes we infer masses of $32~{\rm M_J}$ and $63~{\rm M_J}$. The objects are located at an angular separation of $7.92''$ (1500~AU) and $2.79''$ (520~AU) from HIP81972, respectively. Follow-up spectroscopy is necessary to determine the true nature of these secondaries. Although we are sensitive (but incomplete) to massive planets, we classify the faintest secondaries as background stars (irrespective of their location in the color-magnitude diagram), because of isochronal uncertainty and the large number of faint background stars. 
\item[--] In our combined survey of 199~A and late-B members of Sco~OB2 we detect one confirmed companion star with $12~{\rm mag} \leq K_S \leq 14$~mag in the angular separation range $1''-4''$. In this region we detect no other secondary, while both the ADONIS and NACO observations are complete. This indicates a very low frequency of brown dwarf companions in the separation range $130-520$~AU for late-B and A type stars in Sco~OB2.   
\item[--] Our results are in good agreement with a mass ratio distribution of the form $f_q(q) \propto q^{-0.33}$. We find a brown dwarf companion fraction (for $M \ga 30~{\rm M_J}$) of $0.5 \pm 0.5\%$ for A and late-B stars in Sco~OB2.  After correction for unseen faint companions ($M \la 30~{\rm M_J}$), we estimate a substellar-to-stellar companion ratio of $R=0.06\pm 0.02$. 
\end{itemize}
The number of brown dwarfs among A and late-B members of Sco~OB2 is consistent with an extrapolation of the (stellar) companion mass distribution into the brown dwarf regime, suggesting that the formation mechanism for stars and brown dwarfs is the same. 
The embryo ejection mechanism does not need to be invoked to explain the small number of brown dwarf companions among intermediate mass stars in Sco~OB2.


\begin{acknowledgements}

We thank ESO and the Paranal Observatory staff for efficiently conducting the Service-Mode observations and for their support. We thank Simon Portegies Zwart and the anonymous referee for their constructive criticism, which helped to substantially improve the paper. This publication makes use of data products from the Two Micron All Sky Survey, which is a joint project of the University of Massachusetts and the Infrared Processing and Analysis Center/California Institute of Technology, funded by the National Aeronautics and Space Administration and the National Science Foundation. This research is supported by NWO under project number 614.041.006.

\end{acknowledgements}

\bibliographystyle{aa}
\bibliography{/home/tk/bibliography/bibliography}

\begin{thebibliography}{45}
\expandafter\ifx\csname natexlab\endcsname\relax\def\natexlab#1{#1}\fi

\bibitem[{{Armitage} \& {Bonnell}(2002)}]{armitage2002}
{Armitage}, P.~J. \& {Bonnell}, I.~A. 2002, \mnras, 330, L11

\bibitem[{{Bate} {et~al.}(2003){Bate}, {Bonnell}, \& {Bromm}}]{bate2003}
{Bate}, M.~R., {Bonnell}, I.~A., \& {Bromm}, V. 2003, \mnras, 339, 577

\bibitem[{{Blaauw}(1964)}]{blaauw1964A}
{Blaauw}, A. 1964, \araa, 2, 213

\bibitem[{{Blaauw}(1991)}]{blaauw1991}
{Blaauw}, A. 1991, in NATO ASIC Proc. 342: The Physics of Star Formation and
  Early Stellar Evolution, 125

\bibitem[{{Brown}(2001)}]{brown2001}
{Brown}, A. 2001, Astronomische Nachrichten, 322, 43

\bibitem[{{Brown} {et~al.}(1997){Brown}, {Arenou}, {van Leeuwen}, {Lindegren},
  \& {Luri}}]{brown1997}
{Brown}, A.~G.~A., {Arenou}, F., {van Leeuwen}, F., {Lindegren}, L., \& {Luri},
  X. 1997, The First Results of Hipparcos and Tycho, 23rd meeting of the IAU,
  Joint Discussion 14, 25 August 1997, Kyoto, Japan, meeting abstract., 14

\bibitem[{{Brown} {et~al.}(1999){Brown}, {Blaauw}, {Hoogerwerf}, {de Bruijne},
  \& {de Zeeuw}}]{brown1999}
{Brown}, A.~G.~A., {Blaauw}, A., {Hoogerwerf}, R., {de Bruijne}, J.~H.~J., \&
  {de Zeeuw}, P.~T. 1999, in NATO ASIC Proc. 540: The Origin of Stars and
  Planetary Systems, ed. C.~J. {Lada} \& N.~D. {Kylafis}, 411

\bibitem[{{Chabrier} {et~al.}(2000){Chabrier}, {Baraffe}, {Allard}, \&
  {Hauschildt}}]{chabrier2000}
{Chabrier}, G., {Baraffe}, I., {Allard}, F., \& {Hauschildt}, P. 2000, \apj,
  542, 464

\bibitem[{{Chabrier} {et~al.}(2005){Chabrier}, {Baraffe}, {Allard}, \&
  {Hauschildt}}]{chabrier2005}
{Chabrier}, G., {Baraffe}, I., {Allard}, F., \& {Hauschildt}, P.~H. 2005, in
  Resolved Stellar Populations, ASP Conference Series, eds. Valls-Gabaud \&
  Chavez

\bibitem[{{Chanam{\'e}} \& {Gould}(2004)}]{chaname2004}
{Chanam{\'e}}, J. \& {Gould}, A. 2004, \apj, 601, 289

\bibitem[{{Chen} {et~al.}(2006){Chen}, {Henning}, {van Boekel}, \&
  {Grady}}]{chen2006}
{Chen}, X.~P., {Henning}, T., {van Boekel}, R., \& {Grady}, C.~A. 2006, \aap,
  445, 331

\bibitem[{{Close} {et~al.}(1990){Close}, {Richer}, \& {Crabtree}}]{close1990}
{Close}, L.~M., {Richer}, H.~B., \& {Crabtree}, D.~R. 1990, \aj, 100, 1968

\bibitem[{{Cox}(2000)}]{allen2000}
{Cox}, A.~N. 2000, {Allen's astrophysical quantities} (Springer, New York,
  2000.~Edited by Arthur N.~Cox.)

\bibitem[{{Cunha} \& {Lambert}(1994)}]{cunha1994}
{Cunha}, K. \& {Lambert}, D.~L. 1994, \apj, 426, 170

\bibitem[{{Cutri} {et~al.}(2003){Cutri}, {Skrutskie}, {van Dyk}, {Beichman},
  {Carpenter}, {Chester}, {Cambresy}, {Evans}, {Fowler}, {Gizis}, {Howard},
  {Huchra}, {Jarrett}, {Kopan}, {Kirkpatrick}, {Light}, {Marsh}, {McCallon},
  {Schneider}, {Stiening}, {Sykes}, {Weinberg}, {Wheaton}, {Wheelock}, \&
  {Zacarias}}]{2mass}
{Cutri}, R.~M., {Skrutskie}, M.~F., {van Dyk}, S., {et~al.} 2003, VizieR Online
  Data Catalog, 2246

\bibitem[{{de Bruijne}(1999)}]{debruijne1999}
{de Bruijne}, J.~H.~J. 1999, \mnras, 310, 585

\bibitem[{{de Geus} {et~al.}(1989){de Geus}, {de Zeeuw}, \& {Lub}}]{degeus1989}
{de Geus}, E.~J., {de Zeeuw}, P.~T., \& {Lub}, J. 1989, \aap, 216, 44

\bibitem[{{de Zeeuw} {et~al.}(1999){de Zeeuw}, {Hoogerwerf}, {de Bruijne},
  {Brown}, \& {Blaauw}}]{dezeeuw1999}
{de Zeeuw}, P.~T., {Hoogerwerf}, R., {de Bruijne}, J.~H.~J., {Brown}, A.~G.~A.,
  \& {Blaauw}, A. 1999, \aj, 117, 354

\bibitem[{{Devillard}(1997)}]{devillard1997}
{Devillard}, N. 1997, The Messenger, 87, 19

\bibitem[{{Diolaiti} {et~al.}(2000){Diolaiti}, {Bendinelli}, {Bonaccini},
  {Close}, {Currie}, \& {Parmeggiani}}]{diolaiti2000}
{Diolaiti}, E., {Bendinelli}, O., {Bonaccini}, D., {et~al.} 2000, \aaps, 147,
  335

\bibitem[{{Duquennoy} \& {Mayor}(1991)}]{duquennoy1991}
{Duquennoy}, A. \& {Mayor}, M. 1991, \aap, 248, 485

\bibitem[{{Girardi} {et~al.}(2002){Girardi}, {Bertelli}, {Bressan}, {Chiosi},
  {Groenewegen}, {Marigo}, {Salasnich}, \& {Weiss}}]{girardi2002}
{Girardi}, L., {Bertelli}, G., {Bressan}, A., {et~al.} 2002, \aap, 391, 195

\bibitem[{{Gizis} {et~al.}(2001){Gizis}, {Kirkpatrick}, {Burgasser}, {Reid},
  {Monet}, {Liebert}, \& {Wilson}}]{gizis2001}
{Gizis}, J.~E., {Kirkpatrick}, J.~D., {Burgasser}, A., {et~al.} 2001, \apjl,
  551, L163

\bibitem[{{Hu{\'e}lamo} {et~al.}(2001){Hu{\'e}lamo}, {Brandner}, {Brown},
  {Neuh{\"a}user}, \& {Zinnecker}}]{huelamo2001}
{Hu{\'e}lamo}, N., {Brandner}, W., {Brown}, A.~G.~A., {Neuh{\"a}user}, R., \&
  {Zinnecker}, H. 2001, \aap, 373, 657

\bibitem[{{Jilinski} {et~al.}(2006){Jilinski}, {Daflon}, {Cunha}, \& {de La
  Reza}}]{jilinski2006}
{Jilinski}, E., {Daflon}, S., {Cunha}, K., \& {de La Reza}, R. 2006, \aap, 448,
  1001

\bibitem[{{Kouwenhoven}(2006)}]{kouwenhoventhesis}
{Kouwenhoven}, M.~B.~N. 2006, PhD thesis, University of Amsterdam
  (astro-ph/0610792)

\bibitem[{{Kouwenhoven} {et~al.}(2005){Kouwenhoven}, {Brown}, {Zinnecker},
  {Kaper}, \& {Portegies Zwart}}]{kouwenhoven2005}
{Kouwenhoven}, M.~B.~N., {Brown}, A.~G.~A., {Zinnecker}, H., {Kaper}, L., \&
  {Portegies Zwart}, S.~F. 2005, \aap, 430, 137

\bibitem[{{Lenzen} {et~al.}(1998){Lenzen}, {Hofmann}, {Bizenberger}, \&
  {Tusche}}]{lenzen1998}
{Lenzen}, R., {Hofmann}, R., {Bizenberger}, P., \& {Tusche}, A. 1998, in Proc.
  SPIE Vol. 3354, p. 606-614, Infrared Astronomical Instrumentation, Albert M.
  Fowler; Ed., 606--614

\bibitem[{{Mamajek} {et~al.}(2002){Mamajek}, {Meyer}, \&
  {Liebert}}]{mamajek2002}
{Mamajek}, E.~E., {Meyer}, M.~R., \& {Liebert}, J. 2002, \aj, 124, 1670

\bibitem[{{Mart{\'{\i}}n} {et~al.}(2004){Mart{\'{\i}}n}, {Delfosse}, \&
  {Guieu}}]{martin2004}
{Mart{\'{\i}}n}, E.~L., {Delfosse}, X., \& {Guieu}, S. 2004, \aj, 127, 449

\bibitem[{{McCarthy} \& {Zuckerman}(2004)}]{mccarthy2004}
{McCarthy}, C. \& {Zuckerman}, B. 2004, \aj, 127, 2871

\bibitem[{{Nitschelm}(2004)}]{nitschelm2004}
{Nitschelm}, C. 2004, in ASP Conf. Ser. 318: Spectroscopically and Spatially
  Resolving the Components of the Close Binary Stars, ed. R.~W. {Hidlitch},
  H.~{Hensberge}, \& K.~{Pavlovski}, 291--293

\bibitem[{{Palla} \& {Stahler}(1999)}]{palla1999}
{Palla}, F. \& {Stahler}, S.~W. 1999, \apj, 525, 772

\bibitem[{{Persson} {et~al.}(1998){Persson}, {Murphy}, {Krzeminski}, {Roth}, \&
  {Rieke}}]{persson1998}
{Persson}, S.~E., {Murphy}, D.~C., {Krzeminski}, W., {Roth}, M., \& {Rieke},
  M.~J. 1998, \aj, 116, 2475

\bibitem[{{Portegies Zwart} {et~al.}(2001){Portegies Zwart}, {McMillan}, {Hut},
  \& {Makino}}]{ecology4}
{Portegies Zwart}, S.~F., {McMillan}, S.~L.~W., {Hut}, P., \& {Makino}, J.
  2001, \mnras, 321, 199

\bibitem[{{Poveda} {et~al.}(1982){Poveda}, {Allen}, \& {Parrao}}]{poveda1982}
{Poveda}, A., {Allen}, C., \& {Parrao}, L. 1982, \apj, 258, 589

\bibitem[{{Preibisch} {et~al.}(2002){Preibisch}, {Brown}, {Bridges},
  {Guenther}, \& {Zinnecker}}]{preibisch2002}
{Preibisch}, T., {Brown}, A.~G.~A., {Bridges}, T., {Guenther}, E., \&
  {Zinnecker}, H. 2002, \aj, 124, 404

\bibitem[{{Preibisch} {et~al.}(2003){Preibisch}, {Stanke}, \&
  {Zinnecker}}]{preibisch2003}
{Preibisch}, T., {Stanke}, T., \& {Zinnecker}, H. 2003, \aap, 409, 147

\bibitem[{{Press} {et~al.}(1992){Press}, {Teukolsky}, {Vetterling}, \&
  {Flannery}}]{numericalrecipies}
{Press}, W.~H., {Teukolsky}, S.~A., {Vetterling}, W.~T., \& {Flannery}, B.~P.
  1992, {Numerical recipes in FORTRAN. The art of scientific computing}
  (Cambridge: University Press, |c1992, 2nd ed.)

\bibitem[{{Reipurth} \& {Clarke}(2001)}]{reipurth2001}
{Reipurth}, B. \& {Clarke}, C. 2001, \aj, 122, 432

\bibitem[{{Robin} {et~al.}(2003){Robin}, {Reyl{\' e}}, {Derri{\` e}re}, \&
  {Picaud}}]{besancon}
{Robin}, A.~C., {Reyl{\' e}}, C., {Derri{\` e}re}, S., \& {Picaud}, S. 2003,
  \aap, 409, 523

\bibitem[{{Rousset} {et~al.}(2000){Rousset}, {Lacombe}, {Puget}, {Gendron},
  {Arsenault}, {Kern}, {Rabaud}, {Madec}, {Hubin}, {Zins}, {Stadler},
  {Charton}, {Gigan}, \& {Feautrier}}]{rousset2000}
{Rousset}, G., {Lacombe}, F., {Puget}, P., {et~al.} 2000, in Proc. SPIE Vol.
  4007, p. 72-81, Adaptive Optical Systems Technology, Peter L. Wizinowich;
  Ed., 72--81

\bibitem[{{Shatsky} \& {Tokovinin}(2002)}]{shatsky2002}
{Shatsky}, N. \& {Tokovinin}, A. 2002, \aap, 382, 92

\bibitem[{{Siess} {et~al.}(2000){Siess}, {Dufour}, \& {Forestini}}]{siess2000}
{Siess}, L., {Dufour}, E., \& {Forestini}, M. 2000, \aap, 358, 593

\bibitem[{{Worley} \& {Douglass}(1997)}]{wds1997}
{Worley}, C.~E. \& {Douglass}, G.~G. 1997, \aaps, 125, 523

\end{thebibliography}


\Online
\appendix
\onecolumn
\section{Results of the NAOS/CONICA survey}

\setlength{\LTcapwidth}{1.0\textwidth}
\begin{longtable}{| l | rrr rr | rrr r | cl |}
  \caption{Results from our multi-color binarity study among 22~Sco~OB2 member stars observed with NACO ({\em top part of the table}) and the subset of 9~members with multi-color observations in in the ADONIS survey ({\em bottom part of the table}). The columns show the \textit{Hipparcos} number (for the targets) and the secondary designation, the $J$, $H$, and $K_S$ magnitudes, the angular separation, and the position angle (measured from North to East). Lower limits to the magnitudes are given if an object is not detected in the NACO survey, unless the ADONIS measurement was available (marked with a $\star$). Entries marked with $\star\star$ have no available measurement, e.g., because the object is not in the field of view for that filter. For each primary and companion star we list the absolute $JHK_S$ magnitudes and the mass in columns $7-10$. We additionally provide absolute magnitudes and a mass estimate for the candidate companions, {\em under the assumption} that these are indeed companions. We stress that a significant number of the candidate companions may actually be background stars. The 11th column lists the status of the object (p = primary, c = confirmed companion star, nc = new confirmed companion star, ? = candidate companion star; b = background star). The last column provides additional remarks. A remark ``J'', ``H'', or ``K'' means that the secondary flux in this filter was obtained from the image obtained {\em without} the NDF, using the PSF from the corresponding image that was obtained {\em with} NDF (see \S~\ref{section: componentdetection}). If the secondary status was obtained without color information, an exclamation mark is placed in the last column. The results for the 9~targets with multi-color information in the ADONIS survey are marked with ``ADO''. \label{table: longtable}}\\
  \hline
  Star & \multicolumn{1}{c}{$J$} & \multicolumn{1}{c}{$H$} & \multicolumn{1}{c}{$K_S$} & \multicolumn{1}{c}{$\rho$} & \multicolumn{1}{c}{PA}  & \multicolumn{1}{c}{$M_J$}  & \multicolumn{1}{c}{$M_H$}  & \multicolumn{1}{c}{$M_{K_S}$}  & \multicolumn{1}{c}{Mass}      & \multicolumn{1}{c}{Status} &  \multicolumn{1}{c|}{Remarks} \\
  \hline
       & \multicolumn{1}{c}{mag} & \multicolumn{1}{c}{mag} & \multicolumn{1}{c}{mag}   & \multicolumn{1}{c}{arcsec} & \multicolumn{1}{c}{deg}   & \multicolumn{1}{c}{mag}    & \multicolumn{1}{c}{mag}    &  \multicolumn{1}{c}{mag}       & \multicolumn{1}{c}{M$_\odot$} &        &          \\
  \endfirsthead
  \hline
  \multicolumn{12}{|l|}{\tablename\ \thetable{} -- continued from previous page} \\
  \hline
  Star & \multicolumn{1}{c}{$J$} & \multicolumn{1}{c}{$H$} & \multicolumn{1}{c}{$K_S$} & \multicolumn{1}{c}{$\rho$} & \multicolumn{1}{c}{PA}  & \multicolumn{1}{c}{$M_J$}  & \multicolumn{1}{c}{$M_H$}  & \multicolumn{1}{c}{$M_{K_S}$}  & \multicolumn{1}{c}{Mass}      & \multicolumn{1}{c}{Status} &  \multicolumn{1}{c|}{Remarks} \\
  \hline
       & \multicolumn{1}{c}{mag} & \multicolumn{1}{c}{mag} & \multicolumn{1}{c}{mag}   & \multicolumn{1}{c}{arcsec} & \multicolumn{1}{c}{deg}   & \multicolumn{1}{c}{mag}    & \multicolumn{1}{c}{mag}    &  \multicolumn{1}{c}{mag}       & \multicolumn{1}{c}{M$_\odot$} &        &          \\
  \hline
  \endhead
  \hline
  \multicolumn{12}{|l|}{{Continued on next page}} \\ 
  \hline
  \endfoot
  \hline 
  \endlastfoot
  \hline
  \hline 
  HIP59502       &   6.83   &   6.83   &   6.87   &          &          &   1.86   &   1.87   &   1.91   &   1.80   &   p   &          \\
HIP59502    -1   &   12.35   &   11.83   &   11.64   &   2.94   &   26.39   &   7.39   &   6.86   &   6.68   &   0.14   &   c   &          \\
HIP59502    -2   &   $>15.22$   &   15.26   &   15.37   &   4.76   &   101.87   &      &      &      &          &   b   &   HK       \\
HIP59502    -3   &   $^{\star\star}$   &   $^{\star\star}$   &   13.69   &   9.02   &   309.01   &      &      &      &          &   b   &   K!       \\
\hline
HIP60851       &   6.03   &   6.06   &   6.06   &          &          &   0.94   &   0.97   &   0.97   &   2.63   &   p   &          \\
HIP60851    -1   &   12.81   &   11.62   &   11.46   &   2.07   &   45.30   &      &      &      &          &   b   &   J       \\
HIP60851    -2   &   $>13.33$   &   11.68   &   11.29   &   6.89   &   180.38   &      &      &      &          &   b   &          \\
HIP60851    -3   &   $>13.33$   &   13.63   &   13.69   &   8.16   &   231.46   &   ($>8.24$)   &   (8.54)   &   (8.60)   &   (0.04)   &   ?   &   HK       \\
HIP60851    -4   &   $>13.33$   &   14.82   &   14.80   &   1.61   &   280.38   &      &      &      &          &   b   &   HK       \\
HIP60851    -5   &   $>13.33$   &   15.53   &   14.97   &   8.19   &   69.19   &      &      &      &          &   b   &   HK       \\
HIP60851    -6   &   $>13.33$   &   15.83   &   $^{\star\star}$   &   7.65   &   153.67   &      &      &      &          &   b   &   H!       \\
HIP60851    -7   &   $>13.33$   &   16.67   &   $^{\star\star}$   &   7.47   &   287.03   &      &      &      &          &   b   &   H!       \\
HIP60851    -8   &   $>13.33$   &   16.87   &   16.97   &   5.45   &   76.38   &      &      &      &          &   b   &   HK       \\
\hline
HIP61265       &   7.49   &   7.51   &   7.46   &          &          &   1.85   &   1.87   &   1.81   &   1.82   &   p   &          \\
HIP61265    -1   &   11.98   &   11.66   &   11.38   &   2.51   &   67.15   &   (6.34)   &   (6.02)   &   (5.74)   &   (0.27)   &   ?   &   J       \\
HIP61265    -2   &   15.13   &   14.96   &   14.75   &   3.41   &   167.27   &      &      &      &          &   b   &   J       \\
HIP61265    -3   &   $>15.71$   &   16.30   &   15.29   &   7.00   &   24.46   &      &      &      &          &   b   &          \\
HIP61265    -4   &   $>15.71$   &   16.80   &   16.28   &   6.60   &   31.84   &      &      &      &          &   b   &          \\
HIP61265    -5   &   $>15.71$   &   $>15.90$   &   15.86   &   7.11   &   344.55   &      &      &      &          &   b   &   !       \\
\hline
HIP62026       &   6.28   &   6.32   &   6.31   &          &          &   1.09   &   1.12   &   1.11   &   2.45   &   p   &          \\
HIP62026    -1   &   8.08   &   7.90   &   7.86   &   0.23   &   6.34   &   2.88   &   2.71   &   2.66   &   1.19   &   c   &          \\
\hline
HIP63204       &   6.68   &   6.76   &   6.78   &          &          &   1.48   &   1.55   &   1.57   &   2.05   &   p   &          \\
HIP63204    -1   &   8.72   &   7.85   &   7.50   &   1.87   &   47.44   &      &      &      &          &   b   &          \\
HIP63204    -2   &   8.79   &   8.51   &   8.40   &   0.15   &   236.56   &   3.59   &   3.31   &   3.19   &   1.06   &   c   &          \\
\hline
HIP67260       &   7.03   &   7.00   &   6.98   &          &          &   1.57   &   1.53   &   1.52   &   2.00   &   p   &          \\
HIP67260    -1   &   8.88   &   8.46   &   8.36   &   0.42   &   229.46   &   3.42   &   2.99   &   2.90   &   1.10   &   c   &          \\
HIP67260    -2   &   $^{\star\star}$   &   14.04   &   14.10   &   1.23   &   355.65   &   ($^{\star\star}$)   &   (8.57)   &   (8.63)   &   (0.04)   &   ?   &          \\
HIP67260    -3   &   15.84   &   14.83   &   14.67   &   2.33   &   77.25   &   (10.38)   &   (9.36)   &   (9.20)   &   ($\approx$0.02)   &   ?   &   JHK       \\
\hline
HIP67919       &   6.71   &   6.60   &   6.59   &          &          &   1.63   &   1.52   &   1.51   &   1.97   &   p   &          \\
HIP67919    -1   &   9.98   &   9.38   &   9.10   &   0.69   &   296.56   &   4.89   &   4.30   &   4.02   &   0.75   &   c   &          \\
\hline
HIP68532       &   7.16   &   7.08   &   7.02   &          &          &   1.67   &   1.59   &   1.53   &   1.95   &   p   &          \\
HIP68532    -1   &   10.52   &   9.85   &   9.54   &   3.05   &   288.50   &   5.03   &   4.36   &   4.05   &   0.73   &   c   &          \\
HIP68532    -2   &   11.38   &   10.94   &   10.63   &   3.18   &   291.92   &   5.89   &   5.45   &   5.14   &   0.39   &   c   &          \\
\hline
HIP69113       &   6.17   &   6.30   &   6.37   &          &          &   0.02   &   0.15   &   0.22   &   3.87   &   p   &          \\
HIP69113    -1   &   10.98   &   10.43   &   10.29   &   5.34   &   65.15   &   4.83   &   4.28   &   4.14   &   0.77   &   c   &          \\
HIP69113    -2   &   11.27   &   10.45   &   10.30   &   5.52   &   67.17   &   5.12   &   4.29   &   4.15   &   0.72   &   c   &          \\
\hline
HIP73937       &   6.11   &   6.21   &   6.23   &          &          &   0.65   &   0.75   &   0.77   &   2.94   &   p   &          \\
HIP73937    -1   &   $>8.40$   &   8.46   &   8.37   &   0.24   &   190.58   &   $>2.94$   &   3.00   &   2.91   &   1.11   &   c   &          \\
HIP73937    -2   &   $>11.41$   &   14.32   &   14.71   &   3.56   &   31.24   &      &      &      &          &   b   &   HK       \\
\hline
HIP78968       &   7.47   &   7.42   &   7.42   &          &          &   1.23   &   1.17   &   1.18   &   2.33   &   p   &          \\
HIP78968    -1   &   14.96   &   14.51   &   14.26   &   2.78   &   322.13   &   (8.71)   &   (8.27)   &   (8.01)   &   ($\approx$0.02)   &   ?   &   JHK       \\
\hline
HIP79098       &   5.71   &   5.70   &   5.69   &          &          &   -0.02   &   -0.03   &   -0.04   &   4.30   &   p   &          \\
HIP79098    -1   &   15.67   &   14.14   &   14.24   &   2.37   &   116.63   &      &      &      &          &   b   &   JK       \\
\hline
HIP79410       &   7.20   &   7.14   &   7.09   &          &          &   1.35   &   1.29   &   1.24   &   2.24   &   p   &          \\
HIP79410    -1   &   15.94   &   15.12   &   14.93   &   3.24   &   340.93   &      &      &      &          &   b   &   J       \\
\hline
HIP79739       &   7.17   &   7.16   &   7.08   &          &          &   1.23   &   1.21   &   1.14   &   2.32   &   p   &          \\
HIP79739    -1   &   12.28   &   11.52   &   11.23   &   0.96   &   118.33   &   6.34   &   5.58   &   5.29   &   0.16   &   c   &          \\
\hline
HIP79771       &   7.33   &   7.26   &   7.10   &          &          &   1.39   &   1.31   &   1.15   &   2.14   &   p   &          \\
HIP79771    -1   &   12.00   &   11.28   &   10.89   &   3.67   &   313.38   &   6.06   &   5.33   &   4.94   &   0.19   &   c   &          \\
HIP79771    -2   &   12.39   &   11.79   &   11.42   &   0.44   &   128.59   &   6.44   &   5.85   &   5.47   &   0.13   &   nc   &          \\
\hline
HIP80142       &   6.61   &   6.67   &   6.66   &          &          &   0.41   &   0.47   &   0.46   &   3.33   &   p   &          \\
HIP80142    -1   &   12.01   &   10.59   &   9.51   &   9.23   &   216.16   &      &      &      &          &   b   &   J       \\
HIP80142    -2   &   16.64   &   15.88   &   $^{\star\star}$   &   5.88   &   119.94   &   (10.44)   &   (9.68)   &   ($^{\star\star}$)   &   ($\approx$0.02)   &   ?   &   HJ       \\
\hline
HIP80474       &   6.14   &   5.92   &   5.80   &          &          &   0.27   &   0.05   &   -0.07   &   3.78   &   p   &          \\
HIP80474    -1   &   12.06   &   12.34   &   10.79   &   4.85   &   206.36   &      &      &      &          &   b   &   JHK       \\
\hline
HIP80799       &   7.56   &   7.53   &   7.45   &          &          &   2.04   &   2.01   &   1.93   &   1.86   &   p   &          \\
HIP80799    -1   &   10.60   &   10.04   &   9.80   &   2.94   &   205.02   &   5.08   &   4.51   &   4.27   &   0.34   &   c   &          \\
\hline
HIP80896       &   7.67   &   7.53   &   7.44   &          &          &   2.11   &   1.97   &   1.88   &   1.81   &   p   &          \\
HIP80896    -1   &   11.16   &   10.63   &   10.33   &   2.28   &   177.23   &   5.60   &   5.07   &   4.77   &   0.24   &   c   &          \\
\hline
HIP81949       &   7.38   &   7.40   &   7.33   &          &          &   1.28   &   1.31   &   1.23   &   2.26   &   p   &          \\
HIP81949    -1   &   15.73   &   14.11   &   13.28   &   3.91   &   88.47   &      &      &      &          &   b   &          \\
HIP81949    -2   &   14.34   &   14.28   &   14.06   &   3.48   &   28.46   &      &      &      &          &   b   &          \\
HIP81949    -3   &   $>16.81$   &   15.26   &   14.75   &   5.70   &   292.80   &      &      &      &          &   b   &          \\
HIP81949    -4   &   $>16.81$   &   15.67   &   15.52   &   5.27   &   340.72   &   ($>10.71$)   &   (9.57)   &   (9.42)   &   ($\approx$0.02)   &   ?   &          \\
HIP81949    -5   &   $>16.81$   &   16.52   &   $>15.93$   &   9.63   &   76.17   &      &      &      &          &   b   &   !       \\
HIP81949    -6   &   $>16.81$   &   15.62   &   14.82   &   6.26   &   239.37   &   ($>10.71$)   &   (9.52)   &   (8.73)   &   ($\approx$0.02)   &   ?   &          \\
HIP81949    -7   &   $>16.81$   &   $>16.84$   &   15.59   &   11.72   &   40.80   &      &      &      &          &   b   &   !       \\
HIP81949    -8   &   $>16.81$   &   $>16.84$   &   16.75   &   4.16   &   236.05   &      &      &      &          &   b   &   !       \\
HIP81949    -9   &   $>16.81$   &   $>16.84$   &   16.83   &   3.86   &   105.30   &      &      &      &          &   b   &   !       \\
HIP81949    -10   &   $>16.81$   &   $>16.84$   &   17.10   &   2.38   &   48.01   &      &      &      &          &   b   &   !       \\
HIP81949    -11   &   $>16.81$   &   $>16.84$   &   17.15   &   8.05   &   96.11   &      &      &      &          &   b   &   !       \\
HIP81949    -12   &   $>16.81$   &   $>16.84$   &   17.34   &   8.13   &   36.64   &      &      &      &          &   b   &   !       \\
\hline
HIP81972       &   5.82   &   5.89   &   5.87   &          &          &   -0.56   &   -0.49   &   -0.51   &   4.92   &   p   &          \\
HIP81972    -1   &   11.63   &   10.87   &   10.48   &   2.02   &   313.69   &   (5.25)   &   (4.49)   &   (4.10)   &   (0.67)   &   ?   &          \\
HIP81972    -2   &   11.30   &   10.97   &   10.61   &   7.02   &   258.81   &   (4.92)   &   (4.59)   &   (4.23)   &   (0.68)   &   ?   &          \\
HIP81972    -3   &   12.54   &   11.86   &   11.77   &   5.04   &   213.45   &   6.16   &   5.48   &   5.39   &   0.35   &   c   &   J       \\
HIP81972    -4   &   15.10   &   14.43   &   13.98   &   2.79   &   106.94   &   8.72   &   8.05   &   7.60   &   0.06   &   nc   &   JHK       \\
HIP81972    -5   &   16.11   &   15.63   &   15.26   &   7.92   &   229.27   &   9.73   &   9.25   &   8.88   &   $\approx$0.03   &   nc   &   JHK       \\
HIP81972    -6   &   $>16.58$   &   16.25   &   $>16.61$   &   8.79   &   167.71   &      &      &      &          &   b   &   H!       \\
HIP81972    -7   &   $>16.58$   &   17.12   &   $>16.61$   &   3.58   &   33.65   &      &      &      &          &   b   &   H!       \\
HIP81972    -8   &   $>16.58$   &   17.28   &   $>16.61$   &   7.44   &   265.65   &      &      &      &          &   b   &   H!       \\
\hline
HIP83542       &   5.34   &   4.91   &   5.38   &          &          &   -1.26   &   -1.69   &   -1.22   &   1.10   &   p   &          \\
HIP83542    -1   &   $^{\star\star}$   &   9.72   &   9.90   &   8.86   &   196.21   &   ($^{\star\star}$)   &   (3.12)   &   (3.30)   &   (0.91)   &   ?   &          \\
HIP83542    -2   &   $>15.54$   &   15.65   &   $>12.13$   &   9.84   &   156.45   &      &      &      &          &   b   &   H!       \\
\hline
  \multicolumn{12}{|l|}{{ADONIS targets with multi-color observations}} \\ 
\hline
HIP53701       &   6.30   &   6.37   &   6.48   &          &          &   0.79   &   0.86   &   0.97   &   2.84   &   p   &   ADO       \\
HIP53701    -1   &   9.05   &   8.76   &   8.86   &   3.88   &   75.81   &      &      &      &          &   b   &   ADO       \\
HIP53701    -2   &   13.06   &   12.93   &   13.04   &   6.57   &   120.05   &      &      &      &          &   b   &   ADO       \\
\hline
HIP76071       &   7.05   &   7.10   &   7.06   &          &          &   0.89   &   0.94   &   0.90   &   2.70   &   p   &   ADO       \\
HIP76071    -1   &   $>11.25$   &   11.28   &   10.87   &   0.69   &   40.85   &   $>5.09$   &   5.12   &   4.71   &   0.23   &   c   &   ADO       \\
\hline
HIP77911       &   6.67   &   6.71   &   6.68   &          &          &   0.81   &   0.85   &   0.82   &   2.80   &   p   &   ADO       \\
HIP77911    -1   &   12.68   &   12.20   &   11.84   &   7.96   &   279.25   &   6.82   &   6.34   &   5.98   &   0.09   &   c   &   ADO       \\
\hline
HIP78530       &   6.87   &   6.92   &   6.87   &          &          &   1.08   &   1.13   &   1.08   &   2.48   &   p   &   ADO       \\
HIP78530    -1   &   $>14.50$   &   14.56   &   14.22   &   4.54   &   139.69   &   ($>8.71$)   &   (8.77)   &   (8.43)   &   ($\approx$0.02)   &   ?   &   ADO       \\
\hline
HIP78809       &   7.41   &   7.50   &   7.51   &          &          &   1.65   &   1.74   &   1.75   &   2.03   &   p   &   ADO       \\
HIP78809    -1   &   11.08   &   10.45   &   10.26   &   1.18   &   25.67   &   5.32   &   4.69   &   4.50   &   0.30   &   c   &   ADO       \\
\hline
HIP78956       &   7.52   &   7.54   &   7.57   &          &          &   1.15   &   1.17   &   1.20   &   2.40   &   p   &   ADO       \\
HIP78956    -1   &   9.76   &   9.12   &   9.04   &   1.02   &   48.67   &   3.39   &   2.75   &   2.67   &   1.16   &   c   &   ADO       \\
\hline
HIP79124       &   7.16   &   7.14   &   7.13   &          &          &   1.11   &   1.09   &   1.08   &   2.48   &   p   &   ADO       \\
HIP79124    -1   &   11.38   &   10.55   &   10.38   &   1.02   &   96.18   &   5.33   &   4.50   &   4.33   &   0.33   &   c   &   ADO       \\
\hline
HIP79156       &   7.56   &   7.56   &   7.61   &          &          &   1.44   &   1.44   &   1.49   &   2.09   &   p   &   ADO       \\
HIP79156    -1   &   11.62   &   10.89   &   10.77   &   0.89   &   58.88   &   5.50   &   4.77   &   4.65   &   0.27   &   c   &   ADO       \\
\hline
HIP80238       &   7.45   &   7.45   &   7.34   &          &          &   1.83   &   1.83   &   1.72   &   1.94   &   p   &   ADO       \\
HIP80238    -1   &   7.96   &   7.66   &   7.49   &   1.03   &   318.46   &   2.34   &   2.04   &   1.87   &   1.67   &   c   &   ADO       \\

  \hline
\end{longtable}

\begin{longtable}{| l| c | rrr | }
  \caption{Criteria used to determine whether a secondary is a companion star or a background star. Results are listed for secondaries found around the 22~targets observed with NACO ({\em top part of the table}) and the 9~targets with multi-color observations in the ADONIS dataset ({\em bottom part of the table}). Columns 1 and 2 show the secondary designation and the status of the component as determined in this paper (c = companion star; ? = candidate companion star; b = background star). Columns $3-5$ show the compatibility of the location of the object in the color-magnitude diagrams with the isochrones in terms of $\chi^2$. Confirmed companions have $\chi^2 < 2.30$ and (confirmed) background stars have $\chi^2 > 11.8$. The other secondaries have $2.30 < \chi^2 < 11.8$ and are labeled ``candidate companion''. A substantial fraction of these candidate companions may in fact be background stars. Several faint ($K_S > 14$~mag) secondaries are only detected in one filter (thus have no $\chi^2$), and are all assumed to be background stars. \label{table: criteria}}\\  
  \hline
  Star & Status & $\chi^2_{J-K_S,M_{K_S}}$ & $\chi^2_{H-K_S,M_{K_S}}$ & $\chi^2_{J-H,M_J}$  \\
  \hline
  \endfirsthead
  \hline
  \multicolumn{5}{|l|}{\tablename\ \thetable{} -- continued from previous page} \\
  \hline
  Star & Status & $\chi^2_{J-K_S,M_{K_S}}$ & $\chi^2_{H-K_S,M_{K_S}}$ & $\chi^2_{J-H,M_J}$  \\
  \hline
  \endhead
  \hline
  \multicolumn{5}{|l|}{{Continued on next page}} \\ 
  \hline
  \endfoot
  \hline 
  \endlastfoot
  \hline
  \hline 
  HIP59502    -1   &   c  &   $2.11$    &    $1.09$    &    $0.19$        \\
HIP59502    -2   &   b  &   ---    &    $75.23$    &    ---        \\
HIP59502    -3   &   b  &   ---    &    ---    &    ---        \\
HIP60851    -1   &   b  &   $7.77$    &    $1.29$    &    $15.51$        \\
HIP60851    -2   &   b  &   $>52.81$    &    $0.19$    &    $>45.54$        \\
HIP60851    -3   &   ?  &   ---    &    $7.54$    &    ---        \\
HIP60851    -4   &   b  &   ---    &    $22.34$    &    ---        \\
HIP60851    -5   &   b  &   ---    &    $26.37$    &    ---        \\
HIP60851    -6   &   b  &   ---    &    ---    &    ---        \\
HIP60851    -7   &   b  &   ---    &    ---    &    ---        \\
HIP60851    -8   &   b  &   ---    &    $237.30$    &    ---        \\
HIP61265    -1   &   ?  &   $4.46$    &    $0.02$    &    $3.83$        \\
HIP61265    -2   &   b  &   $13.77$    &    $1.81$    &    $5.29$        \\
HIP61265    -3   &   b  &   ---    &    $13.70$    &    ---        \\
HIP61265    -4   &   b  &   ---    &    $66.72$    &    ---        \\
HIP61265    -5   &   b  &   ---    &    ---    &    ---        \\
HIP62026    -1   &   c  &   $0.91$    &    $0.09$    &    $0.24$        \\
HIP63204    -1   &   b  &   $71.62$    &    $9.64$    &    $17.02$        \\
HIP63204    -2   &   c  &   $1.76$    &    $0.01$    &    $1.38$        \\
HIP67260    -1   &   c  &   $0.01$    &    $0.00$    &    $0.01$        \\
HIP67260    -2   &   ?  &   ---    &    $7.20$    &    ---        \\
HIP67260    -3   &   ?  &   $1.45$    &    $3.59$    &    $5.27$        \\
HIP67919    -1   &   c  &   $0.02$    &    $0.86$    &    $0.39$        \\
HIP68532    -1   &   c  &   $0.93$    &    $1.10$    &    $0.00$        \\
HIP68532    -2   &   c  &   $1.06$    &    $0.02$    &    $1.70$        \\
HIP69113    -1   &   c  &   $1.26$    &    $0.09$    &    $0.63$        \\
HIP69113    -2   &   c  &   $0.56$    &    $0.04$    &    $1.38$        \\
HIP73937    -1   &   c  &   ---    &    $0.01$    &    ---        \\
HIP73937    -2   &   b  &   ---    &    $22.37$    &    ---        \\
HIP78968    -1   &   ?  &   $3.88$    &    $1.26$    &    $0.73$        \\
HIP79098    -1   &   b  &   $8.45$    &    $14.45$    &    $39.65$        \\
HIP79410    -1   &   b  &   $21.58$    &    $23.51$    &    $19.09$        \\
HIP79739    -1   &   c  &   $0.44$    &    $0.09$    &    $0.91$        \\
HIP79771    -1   &   c  &   $1.46$    &    $0.24$    &    $0.52$        \\
HIP79771    -2   &   nc  &   $0.00$    &    $0.04$    &    $0.02$        \\
HIP80142    -1   &   b  &   $164.54$    &    $59.81$    &    $33.27$        \\
HIP80142    -2   &   ?  &   ---    &    ---    &    $2.61$        \\
HIP80474    -1   &   b  &   $5.08$    &    $73.31$    &    $36.05$        \\
HIP80799    -1   &   c  &   $1.11$    &    $0.12$    &    $0.62$        \\
HIP80896    -1   &   c  &   $0.73$    &    $0.02$    &    $0.53$        \\
HIP81949    -1   &   b  &   $66.30$    &    $7.46$    &    $27.51$        \\
HIP81949    -2   &   b  &   $15.93$    &    $0.91$    &    $8.99$        \\
HIP81949    -3   &   b  &   $>13.09$    &    $0.02$    &    $>19.08$        \\
HIP81949    -4   &   ?  &   $>3.88$    &    $5.69$    &    $>5.59$        \\
HIP81949    -5   &   b  &   ---    &    ---    &    ---        \\
HIP81949    -6   &   ?  &   $>9.95$    &    $0.98$    &    $>6.88$        \\
HIP81949    -7   &   b  &   $>5.16$    &    $>11.84$    &    ---        \\
HIP81949    -8   &   b  &   ---    &    ---    &    ---        \\
HIP81949    -9   &   b  &   ---    &    ---    &    ---        \\
HIP81949    -10   &   b  &   ---    &    ---    &    ---        \\
HIP81949    -11   &   b  &   ---    &    ---    &    ---        \\
HIP81949    -12   &   b  &   ---    &    ---    &    ---        \\
HIP81972    -1   &   ?  &   $4.36$    &    $2.27$    &    $0.57$        \\
HIP81972    -2   &   ?  &   $1.42$    &    $1.08$    &    $6.03$        \\
HIP81972    -3   &   c  &   $0.56$    &    $1.65$    &    $0.20$        \\
HIP81972    -4   &   nc  &   $0.51$    &    $0.11$    &    $0.13$        \\
HIP81972    -5   &   nc  &   $0.90$    &    $0.07$    &    $0.40$        \\
HIP81972    -6   &   b  &   ---    &    $>16.02$    &    ---        \\
HIP81972    -7   &   b  &   ---    &    ---    &    ---        \\
HIP81972    -8   &   b  &   ---    &    ---    &    ---        \\
HIP83542    -1   &   ?  &   ---    &    $8.25$    &    ---        \\
HIP83542    -2   &   b  &   ---    &    ---    &    ---        \\
\hline
\multicolumn{5}{|l|}{ADONIS targets with multi-color observations}\\ 
\hline
HIP53701    -1   &   b  &   $12.01$    &    $3.72$    &    $0.80$        \\
HIP53701    -2   &   b  &   $33.11$    &    $8.54$    &    $8.16$        \\
HIP76071    -1   &   c  &   ---    &    $0.57$    &    ---        \\
HIP77911    -1   &   c  &   $0.73$    &    $0.00$    &    $0.74$        \\
HIP78530    -1   &   ?  &   ---    &    $3.10$    &    ---        \\
HIP78809    -1   &   c  &   $0.82$    &    $0.74$    &    $0.02$        \\
HIP78956    -1   &   c  &   $0.01$    &    $0.25$    &    $0.33$        \\
HIP79124    -1   &   c  &   $0.22$    &    $0.83$    &    $1.88$        \\
HIP79156    -1   &   c  &   $0.41$    &    $1.94$    &    $0.45$        \\
HIP80238    -1   &   c  &   $0.43$    &    $0.89$    &    $0.06$        \\

  \hline
\end{longtable}

\begin{longtable}{| l | rrr | rr | l|l |}
  \caption{All companion stars identified in our ADONIS and NACO binarity surveys among A and late-B stars in Sco~OB2 \citep[][and this paper]{kouwenhoven2005}. The columns show the \textit{Hipparcos} number of the primary star, the $JHK_S$ magnitudes, the angular separation, the position angle, the current status of the companion, and the date of observation (dd/mm/yy). If measurements are performed in both the ADONIS and NACO surveys, the NACO data are provided.
    The wide companion of HIP77315 at $\rho=37.37''$ is HIP77317, another member of Sco~OB2. These stars are found to be a common proper motion pair \citep{wds1997}, and were both observed in our ADONIS survey. The confirmed and candidate companions for which the status is determined using their $JHK_S$ photometry, are indicated with ``confirmed'' and ``inconclusive'', respectively. The candidate companions identified by \cite{kouwenhoven2005}, for which the status is determined using the $K_S=12$~mag criterion, and indicated with ``candidate'' here. Background stars are not listed here. \label{table: adonisnaco}}\\
  \hline
  Host primary & $J$ (mag) & $H$ (mag) & $K_S$ (mag) & $\rho$ (``) & PA ($^\circ$) & Companion status & Date \\
  \hline
  \endfirsthead
  \hline
  \multicolumn{8}{|l|}{\tablename\ \thetable{} -- continued from previous page} \\
  \hline
  Host primary & $J$ (mag) & $H$ (mag) & $K_S$ (mag) & $\rho$ (``) & PA ($^\circ$) & Companion status & Date \\
  \hline
  \endhead
  \hline
  \multicolumn{8}{|l|}{{Continued on next page}} \\ 
  \hline
  \endfoot
  \hline 
  \endlastfoot
  \hline
  HIP50520 &          &          &   6.39   &   2.51   &   313.32   &   candidate &            06/06/01        \\
\hline
HIP52357 &          &          &   11.45   &   10.04   &   72.69   &   candidate &           06/06/01         \\
HIP52357 &          &          &   7.65   &   0.53   &   73.01   &   candidate &             06/06/01       \\
\hline
HIP56993 &          &          &   11.88   &   1.68   &   23.07   &   candidate &            06/06/01        \\
\hline
HIP58416 &          &          &   8.66   &   0.58   &   166.12   &   candidate &            06/06/01        \\
\hline
HIP59413 &          &          &   8.18   &   3.18   &   99.83   &   candidate &             06/06/01       \\
\hline
HIP59502 &   12.35   &   11.83   &   11.64   &   2.94   &   26.39   &   confirmed      &    06/04/04   \\
\hline
HIP60084 &          &          &   10.10   &   0.46   &   329.64   &   candidate &           06/06/01         \\
\hline
HIP60851 &          &   13.63   &   13.69   &   8.16   &   231.46   &   inconclusive      &    06/04/04   \\
\hline
HIP61265 &   11.98   &   11.66   &   11.38   &   2.51   &   67.15   &   inconclusive      &    06/04/04   \\
\hline
HIP61639 &          &          &   7.06   &   1.87   &   182.40   &   candidate &            07/06/01        \\
\hline
HIP61796 &          &          &   11.79   &   9.89   &   108.98   &   candidate &           07/06/01         \\
HIP61796 &          &          &   11.86   &   12.38   &   136.77   &   candidate &          07/06/01          \\
\hline
HIP62002 &          &          &   7.65   &   0.38   &   69.24   &   candidate &             08/06/01       \\
\hline
HIP62026 &   8.08   &   7.90   &   7.86   &   0.23   &   6.34   &   confirmed      &    06/04/04   \\
\hline
HIP62179 &          &          &   7.57   &   0.23   &   282.75   &   candidate &            08/06/01        \\
\hline
HIP63204 &   8.79   &   8.51   &   8.40   &   0.15   &   236.56   &   confirmed      &    06/04/04   \\
\hline
HIP64515 &          &          &   6.94   &   0.31   &   165.69   &   candidate &            08/06/01        \\
\hline
HIP65822 &          &          &   11.08   &   1.82   &   303.87   &   candidate &           08/06/01         \\
\hline
HIP67260 &          &   14.04   &   14.10   &   1.23   &   355.65   &   inconclusive      &    28/04/04   \\
HIP67260 &   15.84   &   14.83   &   14.67   &   2.33   &   77.25   &   inconclusive      &    28/04/04   \\
HIP67260 &   8.88   &   8.46   &   8.36   &   0.42   &   229.46   &   confirmed      &    28/04/04   \\
\hline
HIP67919 &   9.98   &   9.38   &   9.10   &   0.69   &   296.56   &   confirmed      &    28/04/04   \\
\hline
HIP68080 &          &          &   7.19   &   1.92   &   10.20   &   candidate &             05/06/01       \\
\hline
HIP68532 &   10.52   &   9.85   &   9.54   &   3.05   &   288.50   &   confirmed      &    28/04/04   \\
HIP68532 &   11.38   &   10.94   &   10.63   &   3.18   &   291.92   &   confirmed      &    28/04/04   \\
\hline
HIP68867 &          &          &   11.61   &   2.16   &   284.76   &   candidate &           08/06/01         \\
\hline
HIP69113 &   10.98   &   10.43   &   10.29   &   5.34   &   65.15   &   confirmed      &    30/04/04   \\
HIP69113 &   11.27   &   10.45   &   10.30   &   5.52   &   67.17   &   confirmed      &    30/04/04   \\
\hline
HIP69749 &          &          &   11.60   &   1.50   &   0.84   &   candidate &             08/06/01       \\
\hline
HIP70998 &          &          &   10.83   &   1.17   &   354.60   &   candidate &           06/06/01         \\
\hline
HIP71724 &          &          &   9.70   &   8.66   &   23.02   &   candidate &             08/06/01       \\
\hline
HIP71727 &          &          &   7.80   &   9.14   &   244.96   &   candidate &            08/06/01        \\
\hline
HIP72940 &          &          &   8.57   &   3.16   &   221.58   &   candidate &            06/06/01        \\
\hline
HIP72984 &          &          &   8.50   &   4.71   &   260.35   &   candidate &            06/06/01        \\
\hline
HIP73937 &          &   8.46   &   8.37   &   0.24   &   190.58   &   confirmed      &    30/04/04   \\
\hline
HIP74066 &          &          &   8.43   &   1.22   &   109.62   &   candidate &            08/06/01        \\
\hline
HIP74479 &          &          &   10.83   &   4.65   &   154.15   &   candidate &           08/06/01         \\
\hline
HIP75056 &          &          &   11.17   &   5.19   &   34.51   &   candidate &            08/06/01        \\
\hline
HIP75151 &          &          &   8.09   &   5.70   &   120.87   &   candidate &            08/06/01        \\
\hline
HIP75915 &          &          &   8.15   &   5.60   &   229.41   &   candidate &            05/06/01        \\
\hline
HIP76001 &          &          &   7.80   &   0.25   &   3.17   &   candidate &              08/06/01      \\
HIP76001 &          &          &   8.20   &   1.48   &   124.82   &   candidate &            08/06/01        \\
\hline
HIP76071 &          &   11.28   &   10.87   &   0.69   &   40.85   &   confirmed      &    02/06/00, 07/06/01    \\
\hline
HIP77315 &          &          &   7.12   &   37.37   &   137.32   &   candidate &           08/06/01         \\
HIP77315 &          &          &   7.92   &   0.68   &   67.01   &   candidate &             05/06/01       \\
\hline
HIP77911 &   12.68   &   12.20   &   11.84   &   7.96   &   279.25   &   confirmed      &    02/06/00, 07/06/01    \\
\hline
HIP77939 &          &          &   8.09   &   0.52   &   119.13   &   candidate &            31/05/00        \\
\hline
HIP78530 &          &   14.56   &   14.22   &   4.54   &   139.69   &   inconclusive      &    02/06/00, 07/06/01    \\
\hline
HIP78756 &          &          &   9.52   &   8.63   &   216.40   &   candidate &            02/06/00        \\
\hline
HIP78809 &   11.08   &   10.45   &   10.26   &   1.18   &   25.67   &   confirmed      &    03/06/00, 07/06/01    \\
\hline
HIP78847 &          &          &   11.30   &   8.95   &   164.02   &   candidate &           03/06/00         \\
\hline
HIP78853 &          &          &   8.45   &   1.99   &   270.39   &   candidate &            08/06/01        \\
\hline
HIP78956 &   9.76   &   9.12   &   9.04   &   1.02   &   48.67   &   confirmed      &    03/06/00, 07/06/01    \\
\hline
HIP78968 &   14.96   &   14.51   &   14.26   &   2.78   &   322.13   &   inconclusive      &    04/05/04   \\
\hline
HIP79124 &   11.38   &   10.55   &   10.38   &   1.02   &   96.18   &   confirmed      &    03/06/00, 07/06/01    \\
\hline
HIP79156 &   11.62   &   10.89   &   10.77   &   0.89   &   58.88   &   confirmed      &    03/06/00, 07/06/01    \\
\hline
HIP79250 &          &          &   10.71   &   0.62   &   180.92   &   candidate &           03/06/00         \\
\hline
HIP79530 &          &          &   8.34   &   1.69   &   219.66   &   candidate &            31/05/00        \\
\hline
HIP79631 &          &          &   7.61   &   2.94   &   127.85   &   candidate &            05/06/01        \\
\hline
HIP79739 &   12.28   &   11.52   &   11.23   &   0.96   &   118.33   &   confirmed      &    19/06/04   \\
\hline
HIP79771 &   12.00   &   11.28   &   10.89   &   3.67   &   313.38   &   confirmed      &    19/06/04   \\
HIP79771 &   12.39   &   11.79   &   11.42   &   0.44   &   128.59   &   confirmed      &    19/06/04   \\
\hline
HIP80142 &   16.64   &   15.88   &          &   5.88   &   119.94   &   inconclusive      &    04/05/04   \\
\hline
HIP80238 &   7.96   &   7.66   &   7.49   &   1.03   &   318.46   &   confirmed      &    02/06/00, 07/06/01    \\
\hline
HIP80324 &          &          &   7.52   &   6.23   &   152.46   &   candidate &            31/05/00, 03/06/00        \\
\hline
HIP80371 &          &          &   8.92   &   2.73   &   140.65   &   candidate &            02/06/00, 03/06/00        \\
\hline
HIP80425 &          &          &   8.63   &   0.60   &   155.77   &   candidate &            08/06/01        \\
\hline
HIP80461 &          &          &   7.09   &   0.27   &   285.64   &   candidate &            31/05/00        \\
\hline
HIP80799 &   10.60   &   10.04   &   9.80   &   2.94   &   205.02   &   confirmed      &    05/05/04   \\
\hline
HIP80896 &   11.16   &   10.63   &   10.33   &   2.28   &   177.23   &   confirmed      &    08/06/04   \\
\hline
HIP81624 &          &          &   7.95   &   1.13   &   224.28   &   candidate &            05/06/01      \\
\hline
HIP81949 &          &   15.62   &   14.82   &   6.26   &   239.37   &   inconclusive      &    04/05/04, 05/05/04, 08/06/04, 25/06/04   \\
HIP81949 &          &   15.67   &   15.52   &   5.27   &   340.72   &   inconclusive      &    04/05/04, 05/05/04, 08/06/04, 25/06/04   \\
\hline
HIP81972 &   11.30   &   10.97   &   10.61   &   7.02   &   258.81   &   inconclusive      &    27/06/04   \\
HIP81972 &   11.63   &   10.87   &   10.48   &   2.02   &   313.69   &   inconclusive      &    27/06/04   \\
HIP81972 &   12.54   &   11.86   &   11.77   &   5.04   &   213.45   &   confirmed      &    27/06/04   \\
HIP81972 &   15.10   &   14.43   &   13.98   &   2.79   &   106.94   &   confirmed      &    27/06/04   \\
HIP81972 &   16.11   &   15.63   &   15.26   &   7.92   &   229.27   &   confirmed      &    27/06/04   \\
\hline
HIP83542 &          &   9.72   &   9.90   &   8.86   &   196.21   &   inconclusive      &    10/09/04   \\
\hline
HIP83693 &          &          &   9.26   &   5.82   &   78.35   &   candidate &             06/06/01       \\

\end{longtable}

\end{document}